\documentclass[12pt,twoside,english]{article}
\usepackage{ifthen}                
\newboolean{boldeqnitalic} 
\setboolean{boldeqnitalic}{true}  

\usepackage{caption}
\usepackage[cp1252]{inputenc}
\usepackage[intlimits] {amsmath}   
\usepackage{defjs2}                
\usepackage{epsfig}
\usepackage{psfig}
\usepackage{graphicx}
\usepackage{psfrag}
\usepackage[usenames]{color}
\usepackage{colortbl}
\usepackage{ifthen}
\usepackage{fancyhdr}
\usepackage{rotating}
\usepackage{multirow}
\usepackage{booktabs}              
\usepackage{amssymb}
\usepackage{amsbsy}
\usepackage{bm}                    

\usepackage{babel}
\usepackage{theorem}
\usepackage{upgreek}  
\usepackage{pifont}   
\usepackage{mathrsfs}
\usepackage{datetime}
\usepackage{tikz}
\usetikzlibrary{fadings}

\usepackage[all]{xy}
\usepackage{rotating}
\usepackage{subfig}
\usepackage[labelfont=bf, justification=justified]{caption}

\usepackage{algpseudocode}
\usepackage{algorithm}

\usepackage{wrapfig}

\definecolor{dunkelgrau}{rgb}{0.8,0.8,0.8}
\definecolor{hellgrau}{rgb}{0.90,0.90,0.90} 
\fancypagestyle{plain}{%
  \fancyhead{}%
  \fancyfoot[c]{\sffamily\thepage}%
}
\makeatletter                           
\def\cleardoublepage{\clearpage\if@twoside \ifodd\c@page\else
  \hbox{}
  \vspace*{\fill}
  \thispagestyle{empty}
  \newpage
  \if@twocolumn\hbox{}\newpage\fi\fi\fi}
\makeatother
\begin{document}
\unitlength1.0cm
\frenchspacing

\thispagestyle{empty}
 
\begin{center}    
	{\bf \large Deep Convolutional Neural Networks Predict} \\[2mm]
	{\bf \large Elasticity Tensors and their Bounds in Homogenization}  
\end{center} 
  
\vspace{8mm}
 
\ce{Bernhard Eidel}

\vspace{4mm}

\ce{\small DFG-Heisenberg-Fellow, Institute of Mechanics, Department Mechanical Engineering} 
\ce{\small University Siegen, 57068 Siegen, Paul-Bonatz-Str. 9-11, Germany} 
\ce{\small $^{\ast}$e-mail: bernhard.eidel@uni-siegen.de, phone: +49 271 740 2224
} 
\vspace{2mm}

\bigskip

%
%
%

%

\begin{center}
{\bf \large Abstract}

\bigskip

{\footnotesize
\begin{minipage}{14.5cm} 
\noindent
In the present work, 3D convolutional neural networks (CNNs) are trained to link random heterogeneous, two-phase materials of arbitrary phase fractions to their elastic macroscale stiffness thus replacing explicit homogenization simulations. In order to reduce the uncertainty of the true stiffness of the synthetic composites due to unknown boundary conditions (BCs), the CNNs predict the stiffness for periodic BCs, its upper bound through kinematically uniform BCs, and its lower bound through stress uniform BCs. The workflow of the homogenization-CNN is described, from the microstructure generation over the CNN design, the operations of convolution, nonlinear activation and pooling as well as optimization up to performance measurements in tests. The CNN predictions are very accurate even for  microstructure samples of the two-phase diamond-SiC coating material. The CNN that covers all three BCs is virtually as accurate as the separate treatment in three different nets. The CNNs of this contribution provide through stiffness bounds an indicator of the proper RVE size for individual snapshot samples. Moreover, they enable statistical analyses for the effective elastic stiffness on ensembles of synthetical microstructures without costly simulations. 
\end{minipage}
}
\end{center}

{\bf Keywords:}
Deep learning; Convolutional neural networks; Homogenization; Structure-property relations; Solid mechanics \hfill 


\section{Introduction} 
\label{sec:intro} 
  
As visual perception in the evolution of mammals is intimately related to the development of their cognitive abilities, mimicking some structural and process properties of visual perception has boosted artificial intelligence (AI) in its deep learning (DL) advancements by convolutional neural networks (CNN), the latter compared to the former in the blink of an eye. 
CNNs in image classification, still one of the major field of their application, disclose their insights in terms of their feature maps in the freely accessible, so-called hidden layers. Not only \emph{what} they have learned, for instance to identify a dog in an image, but also \emph{how} they have learned it, namely, by recognizing the nose, that largely makes a dog a dog in the image classification of \cite{Zeiler.2014}. The broad range of relevant applications of object recognition (most notably in large images) render CNNs important, the disclosure of learned lessons in an image language renders them fascinating for humans, since the sense of vision is likely their strongest. 

For a thorough introduction to deep-learning we refer to \cite{Goodfellow.2016}, for CNNs in particular to \cite{Goodfellow.2016}, Chap.~9, and \cite{Gonzalez.2018}. A comprehensive overview for ANNs with numerous references to original work until 2015 is given in \cite{Schmidhuber.2015}, recent advances in architectures of CNNs are discussed in \cite{Khan.2020}. For a recent presentation of the mathematics of deep learning with roots in learning theory \cite{Cucker.2007} we refer to  \cite{Berner.2021}.

The aim of the present work is to construct and train convolutional neural networks for predictions of the macroscopic elastic properties and their bounds for two-phase composites at arbitrary phase fractions and a wide range of morphologies.  

Composites in their general definition are multiphase materials that consist of at least two different constituents. When the features of their morphology live on much smaller length scales than the structure of interest at large, computational homogenization is instrumental for its favorable accuracy and high efficiency of microstructure sampling in comparison to full numerical simulation (FNS). In homogenization, the appropriate size of a microdomain in the case of non-periodic, random microstructures refers to the notion of representative volume elements (RVE) and effective properties. 
 
According to Hill \cite{Hill.1963} the microstructural features of the RVE must be statistical representative for the heterogeneous material and large enough to be insensitive to the applied boundary conditions which fulfill the requirement of equal, micro-macro energy density. The reference to the statistics of a particular property instead of all (material) properties for the definition of the RVE was introduced by Drugan and Willis \cite{Drugan.1996} resulted in smaller volumes matching the RVE requirement. The consideration of multiple sample and their statistics of apparent properties for the definition of effective properties was introduced by Kanit et al. \cite{Kanit.2003} and Jeulin \cite{Jeulin.2000}; it was shown that increasing a the samples' volume decreases both the bias, the deviation of the mean stiffness for different BCs, and the statistical variance. 
 
Statistical analyses for effective properties are typically based on randomly generated, synthetic samples, since real 3D microstructures from tomography and image acquisition are expensive and therefore limited in numbers. For suchlike snapshot samples the criterion for the RVE and effective properties is based on simulation results for different boundary conditions (BCs), periodic (PBC), kinematical uniform (KUBC), and stress/static uniform (SUBC) boundary conditions. According to Huet \cite{Huet.1990} stiffness deviations between different BCs render the properties ''apparent'', and ''effective'', if the size is sufficiently large for stiffness insensitivity to the applied BCs. Then, the VE can be considered an RVE. For effective properties the inequalities among stiffnesses for different boundary conditions $\mathbb{C}(\text{SUBC}) \leq \mathbb{C}(\text{PBC}) \leq \mathbb{C}(\text{KUBC})$ turn into equalities\footnote{Even for an equality it might turn out that the variance does not vanish, if the statistical distribution was available.} where the relations between positive fourth-order tensors are understood in terms of quadratic forms. 
 
Building links from microstructures to macroscopic properties by CNNs is a newly emerging but very active field of research in machine learning. The type of microstructures ranges from two-phase binaries created by Gaussian filters \cite{Yang.2018}, over spherical and ellipsoidal inclusions \cite{Rao.2020} to short fibre composites \cite{Breuer.2021}. A common denominator of all studies is the application of PBC, typically along with an a posteriori comparison with analytical bounds, of Voigt and Reuss in \cite{Yang.2018,Rao.2020} and with the Mori-Tanaka method in \cite{Breuer.2021}. Here, sharper bounds by KUBC and SUBC would enrich the CNN predictions with respect to the required RVE size and effective properties. 
 
{\bf Novel contribution.} \quad One novel aspect of the present work is to include stiffness bounds for PBC by SUBC and KUBC directly into the CNN instead of an a posteriori comparison with analytical bounds mentioned above. The question arises whether one single CNN for the augmented output of three elasticity tensors can catch up with the accuracy of three distinct CNNs. Here, we venture into unchartered terrain in view of existing results, which have considered one single elasticity tensor at maximum, \cite{Yang.2018,Yang.2019,Rao.2020,Breuer.2021}. Work on CNNs in the related fields of predicting effective conductivities or permeabilities cannot provide answers, since the number of output parameters was restricted to one up to three scalar quantities; to permeability from porous image/volume data \cite{Wu.2018,Tian.2020,Kamrava.2020}, to effective diffusivity \cite{Wu.2019}, to ionic conductivity \cite{Kondo.2017}, and to the triple of porosity, permeability, and tortuosity \cite{Graczyk.2020}.  

Including the stiffness bounds into the CNN predictions enables an interesting quantitative picture already for the full set of training data which exhibits a broad range of morphologies, arbitrary phase fractions at a stiffness contrast of factor 50; how sensitive to the applied BCs are the samples? Which microstructures show the strongest sensitivity, which ones the least? 

A second, novel and challenging aspect is that the CNNs shall be assessed not only for the microstructures stemming from the same generator as for training and validation, but also for the real, two-phase microstructure of a diamond/$\beta$-SiC composite. This material fabricated by chemical vapor deposition (CVD) is of interest for its outstanding mechanical properties such as hardness and wear-resistance. Therefore, diamond/$\beta$-SiC is used, e.g., as protective coatings for vulnerable substrates, but also in biological applications \cite{Yang.2015}. Since the CNNs have seen on their training track only randomly generated synthetic microstructures, but not any sample of this real material, the test goes beyond existing analyses of homogenization CNNs \cite{Yang.2018,Rao.2020, Breuer.2021,Li.2019} in that it addresses the intricate problem to generalize from synthetic to real microstructures without statistical links. 

\section{Generation of Two-Phase Microstructures}
\label{sec:microstructure-generator}
 
The set of cubic volume elements (VE) with different, two-phase microstructures serves as the input of the CNN. The generation of the VEs is carried out in three steps; (i) a random color code is assigned to each voxel in the VE of edge length $100$ voxels, (ii) a Gaussian filter is applied to the voxel number field, (iii) a binarization is carried out with a randomly selected phase fraction between 0 and 100\%. The number of VEs is $M=10^4$.

The three steps shall be briefly described. The 3D Gaussian is given by the equation 
\begin{equation}
\mathcal{N} (\bm x, \bm \mu, \bm C) = \dfrac{1}{\sqrt{(2\pi)^3 |\bm C|}} 
                            \, e^{\displaystyle -\dfrac{1}{2} 
                            	\left( \bm x - \bm \mu \right)^T
                            	\bm C^{-1}  
                            	\left( \bm x - \bm \mu \right) 
                            } \, , 
\end{equation} 
where $\bm x= [x \, y \, z]^T$ contains the coordinates of the considered voxel, 
and $\bm \mu$ represents the mean of the Gaussian $\bm \mu= [\mu_x \, \mu_y \, \mu_z]^T$. The symmetric $(3\times3)$-covariance matrix $\bm C$  
\begin{equation}
\bm C = 
\left[ 
\begin{array}{c c c}
s_x &  s_{xy}  &  s_{xz} \\
&  s_y     &  s_{yz} \\
\text{\scriptsize symm.}   &       &  s_z \\
\end{array}
\right]
\end{equation}
embodies variances $s_x$, $s_y$, $s_z$ on the diagonal, and covariances $s_{xy}$, $s_{xz}$ and $s_{yz}$ off-diagonal. We consider values for the variances $s_{x/y/z} \in [0.5,8]$ along with zero covariances.

\begin{figure}[htbp]
	\begin{minipage}{16.5cm}  
		\centering    
		\subfloat[{0.55/0.65/0.99}] 
		{\includegraphics[height=3.8cm, angle=0]{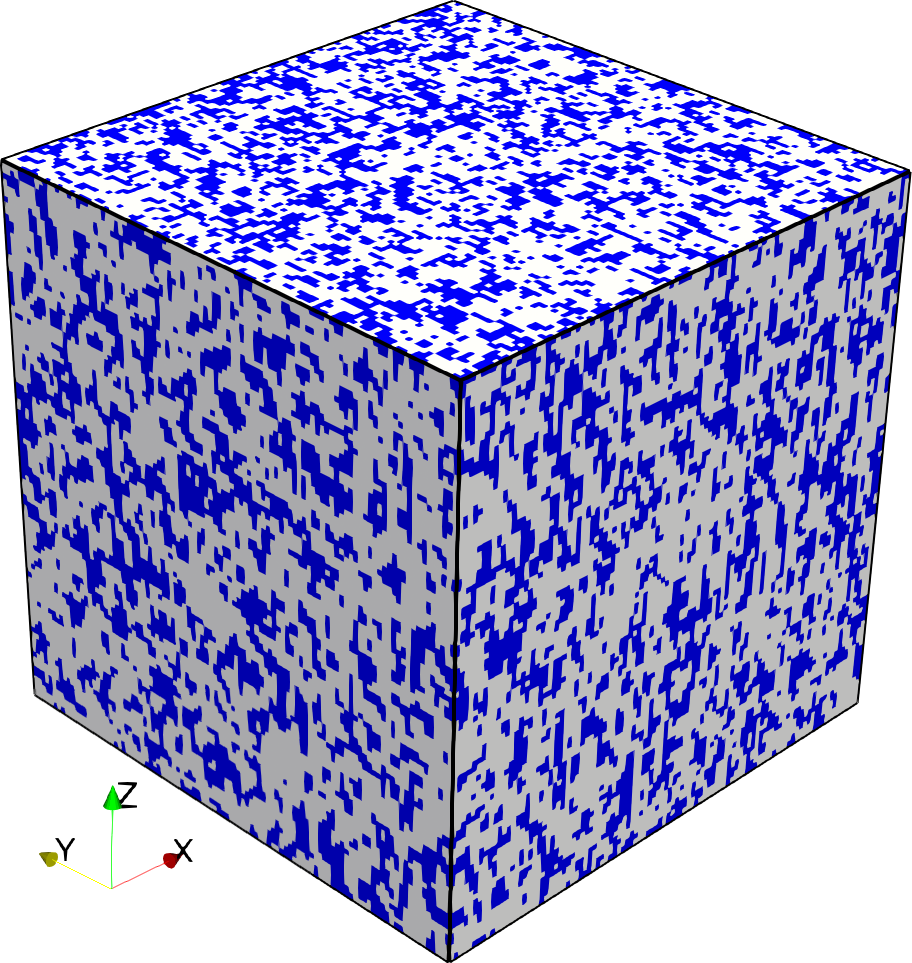}} 
		\hspace*{0.002\linewidth}
		\subfloat[2.31/2.19/2.00]  
		{\includegraphics[height=3.8cm, angle=0]{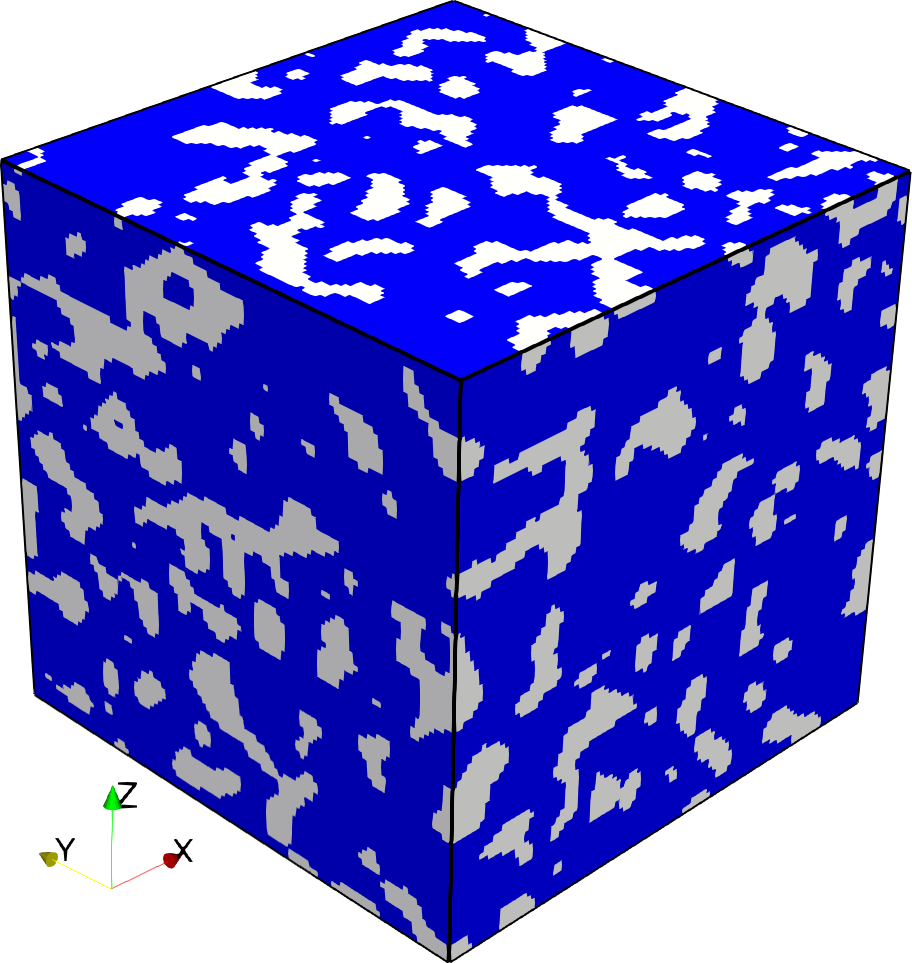}} 
		\hspace*{0.002\linewidth}
		\subfloat[5.85/5.85/5.92]  
		{\includegraphics[height=3.8cm, angle=0]{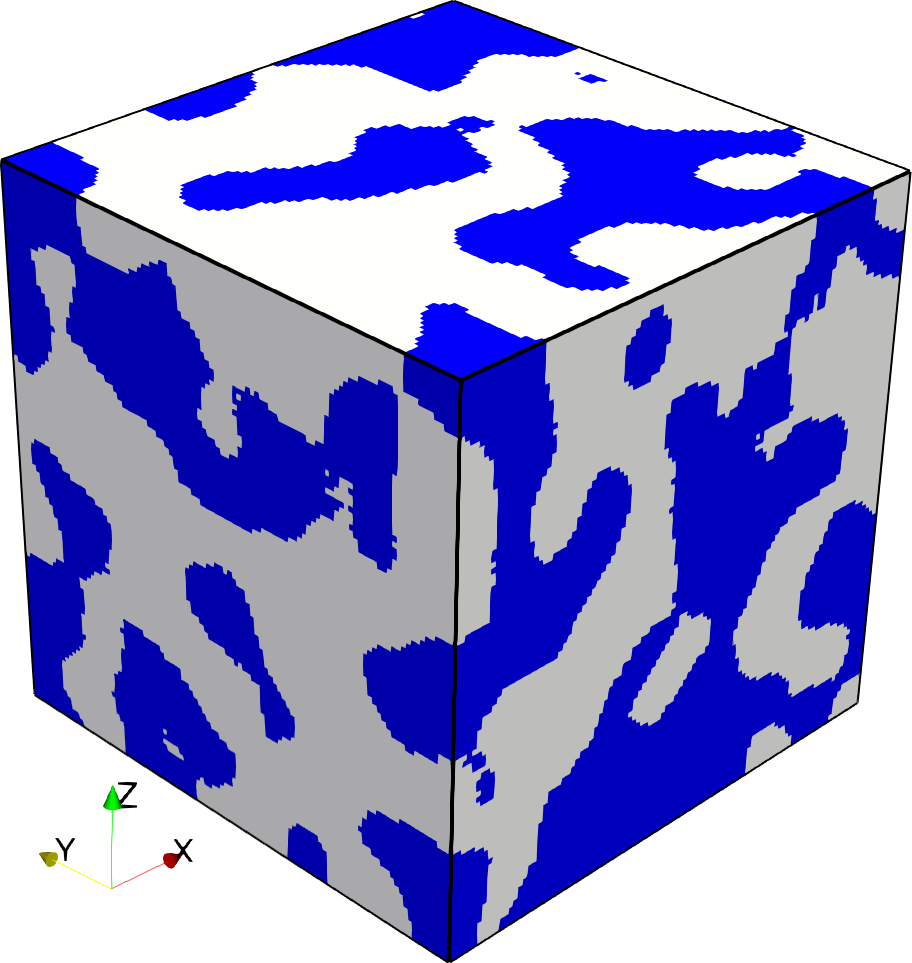}} 
		\hspace*{0.002\linewidth}
		\subfloat[7.73/7.77/7.79]    
		{\includegraphics[height=3.8cm, angle=0]{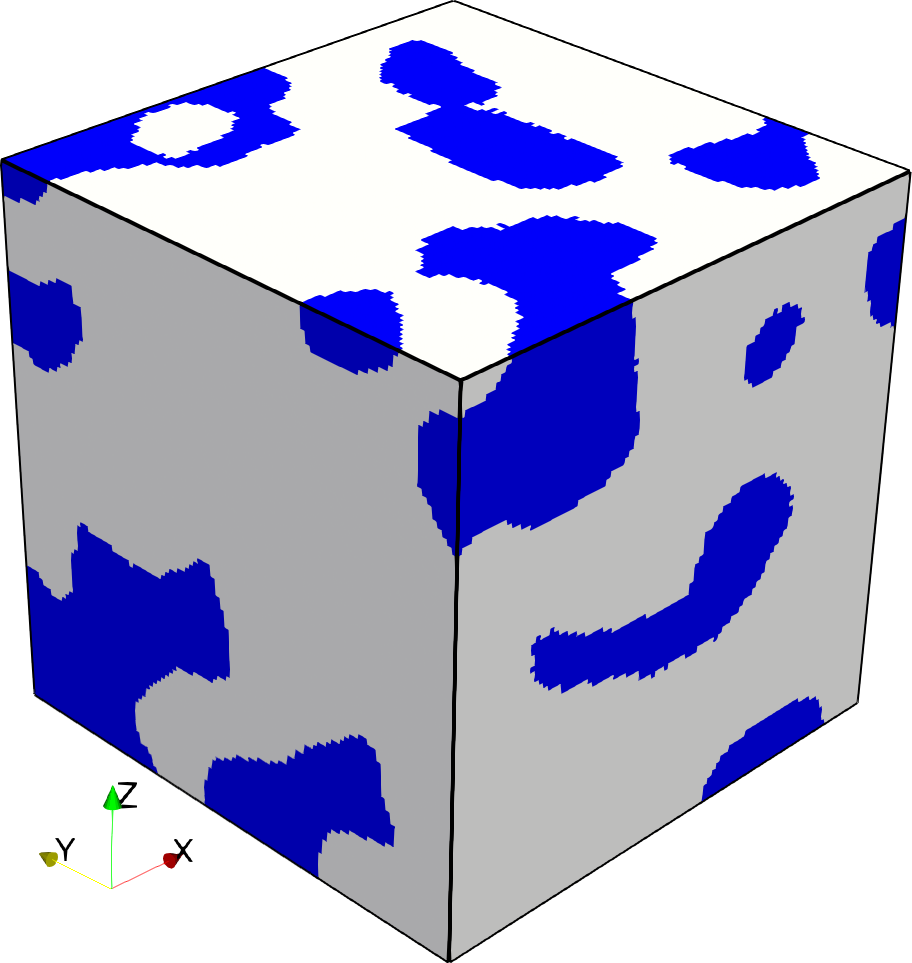}} 
		\\
		\subfloat[0.53/7.98/7.92] 
		{\includegraphics[height=3.8cm, angle=0]{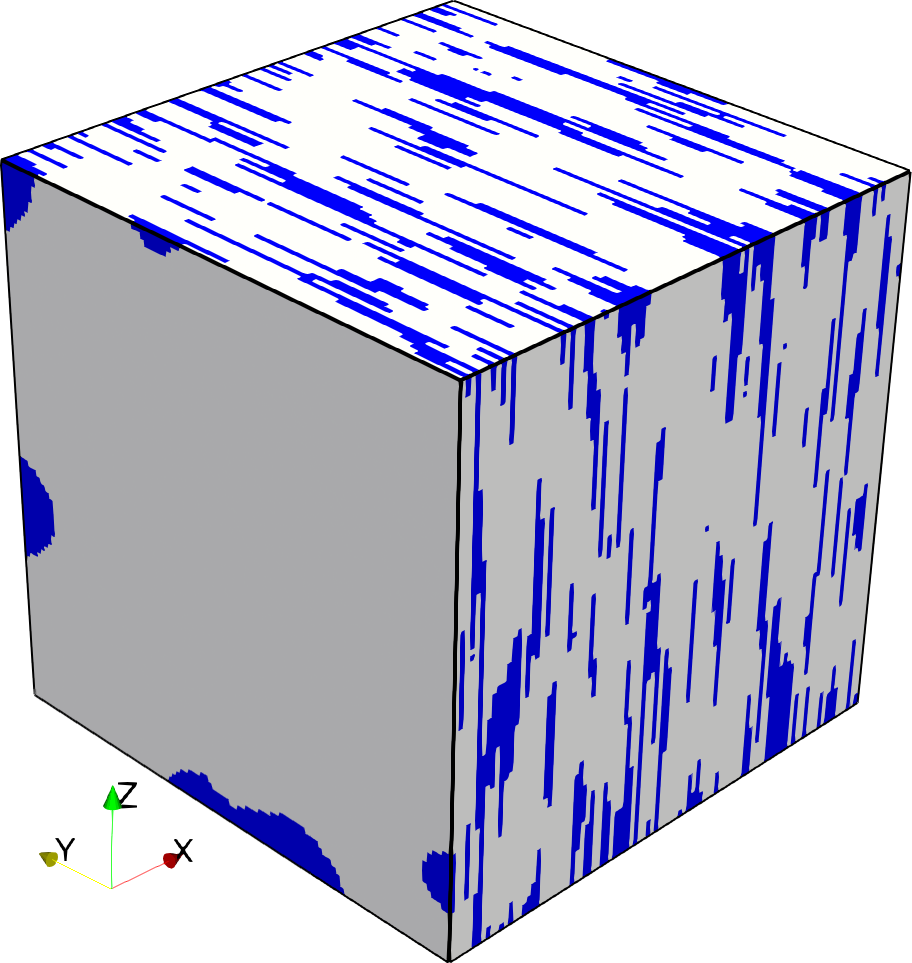}} 
		\hspace*{0.002\linewidth}
		\subfloat[7.71/0.62/7.87]  
		{\includegraphics[height=3.8cm, angle=0]{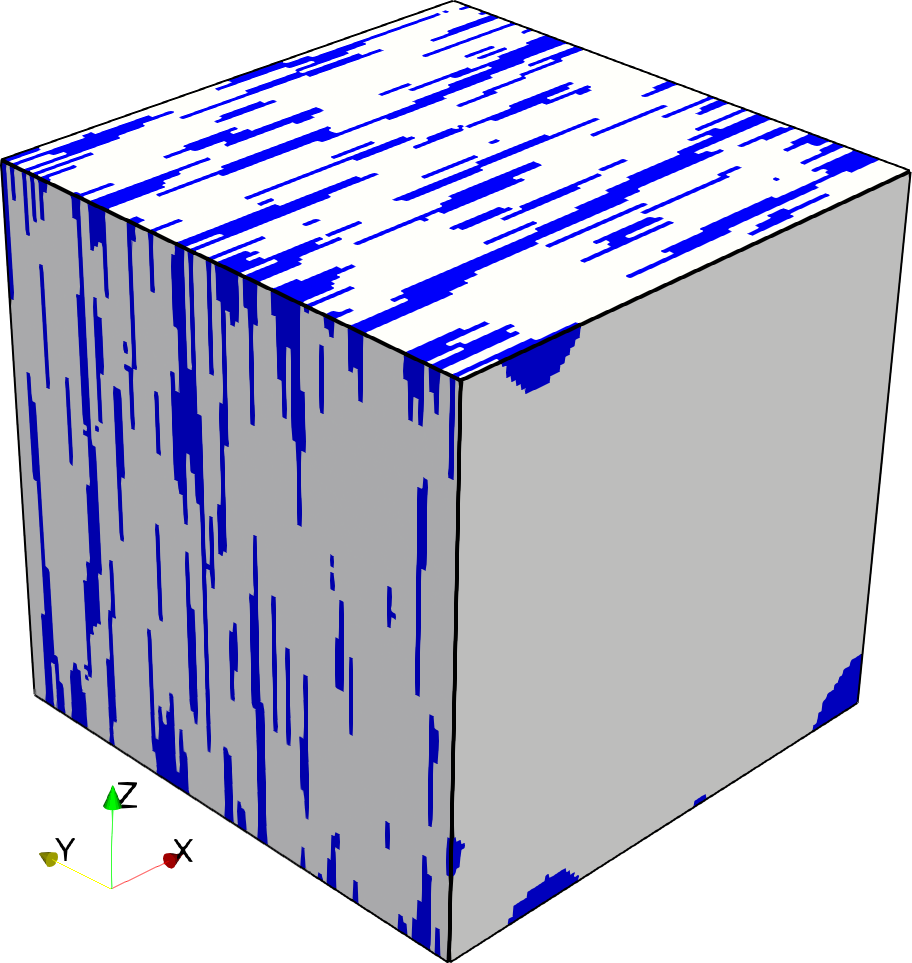}} 
		\hspace*{0.002\linewidth}
		\subfloat[0.51/0.55/5.77]    
		{\includegraphics[height=3.8cm, angle=0]{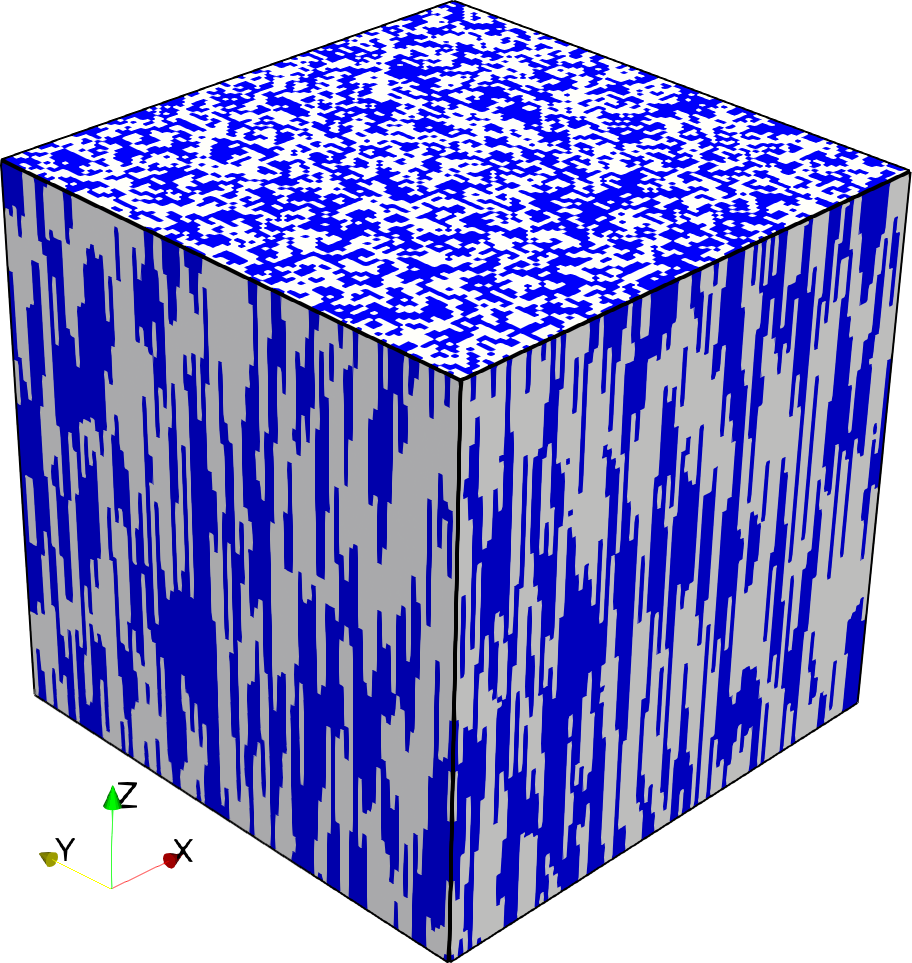}} 
		\hspace*{0.002\linewidth} 
		\subfloat[7.09/0.65/0.60]    
		{\includegraphics[height=3.8cm, angle=0]{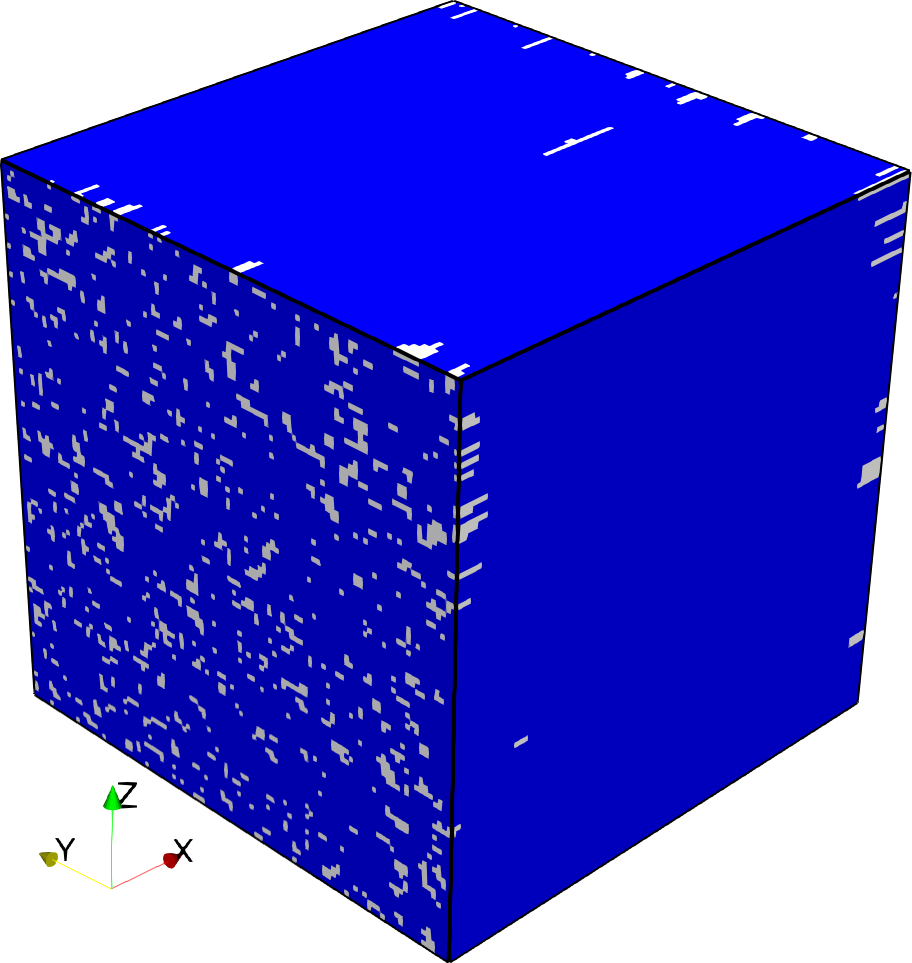}} 
		\\
		\subfloat[1.23/7.51/4.73] 
		{\includegraphics[height=3.8cm, angle=0]{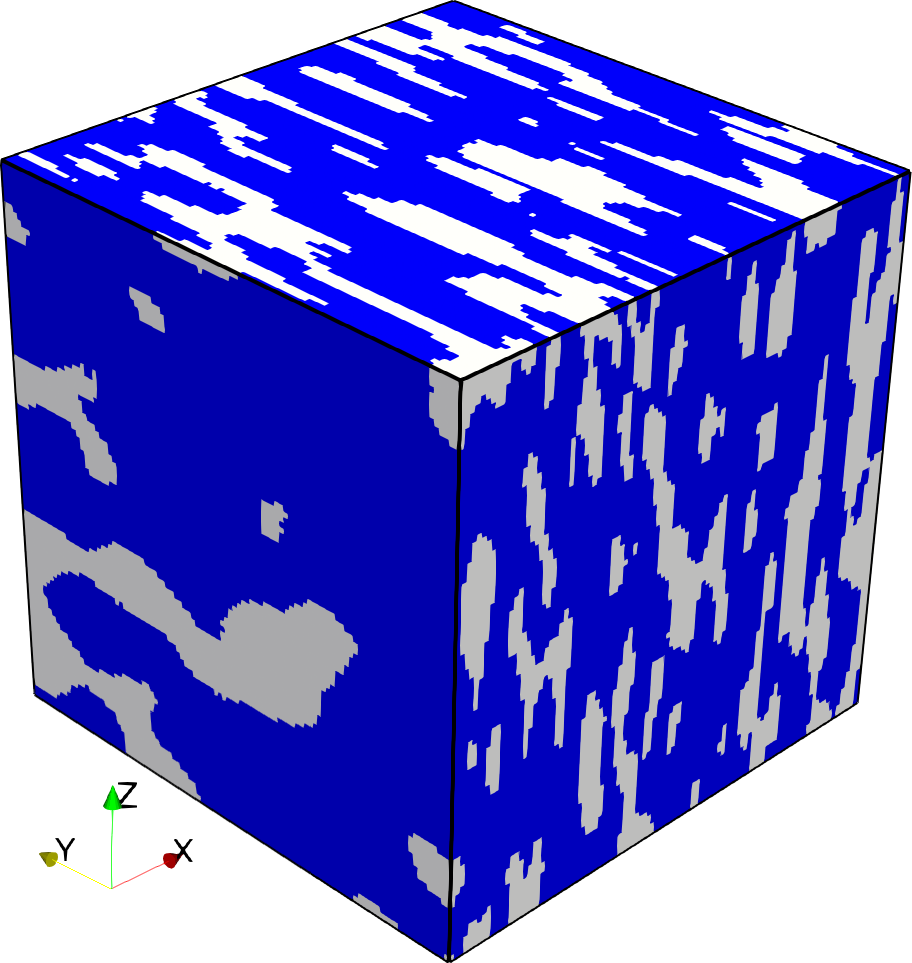}} 
		\hspace*{0.002\linewidth}
		\subfloat[0.5/5.80/5.94] 
		{\includegraphics[height=3.8cm, angle=0]{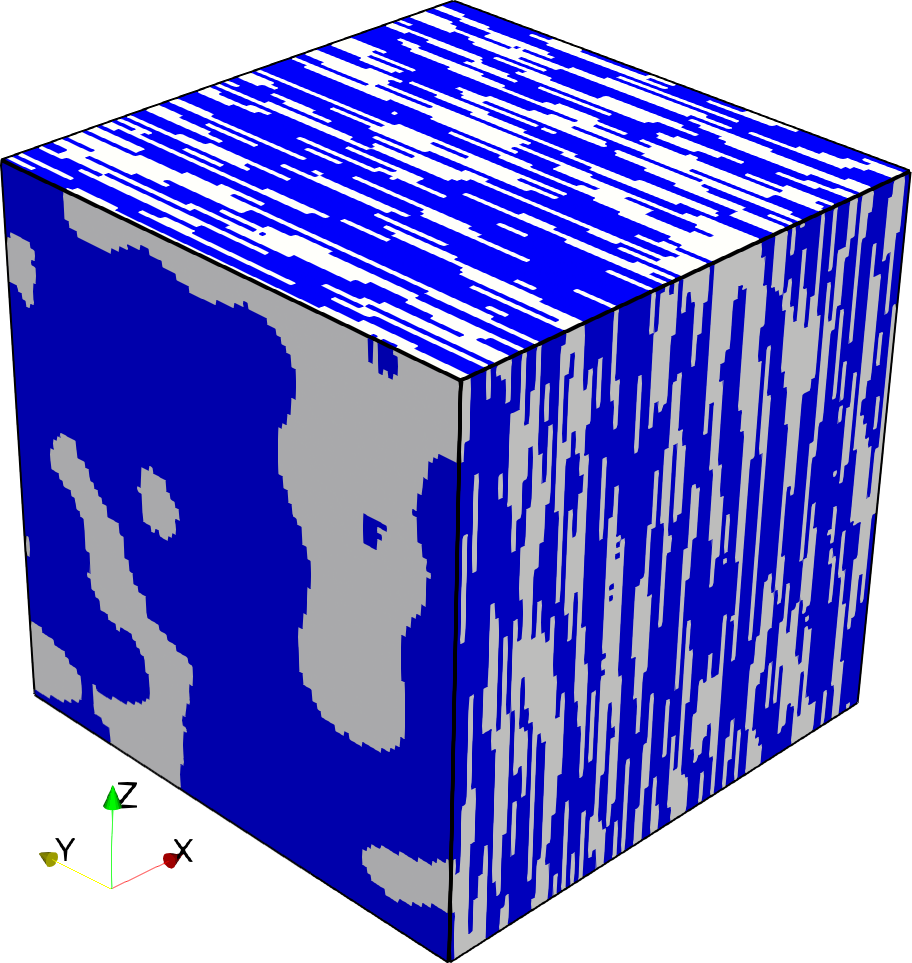}} 
		\hspace*{0.002\linewidth}
		\subfloat[1.17/5.04/7.74] 
		{\includegraphics[height=3.8cm, angle=0]{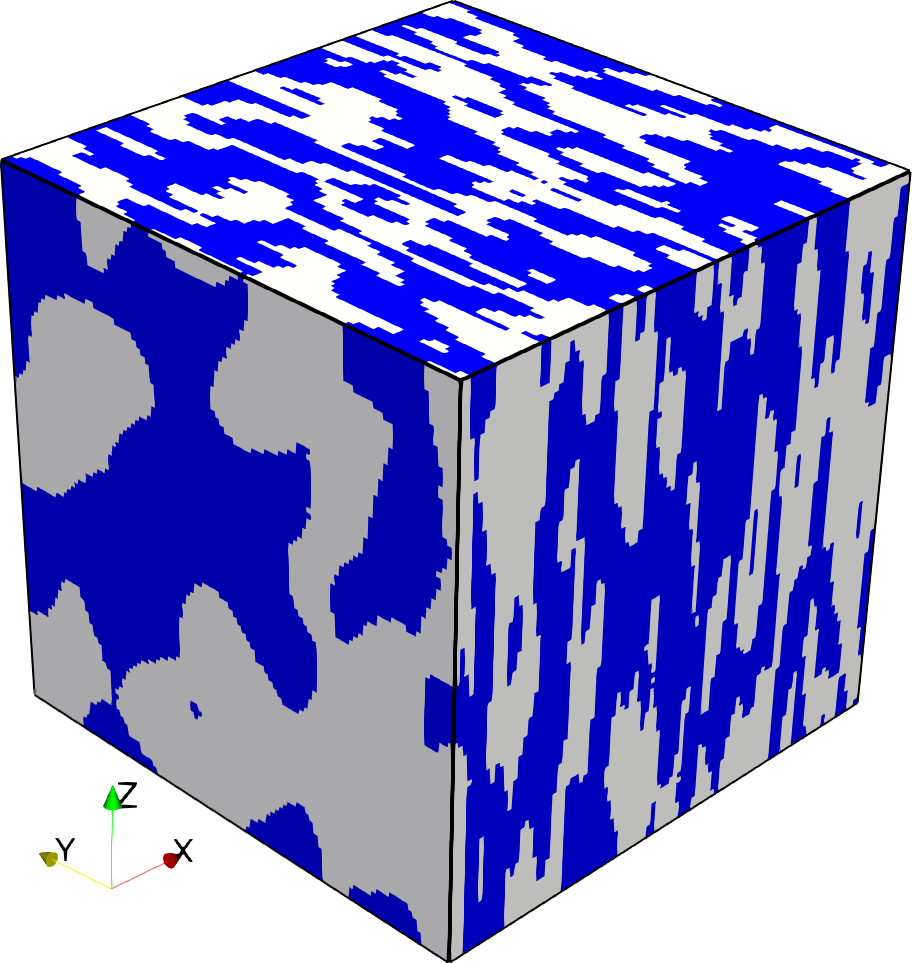}} 
		\hspace*{0.002\linewidth}
		\subfloat[7.89/7.78/5.23]  
		{\includegraphics[height=3.8cm, angle=0]{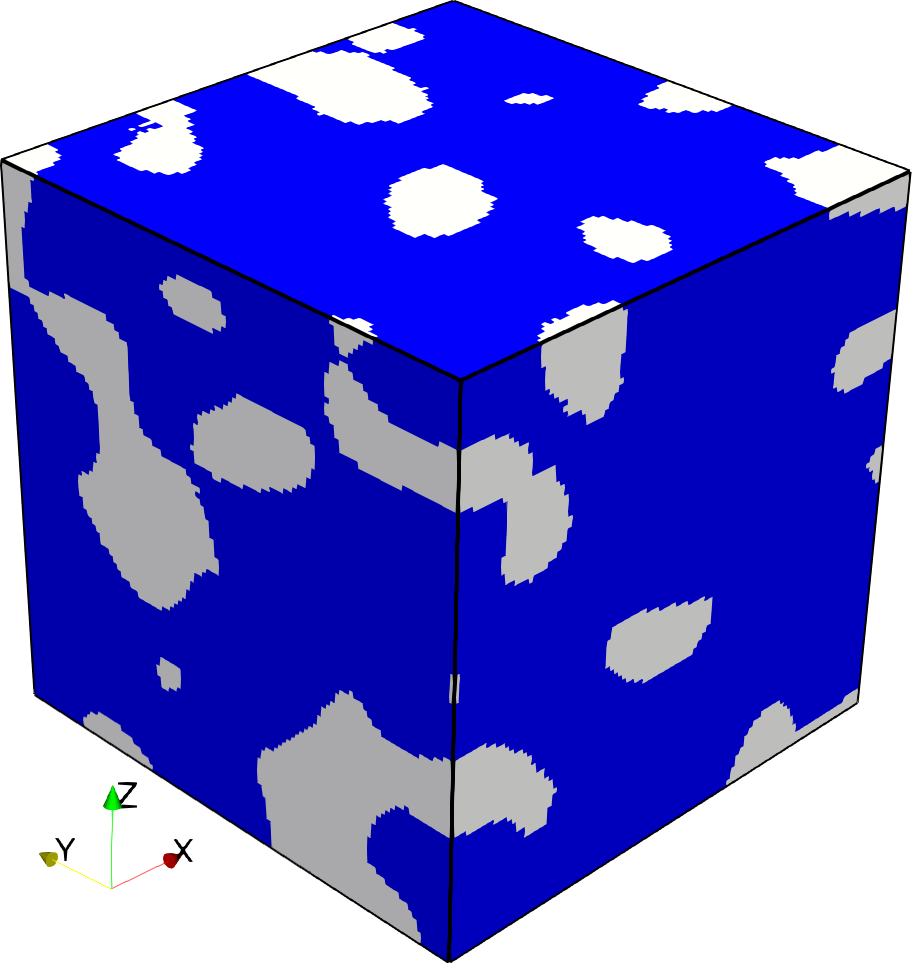}} 
		\hspace*{0.005\linewidth}
	\end{minipage}
	\caption{{\bf Two-phase microstructures.} Snapshot samples for different phase fractions indicate, how the variances (values of $s_x/s_y/s_z$ in the captions) of the Gaussian filter determine the type of microstructure independent of the largely varying phase fractions.
		\label{fig:microstructures}}
\end{figure}

The choice of the interval for variances and the phase fractions generate a large variety of different microstructure characteristics. Some of them are exemplarily visualized in Fig.~\ref{fig:microstructures}. 
 
A homogeneous dispersion of the phases is induced by values of similar magnitude for different variances as displayed in the Subfigs. (a)--(d). The magnitude of the variances determines the granularity of the microstructure. For fine dispersions, the apparent elastic properties can be expected to be isotropic or close to it. Pronounced lamellar structures are generated, when two variances are of the same magnitude and larger than the third one, (e), (f), which renders elastic properties close to orthotropic or cubic symmetry. Pronounced material orientations into a particular $i$-direction of the coordinate system are induced by a large variance value $s_i$ at smaller values for the other variances (g), (h), which results in an elastic behavior close to transversal isotropy. Triples of variances of different size generate more or less pronounced anisotropies, some examples are given in (i)--(l).
  
\section{Computational Homogenization with Bounds}
\label{sec:homogenization}

The elasticity tensors from homogenization simulations render the individual labels of the VEs (in total referred to as ground truth) for the latter training of the CNNs.     

Computational homogenization in terms of two-scale finite element methods solves in each integration point of a macro element the associated microproblem; prominent examples are the FE$^2$ method \cite{Michel.1999,Miehe.1999,Feyel.2000,Kouznetsova.2001,Peric.2011,Schroder.2014,Saeb.2016} and the FE-HMM \cite{Abdulle.2005,Abdulle.2009b,Abdulle.2012}. For the case of the first-order strain-driven computational homogenization the macroproblem drives the boundary value problem (BVP) of the microscale RVE by macroscopic deformation and, vice-versa, the microproblem provides (elastic) stiffness in terms of the macroscopic tangent and averaged microstresses, which implies that a constitutive law merely exists on the microscale.

To put things into perspective, the displacement field $\bm u$ which is the solution to the balance of linear momentum on the microscale $\text{div}\, \bm \sigma= \bm 0$ can be decomposed into displacements induced by a homogeneous infinitesimal strain $\overline{\bm \varepsilon}$ and superimposed microscale fluctuations $\widetilde{\bm w}$ according to $\bm u = \bm \varepsilon \, \bm x = \overline{\bm \varepsilon} \bm x + \widetilde{\bm w}$ as sketched in Fig. \ref{fig:PBCs}. The corresponding additive decomposition of strain and stress read as $\bm \varepsilon = \overline{\bm \varepsilon} + \widetilde{\bm \varepsilon}$ and $\bm \sigma = \overline{\bm \sigma} +  \widetilde{\bm \sigma}$, where the macrostress $\overline{\bm \sigma}$ is calculated as the volumetric mean of the microstresses $\overline{\bm \sigma} = 1/V \int_{\text{RVE}} \bm \sigma \, \text{d}V$.
 
The Hill-Mandel or macrohomogeneity condition \cite{Hill.1963} as the theoretical cornerstone of computational homogenization postulates the equality of the macroscale stress power with the average stress power in the RVE
\begin{equation}
\overline{\bm \sigma} : \dot{\overline{\bm \varepsilon}} = \dfrac{1}{V} \int_{\text{RVE}} \bm \sigma : \dot{\bm \varepsilon} \, \text{d}V \, 
\qquad      \longleftrightarrow   \qquad 
\dfrac{1}{V} \int_{\text{VE}} \bm \sigma : \dot{\bm \varepsilon} \, \text{d}V - \overline{\bm \sigma} : \dot{\overline{\bm \varepsilon}} = 0 \, .
\label{eq:Hill-Mandel-condition1}
\end{equation}

\subsection{Energetically consistent coupling conditions}
\label{subsec:e-consistent-coupling-conditions}

\begin{Figure}[htbp]
	\begin{minipage}{16.0cm}  
		\begin{center}
			\includegraphics[width=11.0cm, angle=0]{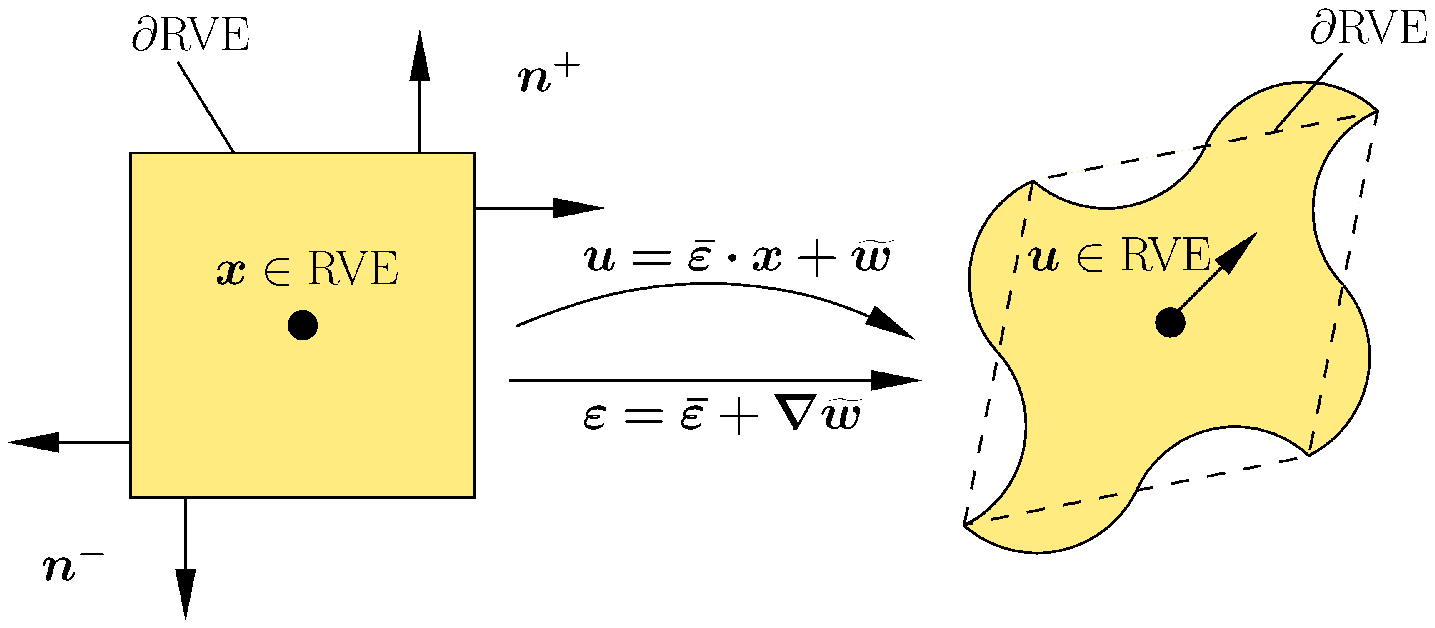}    
		\end{center}
	\end{minipage} 
	\\[8mm]
	\caption{{\bf Coupling conditions.} Periodic coupling conditions on the boundary $\partial \text{RVE}$ with the outward unit normal $\bm n$ of the RVE are displayed as solid lines in the deformed configuration, for linear Dirichlet coupling with $\widetilde{\bm w} \equiv \bm 0$ as dashed lines.
		\label{fig:PBCs}}
\end{Figure}

Constraint conditions consistent to the Hill-Mandel postulate are $\dot{\bm \varepsilon}:= \dot{\overline{\bm \varepsilon}}$ on the whole $\text{RVE}$ (Voigt condition) and $\bm \sigma = \overline{\bm \sigma}$ on the whole $\text{RVE}$ (Reuss condition). Moreover, linear Dirichlet, linear Neumann as well as periodic boundary conditions equally fulfill the condition. They can be obtained from the following results. 

\begin{equation}
\int_{\text{RVE}} {\bm \sigma} : \dot{{\bm \varepsilon}} \, \text{d}V 
=
\overline{\bm \sigma} : \dot{\overline{\bm \varepsilon}} + 
  \int_{\text{RVE}} (\overline{\bm \sigma} + \widetilde{\bm \sigma}) : \dot{\widetilde{\bm \varepsilon}} \, \text{d}V 
=  
\overline{\bm \sigma} : \dot{\overline{\bm \varepsilon}} + \int_{\text{RVE}} \widetilde{\bm \sigma} : \dot{\widetilde{\bm \varepsilon}} \, \text{d}V \, .
\label{eq:transormations-1}
\end{equation}
Inserting \eqref{eq:transormations-1} into \eqref{eq:Hill-Mandel-condition1}, applying the Gauss theorem with the surface tractions $\bm t$ on the boundary $\partial$RVE results in 
\begin{equation}
\int_{\text{RVE}} \widetilde{\bm \sigma} : \dot{\widetilde{\bm \varepsilon}} \, \text{d}V 
=
\int_{\partial\text{RVE}} \widetilde{\bm t} : \dot{\widetilde{\bm w}} \, \text{d}A 
=  
\int_{\partial\text{RVE}} (\bm t - \overline{\bm \sigma} \bm n) \cdot (\dot{\bm u} - \dot{\overline{\bm \varepsilon}} \bm x) \, \text{d}A  = 0\, ,
\label{eq:LinearBCs-Requirement}
\end{equation}
from which the BCs are obtained satisfying the macro-homogeneity condition.  
\begin{enumerate}
\item Kinematically uniform boundary conditions (KUBC)/Dirichlet BCs
\begin{equation}
\dot{\bm u}(\bm x) = \dot{\overline{\bm \varepsilon}} \, \bm x \quad \forall \bm x \in \partial \text{RVE} \, .  
\label{eq:KUBC}
\end{equation}
\item Stress uniform boundary conditions (SUBC)/Neumann BCs 
\begin{equation}
\bm t(\bm x) = \overline{\bm \sigma} \, \bm n \quad \forall \bm x \in \partial \text{RVE} \, . \label{eq:SUBC}  
\end{equation}	
\item Periodic boundary conditions (PBC)  

For the definition of PBC, the boundary $\partial\text{RVE}$ is split into pairwise periodic parts $\partial \text{RVE}^+$ and $\partial \text{RVE}^-$ with a corresponding sign convention for position vectors $\bm x^+$ and $\bm x^-$ and outward unit normal vectors $\bm n^+ = - \bm n^-$ as visualized in Fig.~\ref{fig:PBCs}. Periodic fluctuation displacements $\widetilde{\bm w}^+ = \widetilde{\bm w}^- = \widetilde{\bm w}$ along with \eqref{eq:LinearBCs-Requirement} result in 
\begin{equation}
	\int_{\partial\text{RVE}} \widetilde{\bm t} \cdot \dot{\widetilde{\bm w}} \, \text{d}A 
	=  
	\int_{\partial \text{RVE}^+} \widetilde{\bm t}^+ \cdot \dot{\widetilde{\bm w}} \, \text{d}A 
	+ 
	\int_{\partial \text{RVE}^-} \widetilde{\bm t}^- \cdot \dot{\widetilde{\bm w}} \, \text{d}A 
	=
	\int_{\partial\text{RVE}^-} (\widetilde{\bm t}^+ + \widetilde{\bm t}^-) \cdot \dot{\widetilde{\bm w}} \, \text{d}A  = 0\, ,
	\label{eq:Periodic-BCs}
	\end{equation}
	which implies the additional condition $\widetilde{\bm t}^+ = - \widetilde{\bm t}^-$; in conclusion, it holds 
	\begin{equation}
	\widetilde{\bm w}^+ = \widetilde{\bm w}^- \qquad \mbox{and} \qquad \widetilde{\bm t}^+ = - \widetilde{\bm t}^- \qquad \forall \bm x \in \partial \text{RVE} \, .
	\end{equation}
\end{enumerate}

\subsection{Stiffness relations}
 
Different BCs result in different moduli in Hooke's law of linear elasticity $\overline{\bm \sigma} = \mathbb{C} \, \overline{\bm \varepsilon}$; they are related by inequalities that turn into equalities for effective properties
\begin{equation}
\mathbb{C}(\text{SUBC}) \leq \mathbb{C}(\text{PBC}) \leq \mathbb{C}(\text{KUBC}) \, . 
\label{eq:Stiffness-hierarchy}
\end{equation}
The inequalities in \eqref{eq:Stiffness-hierarchy} between positive fourth-order tensors  are understood in terms of quadratic forms, i.e. $\mathbb{C}_a \geq \mathbb{C}_b \, \, \Leftrightarrow \, \, \bm \varepsilon :  \mathbb{C}_a \, \bm \varepsilon \geq \bm \varepsilon : \mathbb{C}_b \, \bm \varepsilon \, \forall \, \bm \varepsilon \, .$

Volumes of heterogeneous materials with arbitrary microstructure are rarely periodic and therefore do rarely deform according to PBC. For non-periodicity, the true stiffness depends on the continuation of the considered microdomain in all three directions of space. If the stiffness response considerably depends on the applied BCs and the spatial continuation of the VE is available, the postulate for representativeness suggests to increase the VE size. If the VE can not be increased in size, the true stiffness remains uncertain. The uncertainty however is bounded, SUBC and KUBC provide lower and upper bounds to the true apparent stiffness. These bounds do not necessarily bound the effective stiffness. 
 
For the VEs computational homogenization is carried out for the three above mentioned energetically consistent boundary conditions 1.--3.. The corresponding homogenized elasticity tensors $\mathbb{C}(\text{KUBC})$, $\mathbb{C}(\text{PBC})$, and $\mathbb{C}(\text{SUBC})$ are obtained by the microsolver of the Finite Element Heterogeneous Multiscale Method (FE-HMM). For the mathematical foundation of FE-HMM in linear elasticity we refer to \cite{Abdulle.2006} for a finite element formulation and aspects of implementation to \cite{Eidel.2018}. The boundary conditions in the homogenization simulations are fulfilled through the Lagrange multiplier method as described in \cite{Eidel.2018} for PBC, for KUBC and by a novel, simple and efficient formulation for SUBC in \cite{Fischer.2019b}. 

\subsection{Range of validity and limitations}
\label{subsec:Validity-and-limitations-9moduli}

For the material behavior we assume linear elasticity. For general triclinic materials, hence without any symmetry, the elasticity matrix exhibits 21 independent moduli. The matrix form of Hooke's law reads in Voigt notation 

\begin{equation}
\left[\begin{array}{c}
\sigma_{11} \\ \sigma_{22} \\ \sigma_{33} \\ \sigma_{12} \\ \sigma_{23} \\ \sigma_{13}
\end{array}\right]
 =
\displaystyle \underbrace{\left(\begin{array}{c c c c c c}
	C_{11}  &  C_{12}  &  C_{13}  & {\color{blue}C_{14}}  &   {\color{blue}C_{15}} &   {\color{blue}C_{16}}   \\ 
	        &  C_{22}  &  C_{23}  & {\color{blue}C_{24}}  &  {\color{blue}C_{25}} &   {\color{blue}C_{26}}  \\ 
	        &          &  C_{33}  &   {\color{blue}C_{34}}  &   {\color{blue}C_{35}}  &   {\color{blue}C_{36}}  \\ 
         	&          &          &  C_{44}  &   {\color{red}C_{45}}  &  {\color{red}C_{46}} \\
         	&          &          &          &  C_{55}  &  {\color{red}C_{56}} \\
	sym.    &          &          &          &          &  C_{66}   \\   
	\end{array}\right)}_{\substack{\text{Elastic stiffness matrix \,} \mathbf{\mathbb C} }} 
\left[\begin{array}{c}
\varepsilon_{11} \\ \varepsilon_{22} \\ \varepsilon_{33} \\ 2\varepsilon_{12} \\ 2\varepsilon_{23} \\ 2\varepsilon_{13}
\end{array}\right]
\label{eq:3D-triclinic}
\end{equation}
  
The CNN is trained to predict 9 components which implies the treatment of elastic coupling effects (CE) as follows:
\begin{enumerate}
	\item[CE-1] Considered are the direct effect of normal strains on normal stresses through ${C}_{11}, {C}_{22}, {C}_{33}$, the direct effect of shear strains on normal stresses through ${C}_{44}, {C}_{55}, {C}_{66}$, and the Poisson's effect through ${C}_{12}, {C}_{13}, {C}_{23}$. 
	\item[CE-2] Not considered are coupling effects of normal strains on shear stresses through ${C}_{ij}, i=1, 2, 3, j=4, 5, 6$ (in blue color in \eqref{eq:3D-triclinic}). 
	\item[CE-3] Not considered are the coupling effect of shear strain on shear stress in perpendicular planes (corresponding to the Poisson's effect for normal stress and strain in perpendicular directions) through ${C}_{45}, {C}_{46}, {C}_{56}$ (in red color in \eqref{eq:3D-triclinic}). 
\end{enumerate}
  
Considering the above 9 elasticity components and discarding the others is based on two assumptions related to the material symmetries and their representation, \cite{Ting.1996}, \cite{Vannucci.2018}:
\begin{enumerate}
	\item[Asm-1] The considered microstructures are restricted to materials of orthotropic, tetragonal, transversely isotropic or cubic symmetry or are even isotropic.
	\item[Asm-2] The anisotropic materials with the symmetries of Asm-1 are presented and analyzed in their symmetry basis. 
\end{enumerate} 
The symmetry classes of Asm-1 share the same skyline of zero and nonzero entries in the elasticity matrix in the symmetry basis system, but differ in dependencies of the elasticity moduli. In other generic bases the above material classes exhibit in general a fully populated elasticity matrix, where those coupling effects, which vanish in the symmetry basis, are active. A post-processing of the CNN prediction identifying these dependencies can inform about the particular symmetry class. 

The limitation of the present approach in the format of the elasticity matrix is consistent with the type of generated microstructures, since the variation of the variances imply symmetry planes that are parallel to the planes of the generating coordinate system or obtained by an additional axis rotation.  
  
\section{3D Convolutional Neural Networks}

\begin{Figure}[htbp] 
	\begin{minipage}{17cm}  
		\centering   
		\includegraphics[width=16.0cm, angle=0, clip=]{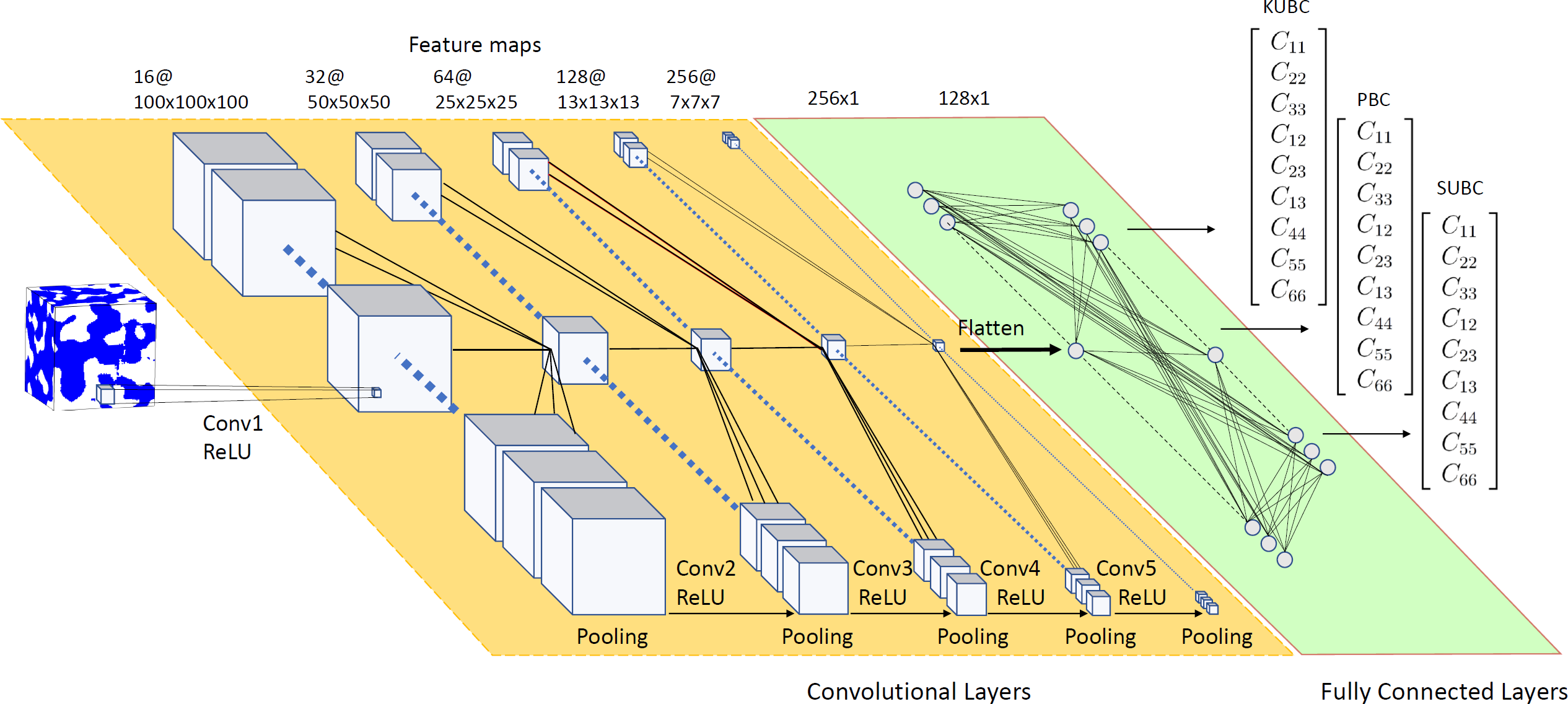} 
	\end{minipage}
	\caption{{\bf 3D-CNN architecture for homogenization with bounds.} From the input layer of a two-phase microstructure to the output layer of elasticity components for the coupling conditions of KUBC, PBC, and SUBC, in between a stack of convolutional layers and fully connected layers (FC). Each convolutional layer contains the operations of activation, nonlinear activation (here by ReLU) and pooling. The transition to the FC block requires a flattening of 3D arrays into a vector.
    \label{fig:CNN}}
\end{Figure}

\subsection{General}

The architecture of the present 3D-CNN for homogenization is displayed in Fig.~\ref{fig:CNN} and briefly described in the caption. The input layer of the CNN consists of a random variable $\bm X$, the voxel data of the VE, the output layer of a random variable $\bm Z$, the elastic moduli. The output is the result of an input-output mapping $\bm Z = \bm Z(\bm X, \bm \Theta)$ depending on the parameters $\bm \Theta=(\bm w, \bm b)$ of the weights $\bm w$ and the biases $\bm b$. In an approach of supervised learning the net is trained to predict the target variable $\bm Y$, also referred to as label, which is equally random since the homogenization computation operates on $\bm X$, $\bm Y= \bm Y(\bm X)$. Since the error function (also called loss or cost function) measures the distance between $\bm Y$ and $\bm Z$, a reasonable choice for the loss is given by the expectation $\mathbb{E}$ of the squared difference of $\bm Y$ and $\bm Z$
\begin{equation} 
\mathcal{L}(\bm \Theta) = \mathbb{E}[(\bm Y-\bm Z)^2]= \mathbb{E}[(\bm Y (\bm X) - \bm Z(\bm X, \bm \Theta))^2] \, . 
\label{eq:LossF=expectedDiffYZ}
\end{equation}
The objective of training is finding the parameters $\bm \Theta$ for which the cost function becomes minimal, hence $\bm \Theta^{\ast} = {\operatorname{arg\,min}}_{\bm \Theta} \, \mathcal{L}(\bm \Theta)$. In practice, optimization is carried out for observations or measurements of $\bm X$ and $\bm Y$ given by $\{(\bm x_i, \bm y_i)\}_{i=1}^M$. The fact that each out of $L$ layers exhibits its individual parameter set $\bm \Theta = \{ \bm \Theta^l\}=\{ \bm w^l, \bm b^l\}_{l=1}^{L}$ is consequential for minimizing the cost function giving rise to the so-called backpropagation. 
 
The processing of input data and their conversion through the hidden layers are described in the following. 

\subsection{Classifying the task}
Before, a comment on classification is in order. Following the scheme of ML approaches \cite{Goodfellow.2016}, the present CNN for homogenization can be understood as a task for a structured output, since the output is a vector that exhibits relationships between its elements. The relationships are established by existing material symmetries, compare Asm-1 in Sec.~\ref{subsec:Validity-and-limitations-9moduli}. For instance, if the elastic properties are close to isotropy, then all components of the elasticity tensors merely depend on the Lam\'{e}-constants $\lambda$ and $\mu$. A second type of relationships between the three vectors in the output of Fig.~\ref{fig:CNN} is established by homogenization applied to the same sample but for different BCs.
 
The task for a structured output is not too far from the task of classification, where the CNN is trained to assign one out of several categories to an input. The output is then either a number which identifies a category or it is a probability distribution over classes. The most prominent example is image identification and classification as well as object recognition in images, the applications for which CNNs have become famous \cite{LeCun.1989,Krizhevsky.2012} and many more.  

An important difference between the tasks of these two cases is that classification requires invariance with respect to translation, rotation and scaling\footnote{Pooling over spatial regions produces invariance to translations \cite{Goodfellow.2016}, a strategy for invariance to rotations is data augmentation; similarly, scaling invariance is supported by adding random crops of input images.}. For the present case of homogenization, these invariances would falsify the predictions for obvious reasons.

\subsection{Convolution}
 
A 3D filter (also referred to as kernel) endowed with the weights obtained from training ''scans'' the phase voxels and applies convolutional operation to produce the feature map. The feature maps reflect the salient features of the CNN training objective. 

Formally, the input $\bm x^{l-1}$ of layer $l$ is subject to an affine transformation of the form
\begin{equation}
 \bm z^{l} = \bm w^{l} * \bm x^{l-1} + \bm b^{l} \, ,
 \label{eq:convolution}
\end{equation}
where $*$ is the convolution symbol, $\bm w^{l}$ denotes the weights of a convolution filter and $\bm b^{l}$ the biases; $\bm w^{l}$ and $\bm b^{l}$ can be incorporated in the parameter set $\bm \Theta^{l}$ as mentioned above. In fact, the application of a convolution can be cast into an affine transformation of the form $\bm z^{l} = \bm W^{l} \bm x^{l-1} + \bm b^{l}$; where $\bm W^{l}$ is a matrix with a particular structure. Since the result $\bm z^{l}$ undergoes activation $\bm a^{l}=f(\bm z^{l})$ by a nonlinear function $f$, where the outcome serves as input in the consecutive convolutional layer, hence $\bm x^{l-1} \rightarrow \bm z^{l} \rightarrow \bm a^{l} =: \bm x^{l}$, the notation is adjusted to this procedure in the following. 

The representation \eqref{eq:convolution}, though correct and frequently used, does hide many relevant properties of convolutional operations. Therefore, we prefer a more transparent form; technically, in 3D convolution a kernel as a 3D array of $R^l$ rows ($0 \leq r \leq R^l-1$), $C^l$ columns ($0 \leq c \leq C^l-1$), and a depth of $D^l$ ($0 \leq d \leq D^l-1$) sums up the product of its weights $w^{lmq}_{rcd}$ at position $(r,c,d)$ with the entry  $x_{(i+{\color{black}r})(j+{\color{black}c})(k+{\color{black}d})}^{(l-1)q}$ of the $q$th feature map of the previous layer $(l-1)$ in terms of \eqref{eq:3d-conv-FeatureMap}
\begin{eqnarray}
\text{3D:} \qquad x_{ijk}^{lm} &=& f \left(   
b^{lm} 
+ \sum_{q=0}^{Q^{(l-1)}-1} \, \, 
\sum_{{\color{black}r}=0}^{R^l-1} 
\sum_{{\color{black}c}=0}^{C^l-1} 
\sum_{{\color{black}d}=0}^{D^l-1} x_{(i+{\color{black}r})(j+{\color{black}c})(k+{\color{black}d})}^{(l-1)q} \, w^{lmq}_{{\color{black}rcd}}  
\right) 
\label{eq:3d-conv-FeatureMap}
\\
\text{2D:} \qquad x_{ij}^{lm} &=& f \left(   
b^{lm} 
+ \sum_{q=0}^{Q^{(l-1)}-1} \, \,
\sum_{{\color{black}r}=0}^{R^l-1} 
\sum_{{\color{black}c}=0}^{C^l-1} 
\phantom{\sum_{{\color{black}d}=0}^{D^l-1}} x_{(i+{\color{black}r})(j+{\color{black}c})}^{(l-1)q} \, w^{lmq}_{{\color{black}rc}}  
\right)  
\label{eq:2d-conv-FeatureMap}
\\
\text{1D:} \qquad x_{i}^{lm} &=& f \, \Bigg(   
b^{lm} 
+ \sum_{q=0}^{Q^{(l-1)}-1} \, \,
\underbrace{\sum_{{\color{black}r}=0}^{R^l-1} 
\phantom{\sum_{{\color{black}c}=0}^{C^l-1}} 
\phantom{\sum_{{\color{black}d}=0}^{D^l-1}}}_{\text{dimensionality}} x_{(i+{\color{black}r})}^{(l-1)q} \, w^{lmq}_{{\color{black}r}}  
\Bigg) \, .
\label{eq:1d-conv-FeatureMap}
\end{eqnarray}
This operation results in the output $x_{ijk}^{lm}$ at position $(i,j,k)$ on the $m$th out of $Q^l$ feature maps in the $l$th layer. The first sum in \eqref{eq:3d-conv-FeatureMap} over $q$ indicates that the obtained feature map is the superposition of the outcome of convolution applied to all feature maps of the previous, hence $(l-1)$th layer.
The weights as well as the biases are obtained in the training process minimizing the cost function. To highlight the dimensionality, we present in \eqref{eq:2d-conv-FeatureMap} the convolution for a 2D image, and in \eqref{eq:1d-conv-FeatureMap} for a 1D sequence of data. Active in convolution are the subscripts in \eqref{eq:3d-conv-FeatureMap}--\eqref{eq:1d-conv-FeatureMap}. 
 
In the present case the input and all feature maps are of cubic shape having the dimension $N^{l} \times N^{l} \times N^{l}$, hence $0 \leq i,j,k \leq N^{l}-1$. Equally of cubic shape are the convolutional filters, which are moreover of fixed size $F$ for all layers, $F \times F \times F$, hence $R^l=C^l=D^l=F$ and $0 \leq r,c,d \leq F-1$.

One convolutional kernel endowed with particular weights and biases is applied across the entire cube and therefore extracts on type of feature. At the beginning, a kernel is placed on the feature map at the location ($i, j, k$) of (0, 0, 0) and carries out convolution as described. Then the kernel is moved by a prescribed increment $S$ referred to as stride for a consecutive convolution.  

By the choice of parameters $F$, $S$ and zero-padding thickness $P$ (adding a zero-layer of thickness $P$ to the boundaries of a feature map for convolution), the feature map size is determined.  
For a 3D CNN with feature maps of cubic shape and the number of voxels/neurons per edge of $N^{l-1}$ in the $(l-1)$th layer, with $F$, $P$, and $S$ as defined, the output has dimension $N^{l}$ in each direction of space with 
\begin{equation}
N^{l} = \dfrac{N^{l-1}-F + 2P}{S}+1\, .
\label{eq:Dimension-after-convolution}
\end{equation}
In the present case of $F=3$ along with zero-padding $P=1$ and stride $S=1$, it holds $N^{l}=N^{l-1}$, the feature map size is unaltered by convolution. This property conceptually enables nets of arbitrary depth.

{\bf Nonlinear activation.} \quad Equations~\eqref{eq:3d-conv-FeatureMap}--\eqref{eq:1d-conv-FeatureMap} contain two out of three operations which we understand as part of one convolutional layer \cite{Goodfellow.2016}; the first one is an affine transformation referred to as linear activation, which is the term in the brackets, the second is a nonlinear activation by a function $f$, which is frequently taken to be a ReLU (rectified linear unit ReLU$(z):=\text{max}(z,0)$ in order to avoid vanishing gradients. The third stage is pooling which is described in Sec.~\ref{subsec:pooling}.
  
Some comments on common CNN properties are in order:
\begin{enumerate}
    \item[C-1] {\bf Weight sharing; locality.} For all the convolutional operations of a filter applied to a particular feature map in the previous layer the weights are the same. (The weights in \eqref{eq:3d-conv-FeatureMap} do not depend on $ijk$.) This is referred to as weight sharing. The property that a neuron in a layer $l$ is only connected to (a finite set of) corresponding neurons in the lower layer $l-1$ is called locality. (In terms of the indices in \eqref{eq:3d-conv-FeatureMap}: $ijk \leftarrow (i+r)(j+c)(k+d)$). Weight sharing implies a relatively small number of parameters in convolutional layers compared to FC layers, locality renders convolution an efficient scheme. 
    \item[C-2] {\bf First analysis, then synthesis.} The convolution in the lowest layer acts on one single ''feature map'', in  \eqref{eq:3d-conv-FeatureMap} the sum over $q$ boils down to one single element, which is the VE in the input layer as displayed in Fig.~\ref{fig:CNN}. The VE's microstructure is filtered for its features, which is an analytical process. Consecutive convolutions in higher layers $l\geq2$ are superpositions of all feature maps of the previous layer (sketched in Fig.~\ref{fig:CNN} for the central feature map) as described by the first sum over $q$ in \eqref{eq:3d-conv-FeatureMap}, which can be understood as a synthesis. Notice that for a convolution in higher layers not only multiple feature maps are involved but also the weights in convolving each feature map are altered (superscript $q$ in the weights). 
	\item[C-3] {\bf A design principle.} As a general design rule of CNNs the number of kernels increases for higher layers, so does the number of feature maps $Q^l$, while in higher layers the size of feature maps $N^{l}$ decreases. It is the pooling operation between convolution layers that compresses the feature maps.  
\end{enumerate} 
 
\subsection{Pooling}
\label{subsec:pooling}
In view of \eqref{eq:Dimension-after-convolution}, size reduction can be realized by convolution. Frequently (and throughout in the present work) convolution along with nonlinear activation is followed by pooling which carries out downsizing the feature maps. 

Average pooling reports the average output within a cubic neigborhood. For an input size of $N \times N \times N$ voxels/neurons and a pooling kernel size of $G \times G \times G$ entries, average pooling yields the down-sized output at row-column-depth $ijk$ of the feature map $m$ at layer $l$
\begin{equation}
 x_{ijk}^{lm} = \dfrac{1}{G^3} \sum_{r=0}^{G-1} \sum_{c=0}^{G-1} \sum_{d=0}^{G-1}
                             x^{lm}_{(G\,i+r)(G\,j+c)(G\,k+d)} \, . 
 \label{max-pooling}
\end{equation}
Since pooling applies to the outcome of convolution, the indices for layer $l$ and feature map $m$ are unaltered. 

The corresponding case of 3D max(imum) pooling, which reports the maximum output within a cubic neigborhood, leads to the output at position $ijk$ of
\begin{equation}
x_{ijk}^{lm} = \max\limits_{r,c,d \in \{0,1,\ldots, G-1\}} x^{lm}_{(G\,i+r)(G\,j+c)(G\,k+d)} \, .
\label{average-pooling}
\end{equation}
As an example, the feature map size is reduced to one eights for $G=2$. 

\subsection{Fully connected layers}
As displayed in Fig.~\ref{fig:CNN} the stack of convolutional layers is followed by a stack of fully connected (FC) layers, where each neuron is connected to all neurons in the consecutive layer. Since the FC layers are 1D, the 3D arrays of the feature maps need to be flattened at the interface of convolutional to fully connected layers. Notice that the FC layers for their full connectivity largely determine the overall size of a CNN.

In FC layers, the value $x_{i}^{l}$ of each neuron $i$ out of $N^l$ in layer $l$ is computed as the sum over the weighted neuron values in the previous layer $l-1$
according to 
\begin{equation}
z_{j}^{l} =  b_j^{l} \, + \,
 \sum_{i=0}^{N^{l-1}-1} \, 
x_{i}^{l-1} \, w^{l}_{ij} \, ,
\label{eq:FC-linearMap}
\end{equation}
where, as for the convolutional layers, the affine transformation of \eqref{eq:FC-linearMap} can be followed by a nonlinear activation $f$ of the output $z_{j}^{l}$ written as $a_{j}^{l} =  f(z_{j}^{l})$.
Similar to the convolutional layers, the outcome of an FC layer serves as input in the consecutive layer, hence $\bm x^{l-1} \rightarrow \bm z^{l} \rightarrow \bm a^{l} =: \bm x^{l}$. At the highest FC layer, $l=L-1$, the neurons are mapped to the output, it holds $x_j^{L-1}=z_j$. In the present task for a structured output, the output layer consists either of 9 elastic moduli for a CNN designed for one case of BCs, or of a vector of length 27 for one single CNN covering three different BCs. For the final mapping into the output the linear activation of \eqref{eq:FC-linearMap} is used. For object classification this mapping is typically carried out by Softmax; it provides probabilities for detected objects belonging to different classes. 

\subsection{Optimization}
\label{subsec:optimization}

Supervised learning of a CNN is realized by minimizing the loss function which measures the distance of the CNN prediction to the target, which is here the macroscopic elastic stiffness obtained by homogenization simulation. 
 
The CNN's input in terms of VEs as well as its output are understood as known measurements of the random variables $(\bm X,\bm Y)$. The measurements are given by $\{(\bm x_i, \bm y_i)\}_{i=1}^M$. 
The corresponding loss function \eqref{eq:LossF=expectedDiffYZ} as a mean squared error MSE reads as
\begin{equation} 
 \mathcal{L}(\bm \Theta) = \dfrac{1}{M} \sum_{I=1}^M (\bm y_I - \bm z_I(\bm \Theta, \bm x_I))^2 \, .
 \label{eq:Loss_MSE}  
 \end{equation}
For optimization and testing, the total data set $\{(\bm x_I, \bm y_I)\}_{I=1}^M$ is decomposed into three sets, 
\begin{enumerate}
	\item[(i)] the training set $\mathcal{T}=\{(\bm x_I,\bm y_I)\}_{I=1}^{M_{\mathcal{T}}}$ with $M_{\mathcal{T}}=\text{card}(\mathcal{T})$, 
	\item[(ii)] the validation set $\mathcal{V}=\{(\bm x_I,\bm y_I)\}_{I=1}^{M_{\mathcal{V}}}$ with $M_{\mathcal{V}}=\text{card}(\mathcal{V})$, and 
	\item[(iii)] the testing set for assessment $\mathcal{A}=\{(\bm x_I,\bm y_I)\}_{I=1}^{M_{\mathcal{A}}}$ with $M_{\mathcal{A}}=\text{card}(\mathcal{A})$ \, , 
\end{enumerate}	
along with $M=M_{\mathcal{T}}+M_{\mathcal{V}}+M_{\mathcal{A}}$, frequently in a ratio of 70\% to 20\% to 10\%.
  
Optimization is carried out for the loss function operating on $\mathcal{T}$ and $\mathcal{V}$ in two steps:
\begin{enumerate}
	\item Minimize the loss $\mathcal{L}(\bm \Theta)$ of \eqref{eq:Loss_MSE} for the training set $\mathcal{T}$, by finding $(\bm \Theta^{\ast}) = \operatorname{arg\,\underset{\bm \Theta}{min}}\, \mathcal{L}(\bm \Theta)$. 
	\item Evaluate the loss $\mathcal{L}(\bm \Theta^{\ast})$ for the validation set $\mathcal{V}$, at the currently optimal parameter values $(\bm \Theta^{\ast})$.
\end{enumerate} 
The procedure of 1. and 2. is the standard method in DNNs to detect overfitting in optimization. It manifests in that training data can be accurately reproduced by the CNN, but new data lead to a large gap to the training accuracy. 

The training set $\mathcal{T}$ is decomposed into mini-batches. The batch size is the number of samples that will be propagated through the network in one forward pass inducing an update of the parameters $\bm \Theta^{\ast}$. The number of forward passes to pipe the full set of training data $M_{\mathcal{T}}$ through the network is the number of iterations per epoch. 
Large batch sizes are not only memory demanding but also lead to a significant degradation of the net's ability to generalize in tests. This observation made in practice was explained by the fact that large-batch methods tend to converge to sharp minimizers in the landscape of the loss function and get trapped therein, whereas smaller batch size methods consistently converge to flat minimizers \cite{Keskar.2017}.

\subsubsection{Gradient descent and backpropagation}  

For gradient-based minimization methods, the parameters are corrected through gradient descent according to 
\begin{equation}
\bm \Theta^l \leftarrow \bm \Theta^l - \alpha \dfrac{\partial \mathcal{L}}{\partial \bm \Theta^l} \,  
\label{eq:parameter-correction}
\end{equation}
with the learning rate $\alpha$. For gradient descent, recall that the final output is the result of activation in the highest of all layers $\bm a^{L}=f_{L}(\bm \Theta,\bm x)$ and that the first activation takes place in layer $l=1$, where $l=0$ is the input layer. This implies a representation of the loss function for a mini-batch $\mathcal{T}_{\text{batch}} \subset \mathcal{T}$ of size $M_{\text{batch}} \ll M_{\mathcal{T}}$ 
\begin{equation}
   \mathcal{L}(\bm \Theta) = \dfrac{1}{M_{\text{batch}}} \sum_{I=1}^{M_{\text{batch}}} \left(\bm y (\bm x_I) - \bm a^{L} (\bm \Theta, \bm x_I)\right)^2 \, . 
\end{equation}
To obtain the gradients for the descent, the partial derivatives of $\mathcal{L}$ with respect to the parameters $\bm \Theta = \{(\bm w^l,\bm b^l)\}_{l=1}^{L}$ are required for each layer. The chain rule enables to determine the partial derivatives $\partial \mathcal{L}/ \partial \bm \Theta^l$, from the highest to the lowest layer, $l=L, \ldots, 1$, which coins the name backpropagation
\begin{eqnarray}
  \dfrac{\partial \mathcal{L}}{\partial \bm \Theta^{L}}  &=& 
  \dfrac{\partial \mathcal{L}}{\partial f_{L}} 
  \dfrac{\partial f_{L}}{\partial \bm \Theta^{L}} \, ,
  \\
  \dfrac{\partial \mathcal{L}}{\partial \bm \Theta^{L-1}}  &=& 
  \dfrac{\partial \mathcal{L}}{\partial f_{L}} 
  \dfrac{\partial f_{L}}{\partial f_{L-1}}
  \dfrac{\partial f_{L-1}}{\partial \bm \Theta^{L-1}} \, ,
  \\[2mm] 
  &\ldots& \nonumber 
  \\[2mm] 
  \dfrac{\partial \mathcal{L}}{\partial \bm \Theta^{l}}  &=& 
  \dfrac{\partial \mathcal{L}}{\partial f_{L}} 
  \dfrac{\partial f_{L}}{\partial f_{L-1}} 
  \hdots 
  \dfrac{\partial f_{l+1}}{\partial f_{l}} 
  \dfrac{\partial f_{l}}{\partial \bm \Theta^{l}}\, ,
\end{eqnarray}
where $\partial f_{l+1}/\partial f_{l}$ are derivatives of the output with respect to the input of layer $l+1$ (valid for both convolutional and FC layers), and $\partial f_{l}/\partial \bm \Theta^{l}$ are derivatives of the output of the layer $l$ with respect to its parameters. 

Detailed presentations of backpropagation can be found in, e.g., Sec.~12.5. of \cite{Gonzalez.2018} and with an account of stochastic gradient descent in \cite{Higham.2019}.
 A review of the historical development of backpropagation is given in Sec.~5.5. of \cite{Schmidhuber.2015} with references to precursor work of the most prominent papers \cite{Rumelhart.1986,LeCun.1989}.

\section{Results}
\label{sec:ResultsDiscussion}

\begin{wrapfigure}[16]{r}{7.5cm}
	\vskip -8pt
	\centering
	\includegraphics[width=6.0cm, angle=0, clip=]{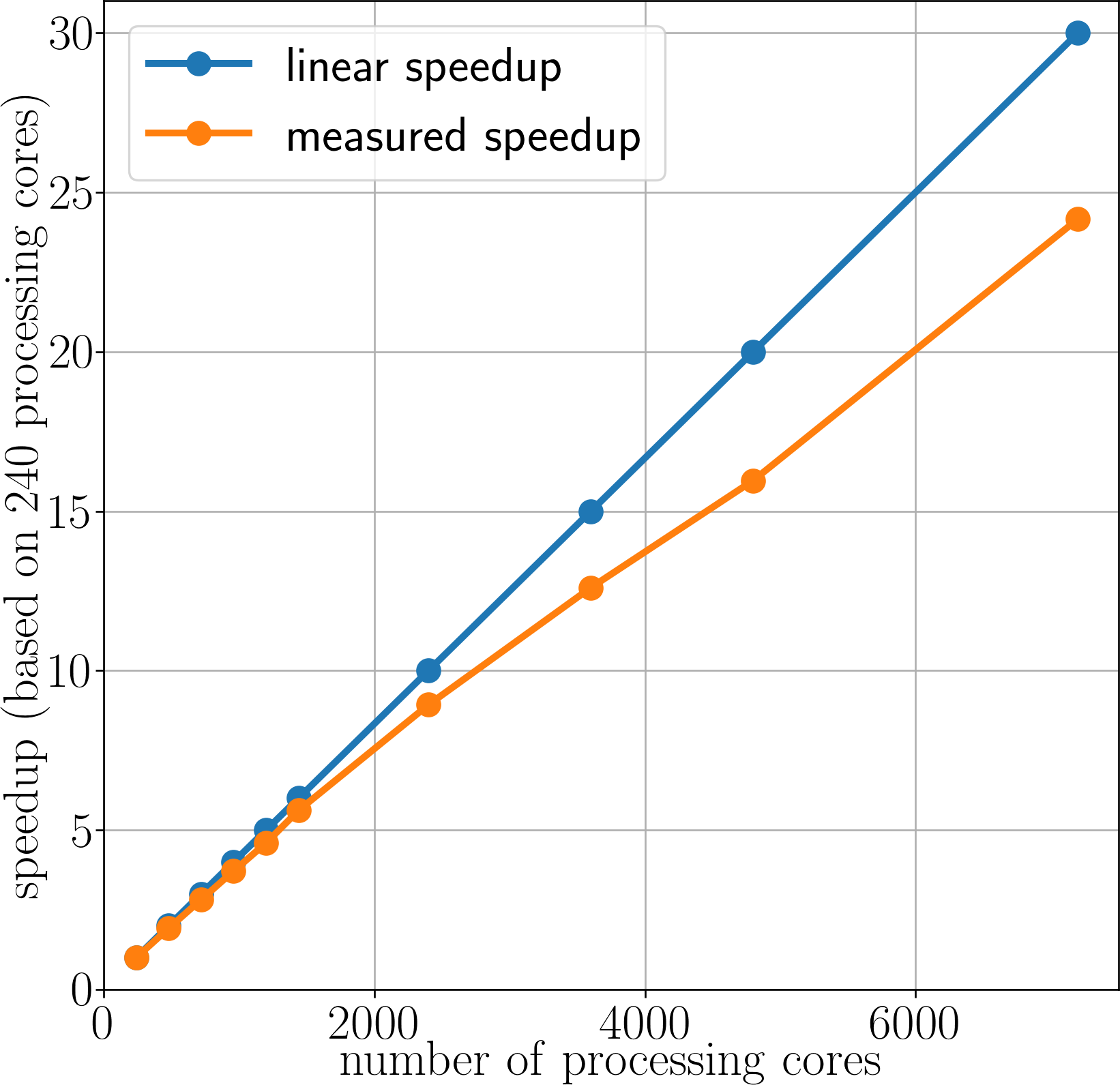} 
	\caption{{\bf Scaling.} Speed-up by MPI-parallelization of homogenization computations. 
	 \label{fig:Speed-up}} 
\end{wrapfigure}
After the microstructure generation described in Sec.~\ref{sec:microstructure-generator} the computation of the homogenized elasticity tensors of each VE for different BCs is carried out by an MPI-parallelized finite element code. The scaling behavior is displayed in Fig.~\ref{fig:Speed-up} for a reference speed obtained for 240 processing cores. Corresponding sets of linear equations are solved by the direct solver PARDISO of the Intel Math Kernel Library (MKL) 2020 and by the iterative solver GMRES along with a preconditioner, both from the PETSc library. In the simulations, the finite element discretization adopts the original, uniform voxel grid, although considerable computational savings at moderate accuracy losses can be realized by resolution coarsening, by adaptive, octree-based mesh coarsening, or combinations thereof \cite{Fischer.2020}. 
 
\subsection{Pre-analysis on stiffness bounds}
\label{subsec:Pre-analysis}
 
We wish to know, how strong elastic stiffness depends on the applied boundary condition, which enables conclusions on the suitability of individual VEs to serve as an RVE. The Young's modulus of the stiff phase is 100 GPa, of the compliant phase 2 GPa, hence a contrast of factor 50. Poisson's contraction is set to $\nu=0.3$ for both phases. 

The point clouds in the diagrams of Fig.~\ref{fig:SUBC-PBC-KUBC-consistency} (a)--(d) show the ratios of stiffness for different BCs in exemplary terms of components ${C}_{11}$ and ${C}_{13}$, which are representative for other elasticity coefficients. They reveal the following characteristics:

\begin{figure}[htbp]
	\begin{minipage}{16.5cm}  
		\centering   
		\subfloat[component ${C}_{11}$] 
		{\includegraphics[width=6.4cm, angle=0, clip=]{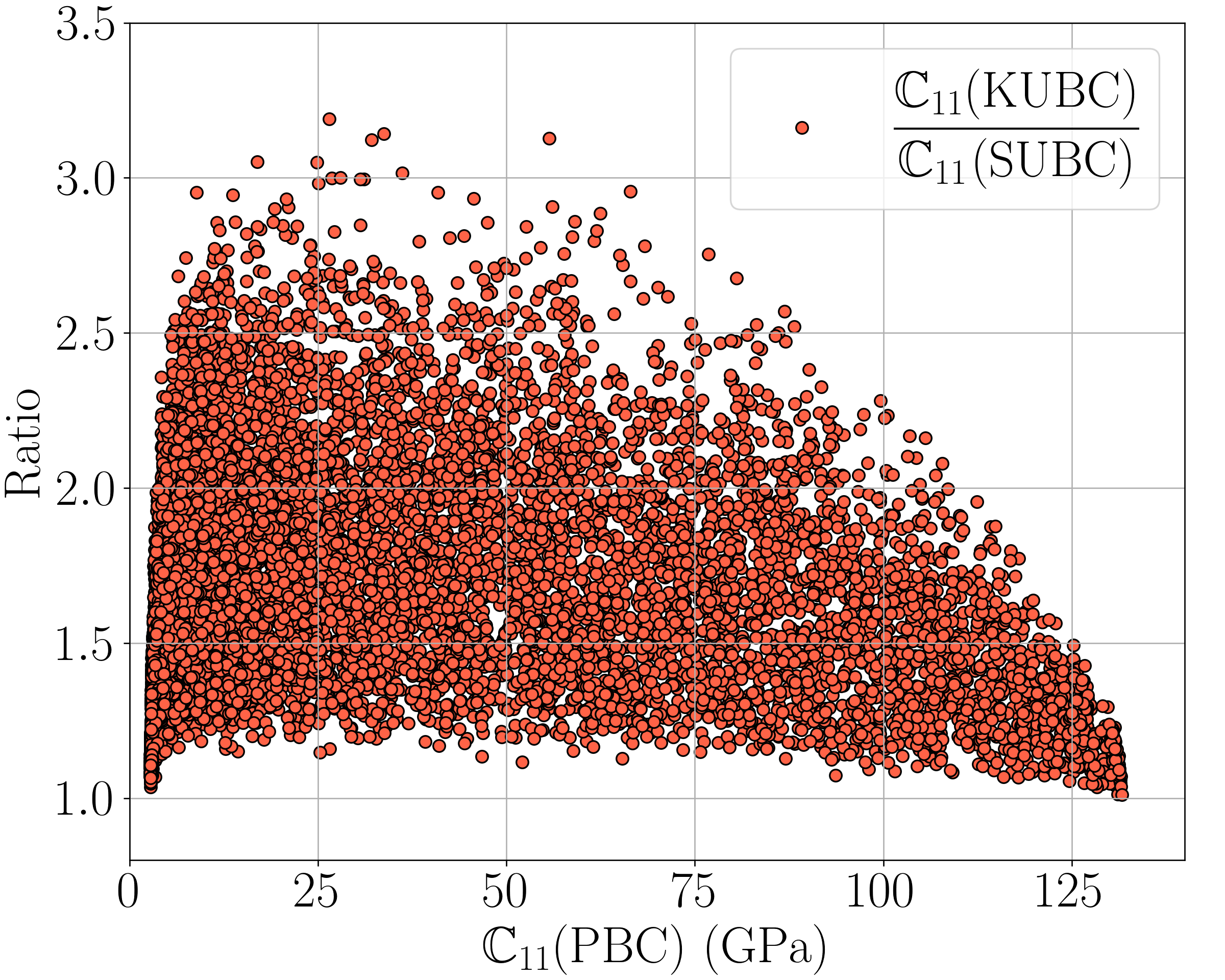}}
		\hspace*{0.05\linewidth}
		\subfloat[component ${C}_{13}$] 
		{\includegraphics[width=6.4cm, angle=0, clip=]{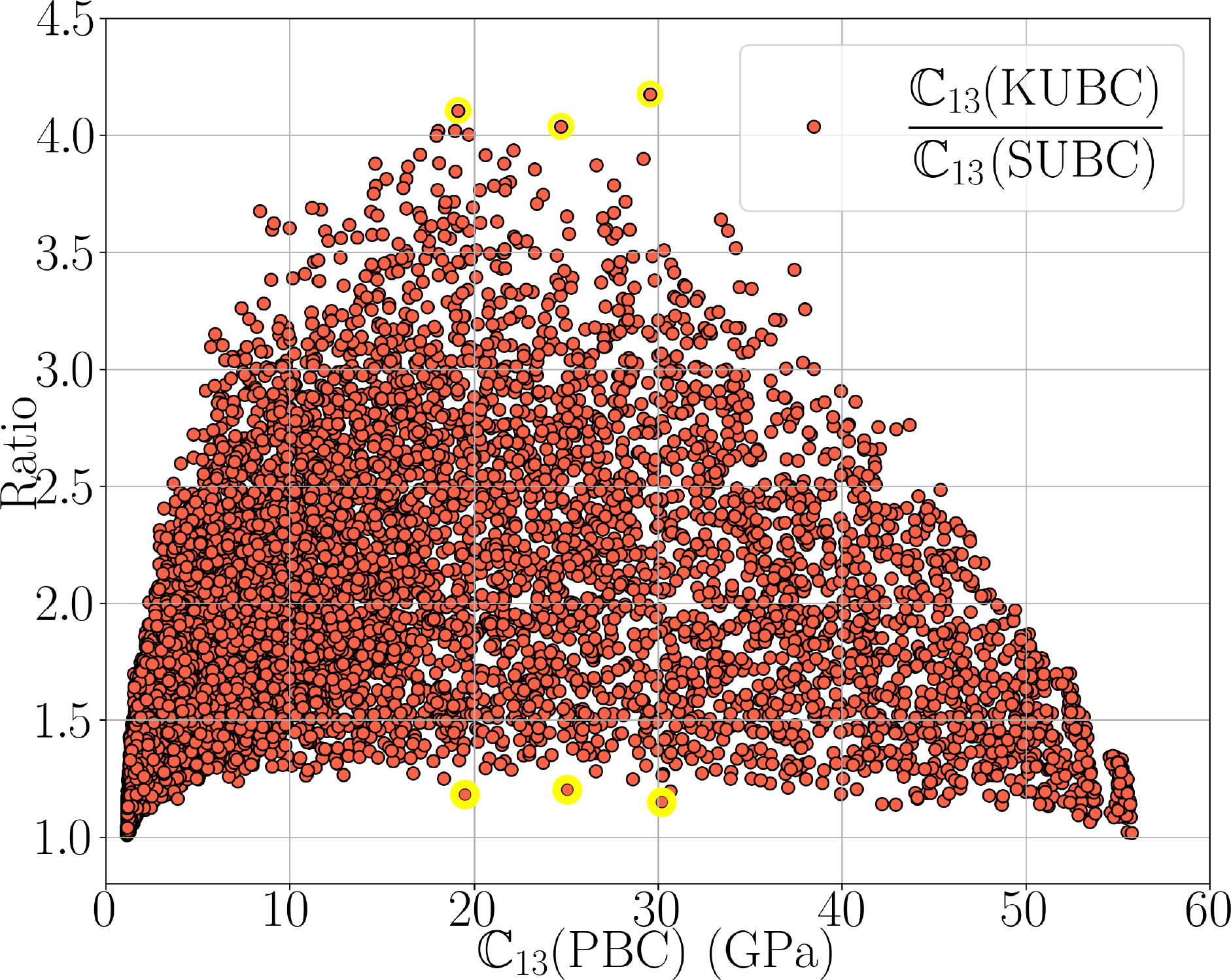}}		
		\\
		\subfloat[component ${C}_{11}$] 
		{\includegraphics[width=6.4cm, angle=0, clip=]{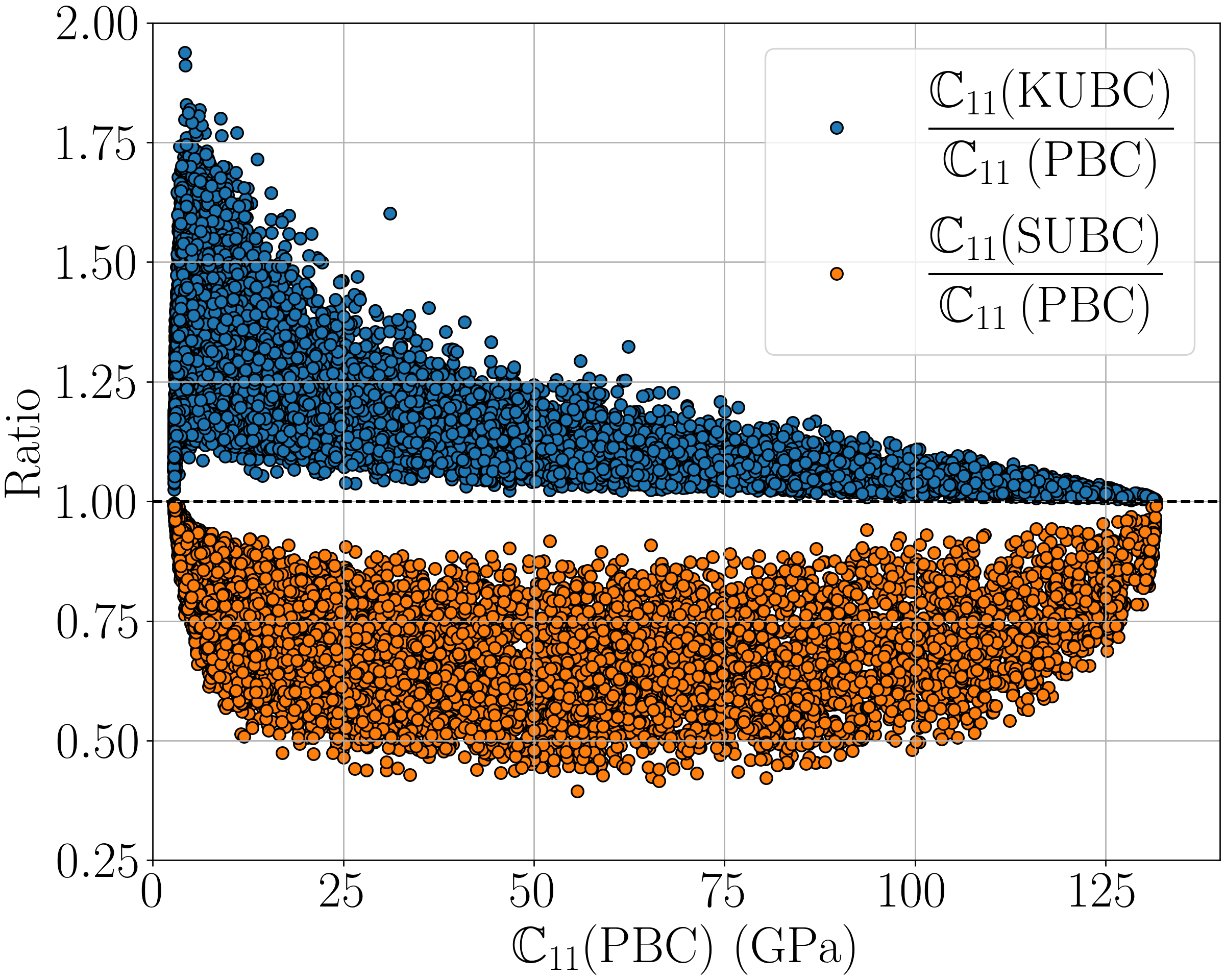}}
		\hspace*{0.05\linewidth}
		\subfloat[component ${C}_{13}$] 
		{\includegraphics[width=6.4cm, angle=0, clip=]{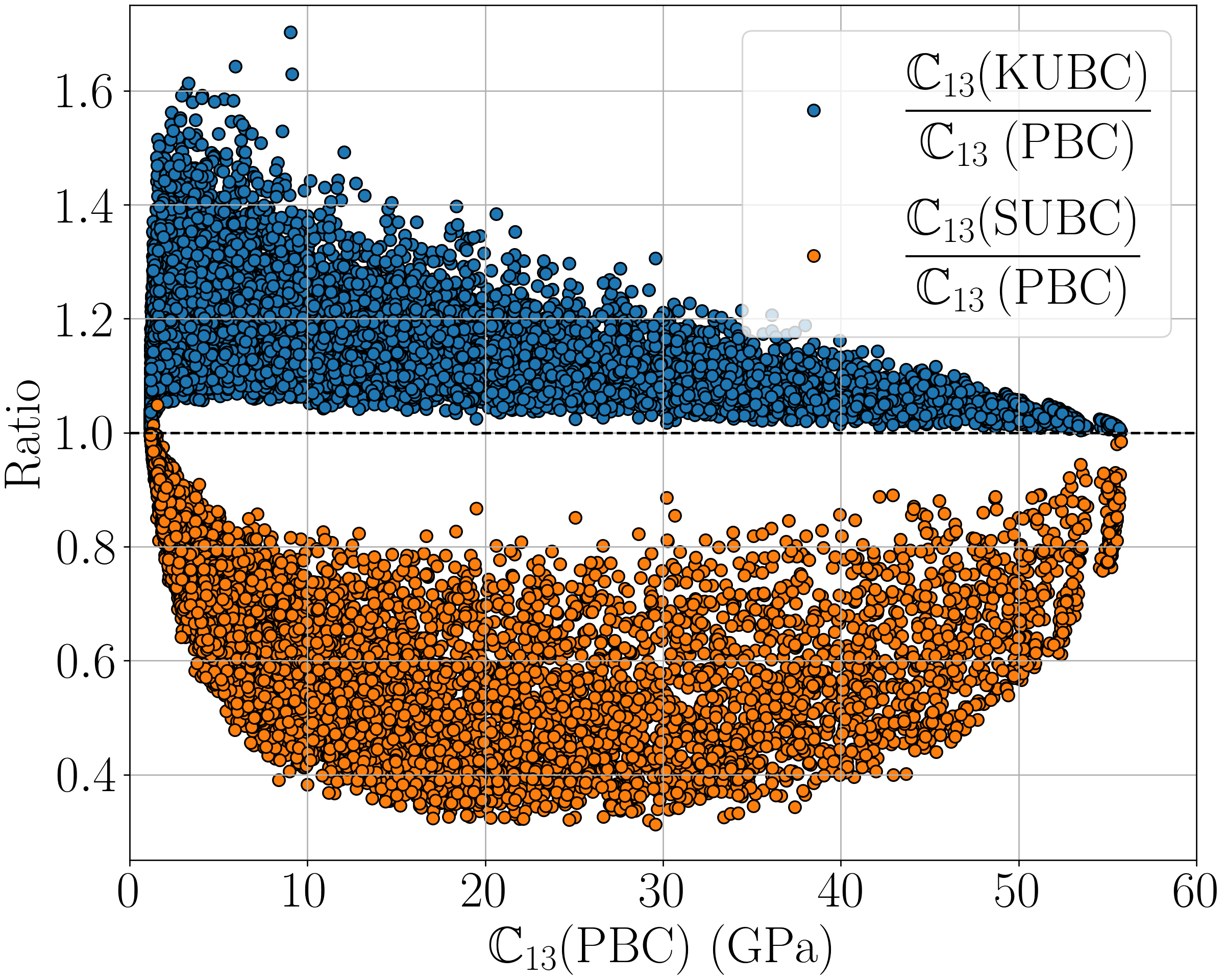}}
		\\
		\subfloat[7.6/5.6/7.4] 
		{\includegraphics[width=2.4cm, angle=0, clip=]{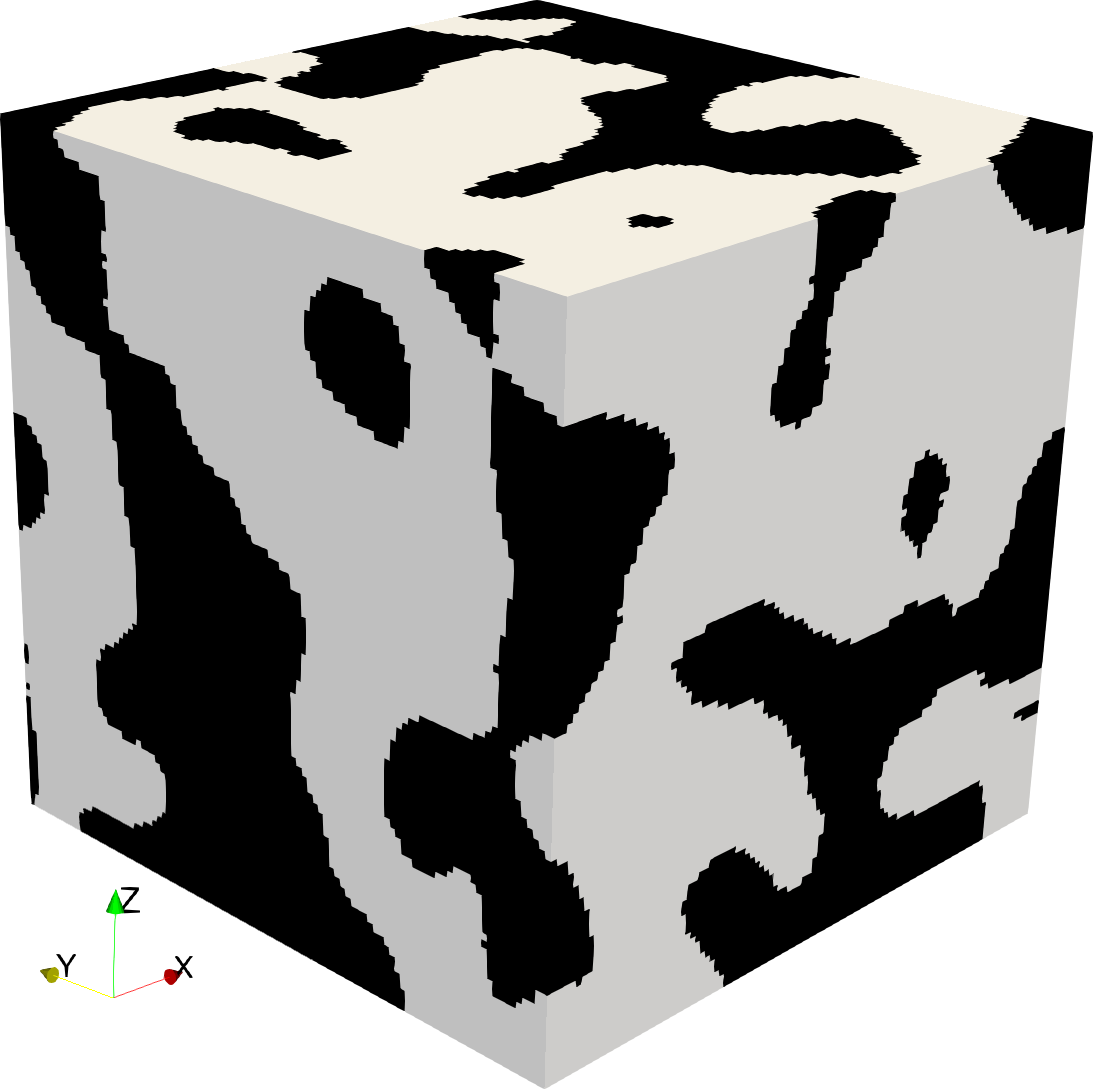}}
		\hspace*{0.001\linewidth}
		\subfloat[7.5/5.8/6.5] 
		{\includegraphics[width=2.4cm, angle=0, clip=]{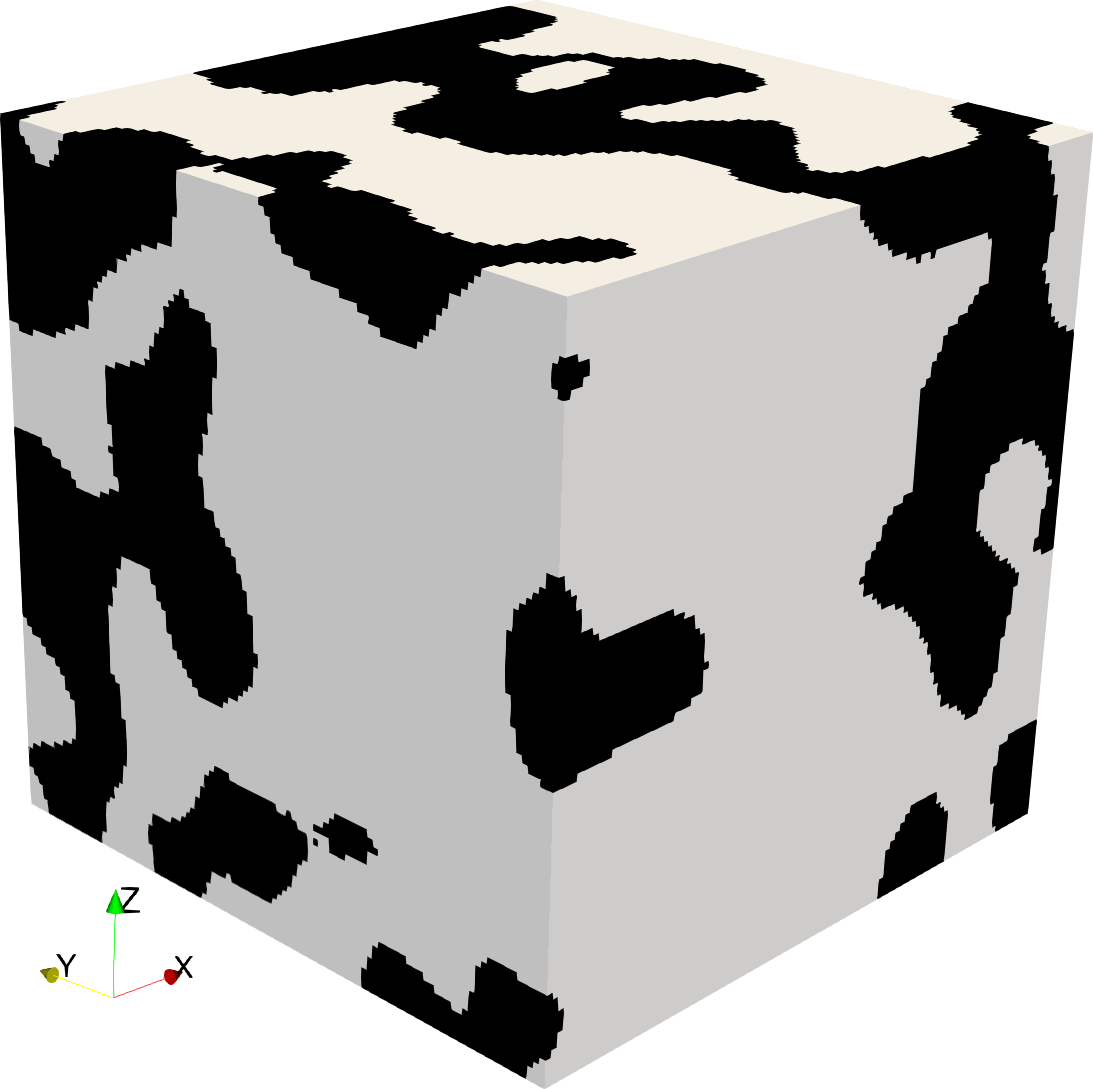}}
		\hspace*{0.001\linewidth}
		\subfloat[7.9/6.8/7.1] 
		{\includegraphics[width=2.4cm, angle=0, clip=]{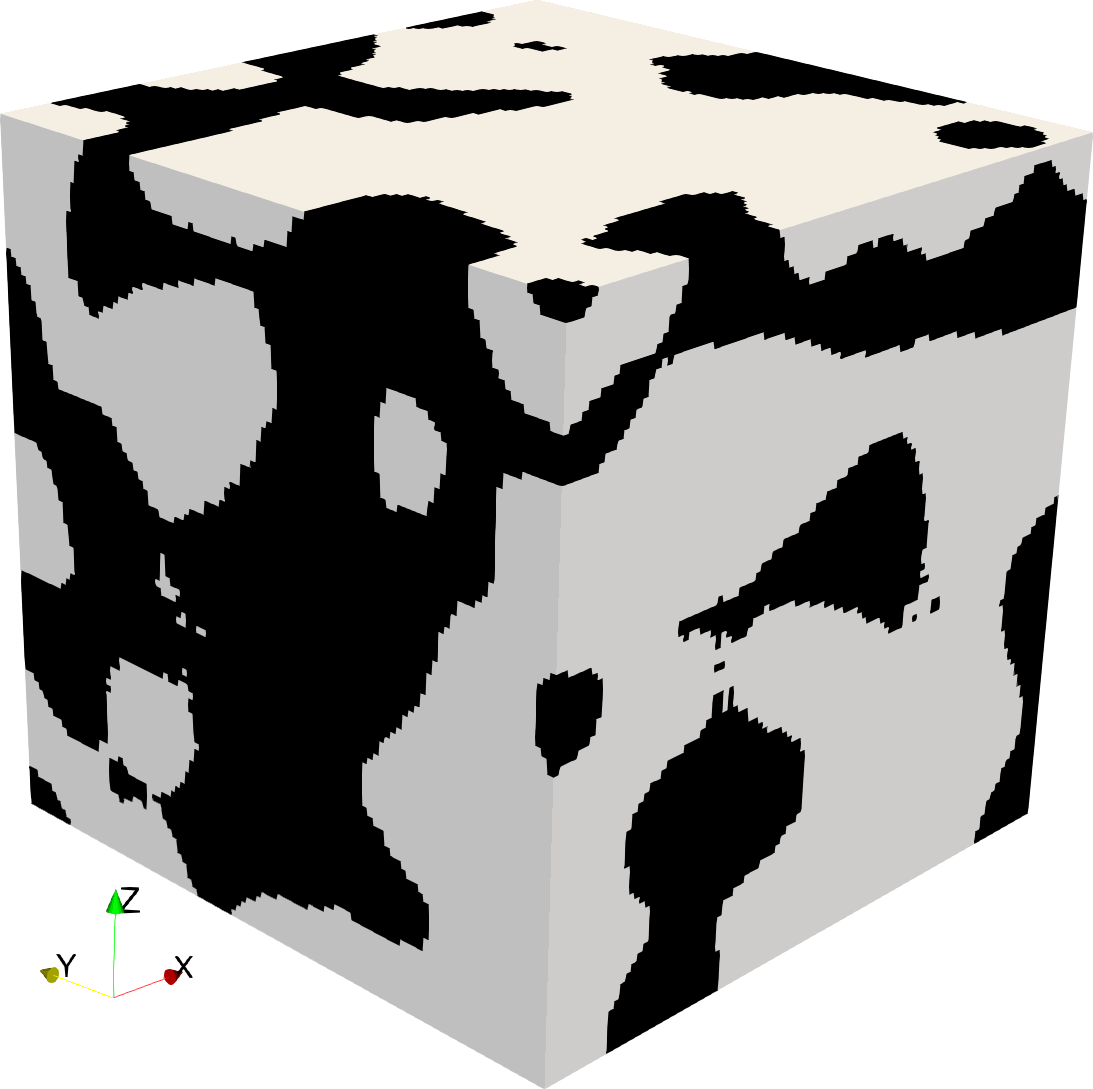}}
		\hspace*{0.01\linewidth}
		\subfloat[0.6/0.7/0.9] 
		{\includegraphics[width=2.4cm, angle=0, clip=]{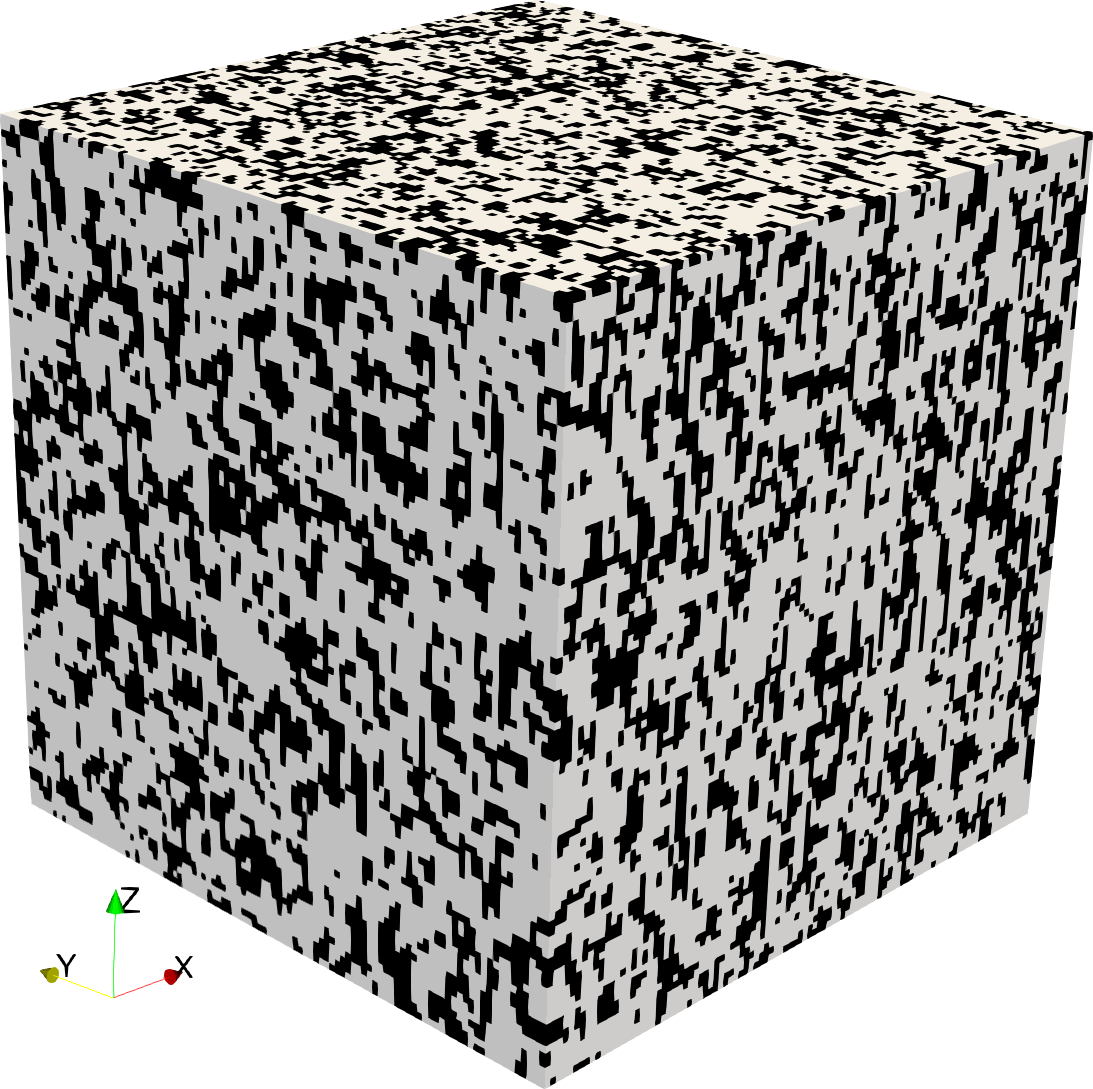}}
		\hspace*{0.001\linewidth}
		\subfloat[0.5/1.4/0.8] 
		{\includegraphics[width=2.4cm, angle=0, clip=]{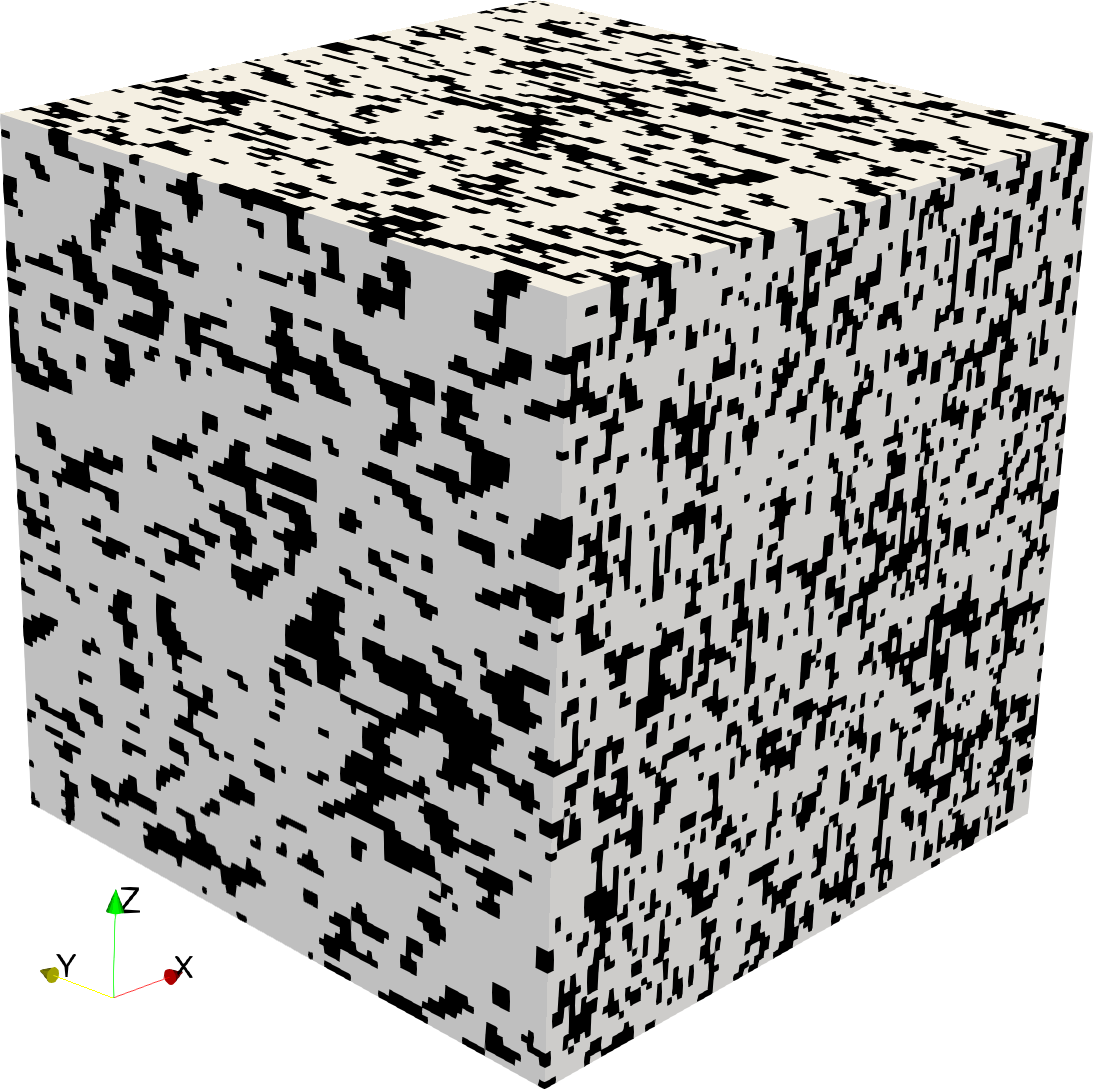}}
		\hspace*{0.001\linewidth}
		\subfloat[1.7/1.1/0.5] 
		{\includegraphics[width=2.4cm, angle=0, clip=]{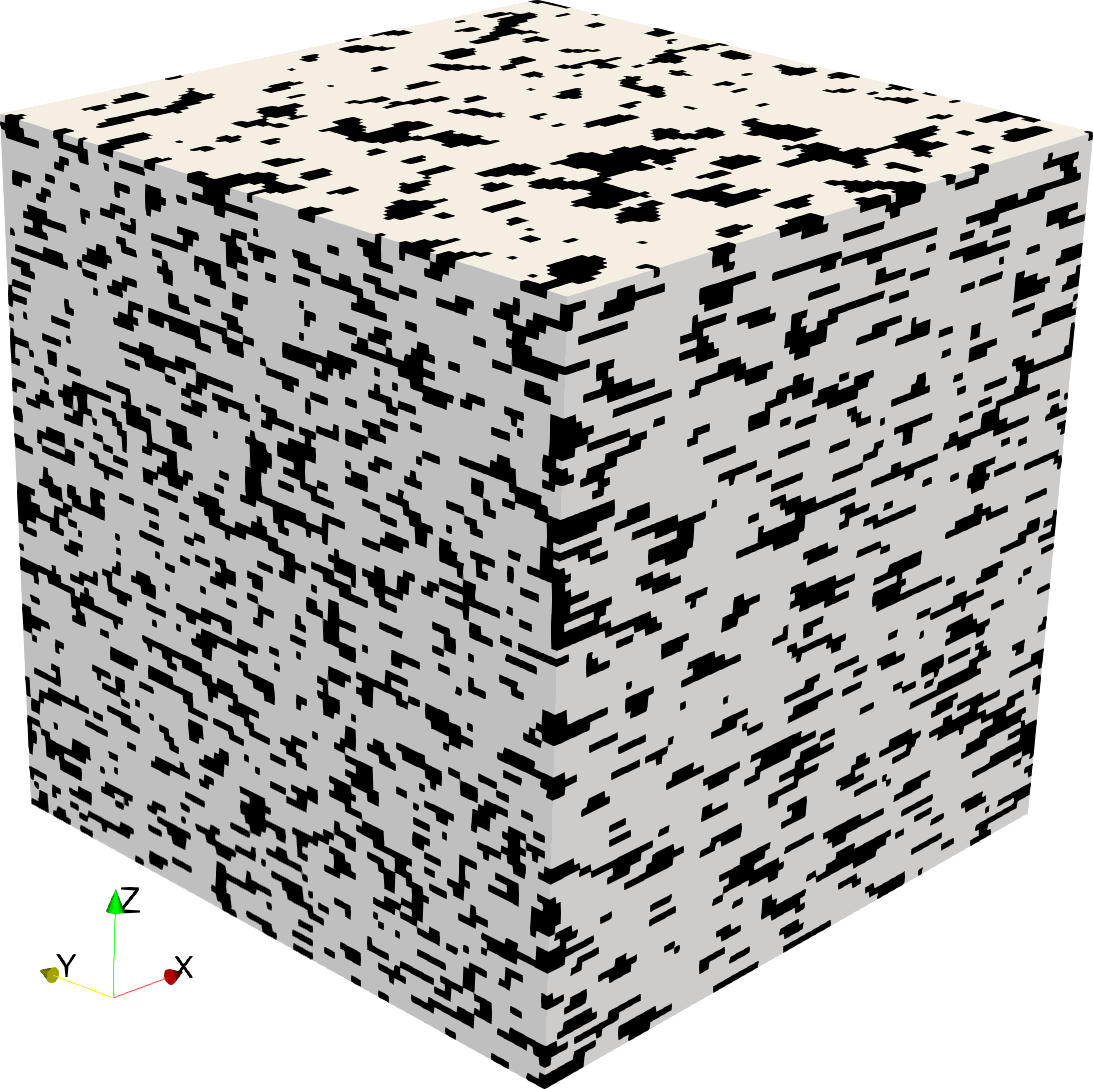}}
	\end{minipage}
	\caption{{\bf Scatter of stiffness spread for different BCs.} In (a) and (c) the stiffness ratio for KUBC/SUBC over PBC stiffness, in (b) and (d) the stiffness ratio for KUBC/PBC (blue) and SUBC/PBC (orange) over PBC for the full dataset of 10$^4$ VEs. For intermediate PBC stiffness in terms of $C_{13}$, the VEs in (e)--(g), highlighted in (b), have the largest KUBC/SUBC stiffness ratio, the VEs in (h)--(j), equally marked in (b), have the smallest KUBC/SUBC ratio. The values of the variances $s_x/s_y/s_z$ are given in the captions of Subfigs.~(e)--(j).
		\label{fig:SUBC-PBC-KUBC-consistency}}
\end{figure} 

\begin{itemize}
	\item[(i)] The stiffness of the majority of VEs significantly depends on the applied BCs. The KUBC stiffness exceeds the SUBC stiffness by a factor of 4 at maximum. The ratio of  KUBC to PBC stiffness exhibits factor 2 at maximum for compliant composites and monotonously decreases for increasing overall stiffness, the minimal ratio of PBC to SUBC stiffness exhibits factor 0.3 at intermediate overall stiffness of the composites. \\[-6mm]
	\item [(ii)] Only in the limit of mono-phase systems, hence of highest or lowest stiffness, the VEs exhibit BC-invariance of stiffness. \\[-6mm]
	\item [(iii)] For intermediate phase fractions there is a white space between the point clouds to the case of BC-invariance (stiffness ratios equal to one) in Fig.~\ref{fig:SUBC-PBC-KUBC-consistency} (a)--(d). This white space follows from minimal values of variances close to 0.5 in the virtual microstructure generation. The VEs coming closest to BC-invariance all exhibit small variance values, the VEs showing strongest sensitivity to BCs exhibit large variance values as underpinned in the captions of Fig.~\ref{fig:SUBC-PBC-KUBC-consistency} (e)--(j). This suggests that for $s\rightarrow 0$ the gap of the white space is decreased or even closed, given that the voxel resolution is consistent to that limit. The error due to the finite voxel resolution of image-based microstructures is introduced into the error framework of two-scale finite element methods for homogenization in \cite{Eidel.2021}. 
\end{itemize} 

In conclusion, the deviation of PBC stiffness from the lower or the upper bound indicates that the majority of samples can not serve as an RVE for their small size. As a consequence, the application of PBC implies an error of the apparent stiffness of uncertain magnitude, which is at least bounded by the SUBC and KUBC results. These bounds, however, do not necessarily apply to a true RVE of sufficient size.   
 
Two comments are in order:
\begin{enumerate}
	\item[C-4] The full dataset of VEs by its construction shall empower the net with predictions for various classes of different microstructures. The statistics of the ensemble is available, but of course not specific enough to predict for a single snapshot-VE the effective elastic properties by a statistical analysis as proposed by Kanit et al. \cite{Kanit.2003}. Quite in contrast, each VE can be understood as a realization of an unknown ensemble with its own statistics. For that reason, the conclusion for the size requirement of the RVE in the present context must be based on the criterion of boundary-invariance of stiffness. Similarly, costly 3D image acquisition by tomography typically provides only a small number of snapshot specimens in voxel resolution instead of a statistical ensemble \cite{Andra.2013,Gote.2021}.
    \item[C-5] The criterion of boundary-insensitivity is of course blind for periodicity. If, for the upper bound of stiffness, KUBC are applied to a VE of edge length $\delta$, which is larger than the periodic unit cell length $\epsilon$, the convergence to boundary insensitivity, i.e. of KUBC and PBC towards each other, scales with $1/\delta$ according to \cite{E.2005} (Thm. 1.2). The corresponding error is the modeling error  in the FE-HMM framework of errors. It has its origin in the boundary layer which feels the rigidity of the Dirichlet constraint. With increasing distance to the boundary the periodic solution is continuously recovered. In this scenario of convergence, the SUBC results are not considered.
\end{enumerate}
 
\subsection{The present CNN architecture, hyperparameters and implementation}

The present CNN exhibits the architecture displayed in Fig.~\ref{fig:CNN}, which adopts the architecture of the pioneering work of \cite{Yang.2018} to a large extent. The exception is the size of the FC layers. We choose for the FC block two layers of widths 256 and 128 instead of 2048 and 1024 of reference \cite{Yang.2018} and thereby reduce the CNN size by more than 82\%. The convolutional block consists of 5 convolutional layers, each one endowed with an increasing number of kernels, from 16 to 256. The convolutional kernel size is constantly $3 \times 3 \times 3$, strides are $1 \times 1 \times 1$. Nonlinear activation is carried out by ReLU. Average pooling and, as an alternative, max pooling with kernel size $2 \times 2 \times 2$ are carried out after each convolution. For regularization, the $L_2$ value is set to 0.001.    

For the input layer, $100^3$ neurons are used according to the number of voxels in each VE. The voxel data are transformed from binary values 0 and 1 to -0.5 and 0.5 following the proposal of \cite{Yang.2018}. It turned out to improve activation and overall performance of the convolution layers. 

The implementation is realized with Keras 2.4.0 as part of Tensorflow 2.4.1 using Python 3.7.
The training simulations were carried out on NVIDIA TESLA V100-GPUs with 32 GB memory for each GPU.  

\subsection{Training}

\begin{figure}[htbp]
	\begin{minipage}{16.8cm}  
		\centering
		\subfloat[KUBC] 
		{\includegraphics[width=6.0cm, angle=0, clip=]{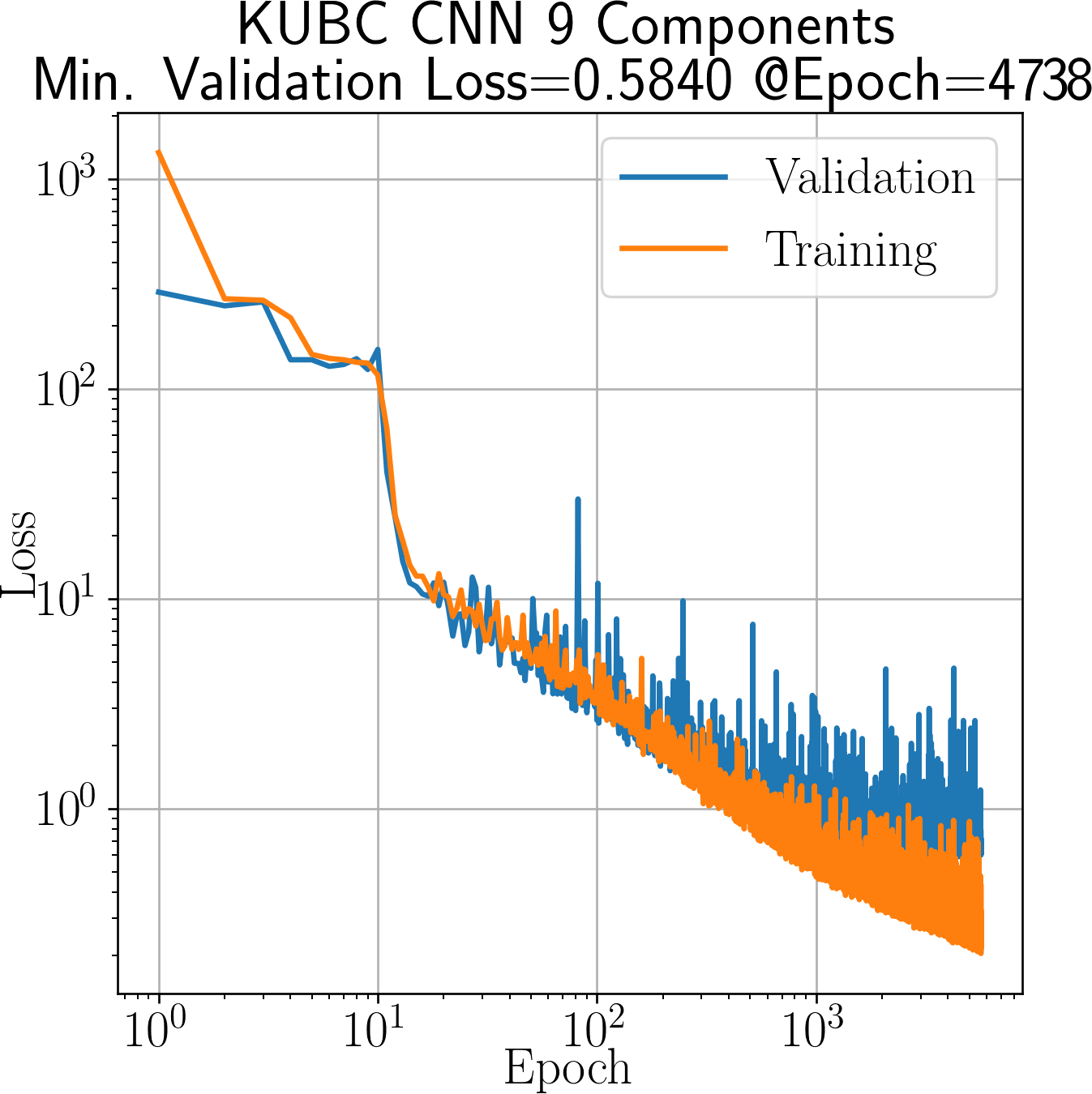}}	
		\hspace*{0.05\linewidth}	
		\subfloat[PBC] 
		{\includegraphics[width=6.0cm, angle=0, clip=]{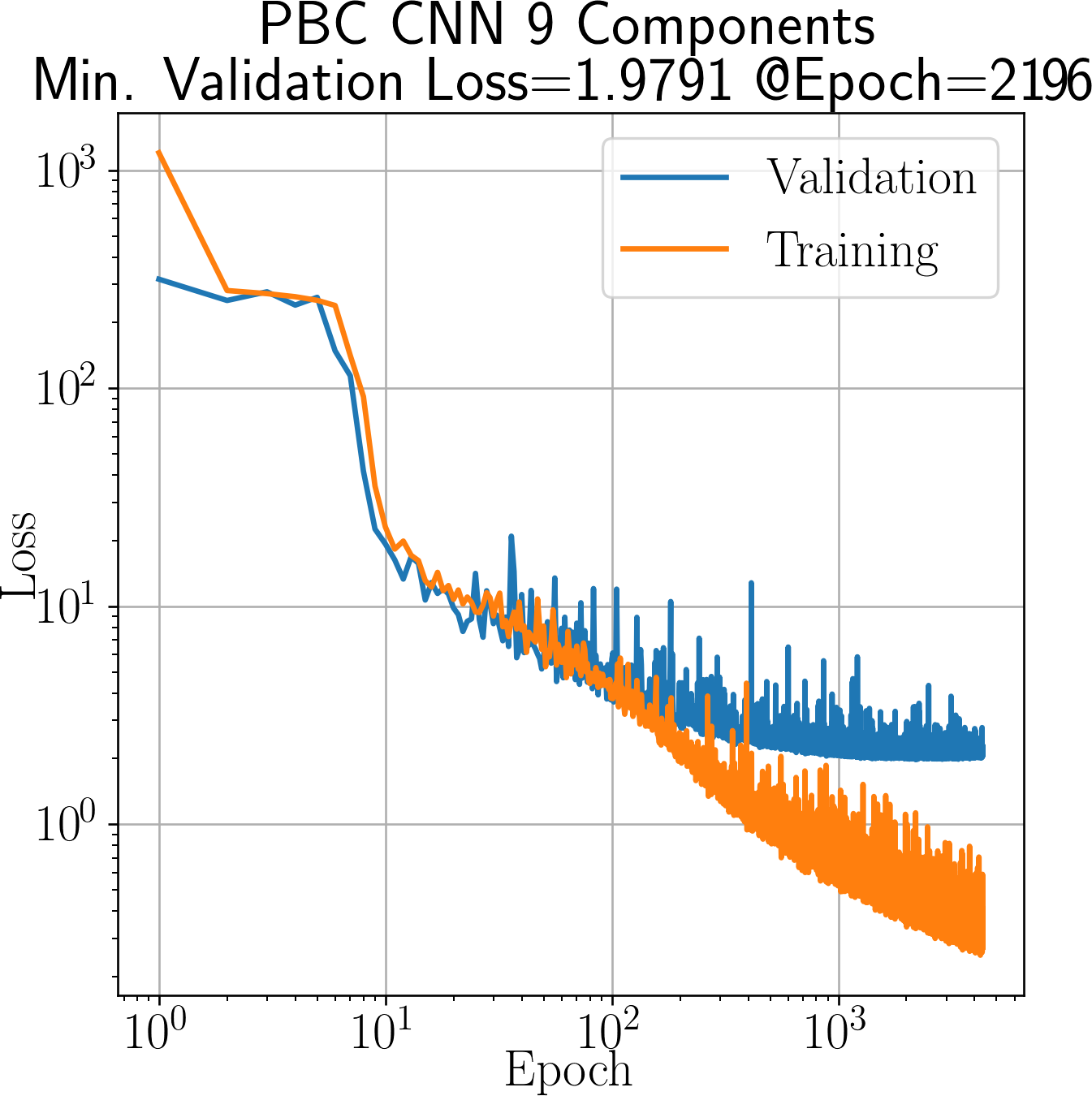}}			
		\\
		\subfloat[SUBC] 
		{\includegraphics[width=6.0cm, angle=0, clip=]{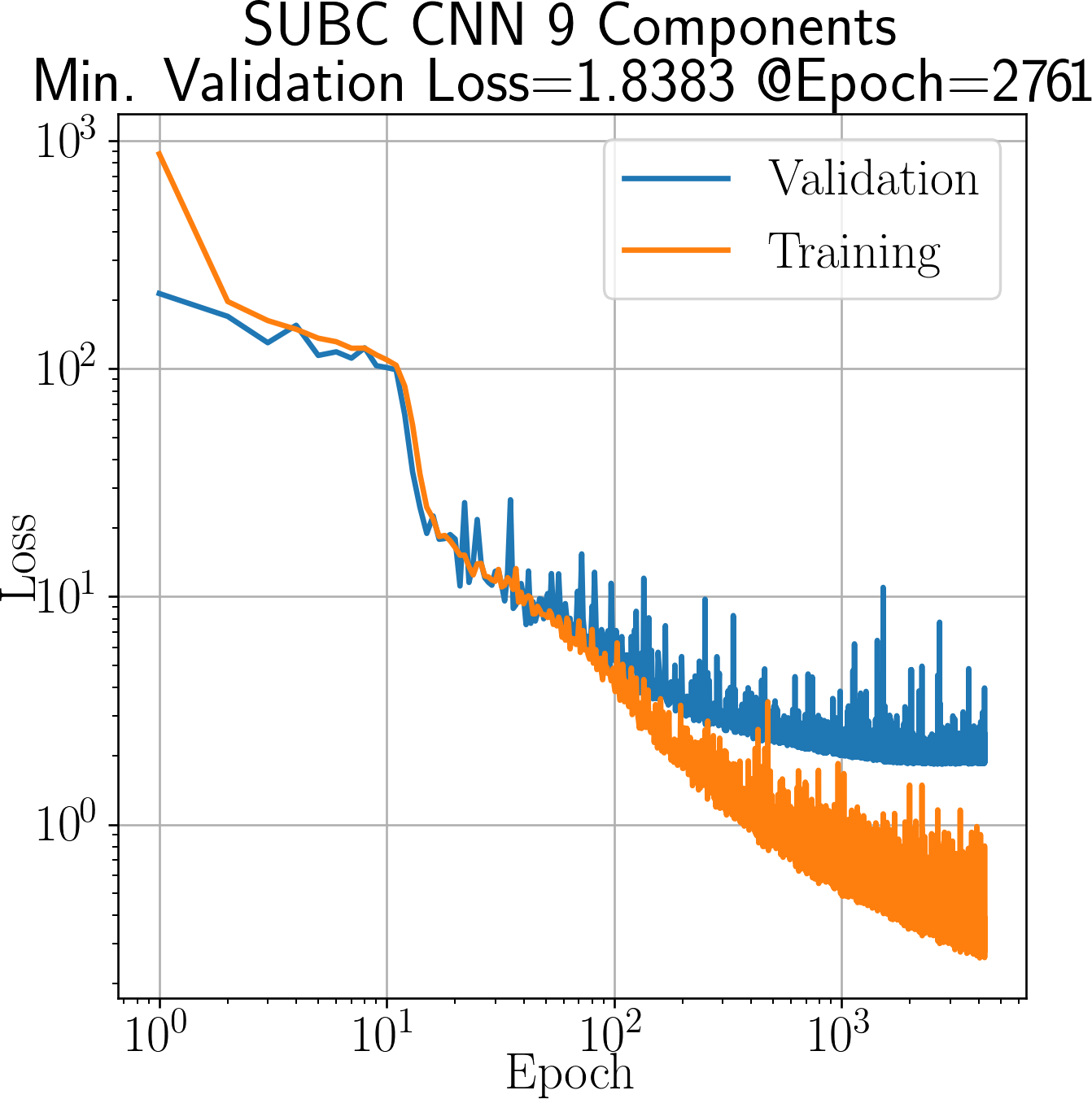}}  
		\hspace*{0.05\linewidth}
		\subfloat[KUBC, PBC, SUBC] 
		{\includegraphics[width=6.0cm, angle=0, clip=]{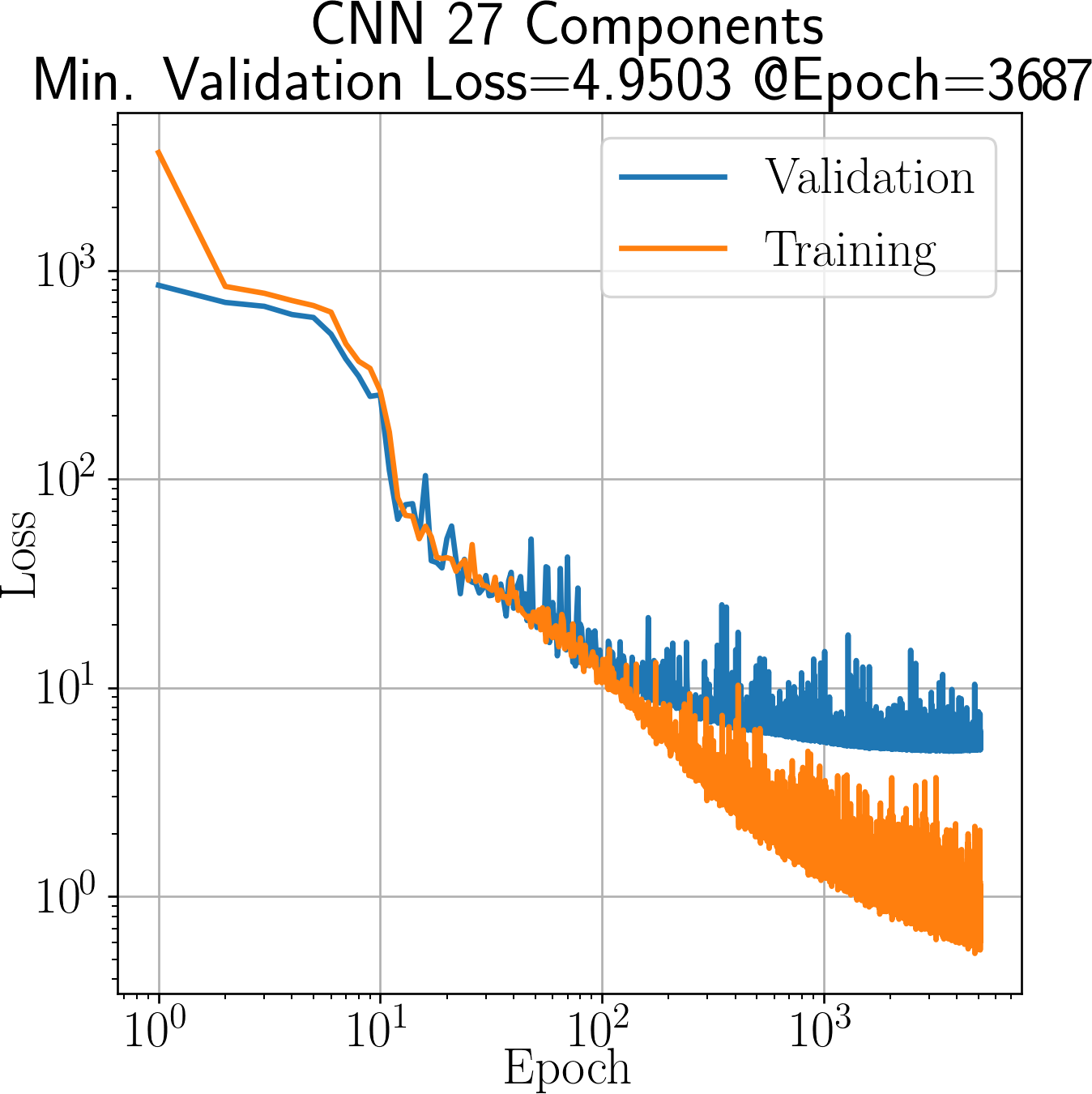}}  
	\end{minipage}
	\caption{{\bf Training and validation losses.} Losses over epochs for different BCs, (a) for KUBC, (b) for PBC, (c) for SUBC, and (d) for the three BCs in one CNN. The losses refer to the cost function (GPa) for the training data and the validation data. \label{fig:Training-LossesOverEpoches}} 
\end{figure}

For optimization, the Adam algorithm \cite{Kingma.2014} is used with the default parameters of the Keras library except of the learning rate $\alpha$ ($\alpha=0.0001$ instead of the default $\alpha=0.001$).  

The set of 10$^4$ VEs is randomly decomposed into sets for training, validation and testing in the ratio of 70:20:10$\%$. For the reasons mentioned in Sec.~\ref{subsec:optimization} the training set is split into mini-batches of 32 VEs, which results in 7000/32$\rightarrow$219 iterations per training epoch. 
 
Figure \ref{fig:Training-LossesOverEpoches} shows the losses over the number of training epochs for the four different CNNs. The validation losses decrease monotonically, and turn into an increase thus indicating the start of overfitting. The hyperparameters are taken at the sweet spot of the minimum of the validation losses, the corresponding epochs are listed in the diagrams (a)--(d) of Fig.~\ref{fig:Training-LossesOverEpoches}. It turns out that average pooling is more accurate than max popling as shown in Tab.~\ref{fig:V-losses}. Dropout in the FC layers with a dropout rate of 0.2 does not improve the results. 
 
\begin{table}
	\centering
	\renewcommand{\arraystretch}{1.2} 
	\resizebox{0.55\columnwidth}{!}{ 
		\begin{tabular}{rlcccc}
			\hline
			&     & \multicolumn{4}{c}{CNN-case} \\
     Pooling &     & KUBC & PBC & SUBC & all-BCs \\
			\hline	
     Average  & Loss & 0.5840  & 1.9791  & 1.8383  & 4.9503 \\
              &  Epoch      & 4738    & 2196    & 2761    &  3687  \\
    \hline
     Max      & Loss & 1.3688  & 2.6781  & 3.1659  & 7.8501 \\
              &  Epoch      & 2323    & 1261    & 950    &  1286  \\
			\hline	
		\end{tabular}
	}
	\caption{{\bf Validation losses.} Minimal validation losses (GPa) with corresponding epochs for average and max pooling. \label{fig:V-losses}}
\end{table}

 
\subsection{Accuracy}  

\subsubsection{The mean absolute stiffness error (MASE)}
The CNN performance is evaluated by calculating the mean absolute stiffness error (MASE) \cite{Yang.2018} for each tensor component, and an average MASE combining the results for all tensor components 
\begin{equation}
\text{MASE}({C}_{ij}) := 
\dfrac{1}{M_T} \sum_{k=1}^{M_T} \dfrac{\left| {C}^{k(\text{target})}_{ij} - {{C}}^{k(\text{prediction})}_{ij} \right|}{\text{\small average}\{{C}_{ij}^{(\text{target})}\}}  \, , \quad 
\overline{\text{MASE}} := \dfrac{1}{9} \sum_{I=1}^{9} \text{MASE}_{I} \, ,
\label{eq:MASE}
\end{equation}
where $M_T$ denotes the total number of VEs in the test set, ${C}^{k(\text{target})}_{ij}$ and ${C}^{k(\text{prediction})}_{ij}$ are the computed apparent stiffness components and CNN-predicted stiffness components for the $k$th VE, respectively. A non-dimensionalization is carried out by the average of ${C}_{ij}$ of all the VEs in $M_T$. 
  
Table \ref{fig:CNN-MASE} provides the values for testing-MASE for each component of the homogenized elasticity tensor separately as well as the averages $\overline{\text{MASE}}$. The results indicate the following characteristics: 
\begin{itemize}
	\item The predictions are very accurate, testing-MASE is below two percent throughout. 
	\item Testing-MASE increases for decreasing stiffness of the applied BCs; they are smallest for KUBC (MASE $< 1\%$), largest for SUBC, for PBC in between. 
	\item Only very minor accuracy losses for one CNN for all BCs compared to the case of three distinct CNNs. This is true for all BCs, but most pronounced for PBC and SUBC, which is best seen for the average of testing-MASE.
\end{itemize}

\begin{table}
	\centering
	\renewcommand{\arraystretch}{1.2} 
	\resizebox{0.82\columnwidth}{!}{ 
		\begin{tabular}{cc|ccccccccc c}
			\hline
			&  & \multicolumn{9}{c}{Testing-MASE (\%)} & $\overline{\text{MASE}}$ (\%) \\
			{Case}  & BC & ${C}_{11}$ & ${C}_{22}$ & ${C}_{33}$ &
			${C}_{12}$ & ${C}_{13}$ & ${C}_{23}$ &
			${C}_{44}$ & ${C}_{55}$ & ${C}_{66}$ & ${C}_{ij}$\\
			\hline	
			& KUBC & 0.56  & 0.52  & 0.53  & 0.67  & 0.72 & 0.71 & 0.54  & 0.58 & 0.55 & 0.60  \\
			\cmidrule{2-12}
			{3 CNNs} & PBC & 1.11  & 1.09  & 1.14  & 1.29  & 1.32 & 1.22 & 1.09  & 1.12 & 1.15 & 1.17  \\ 
			\cmidrule{2-12}
			& SUBC & 1.51  & 1.51  & 1.53  & 1.71  & 1.64 & 1.75 & 1.73 & 1.57 & 1.72 & 1.63  \\
			\hline
			& KUBC & 0.83  & 0.75  & 0.79  & 0.96  & 1.00 & 0.95 & 0.77 & 0.83 & 0.78 & 0.85  \\
			{1 CNN} & PBC & 1.23  & 1.13  & 1.13  & 1.34  & 1.40 & 1.36 & 1.15 & 1.19 & 1.18 & 1.23  \\
			& SUBC & 1.62  & 1.56  & 1.50  & 1.79  & 1.70 & 1.76 & 1.69 & 1.68 & 1.72 & 1.67   \\
			\hline	
		\end{tabular}
	}
	\caption{{\bf CNN Accuracy.} Testing-MASE for each tensor component for various CNNs. \label{fig:CNN-MASE}}
\end{table}

\subsubsection{The relative error in boxplots}
\begin{figure}[htbp]
	\begin{minipage}{16.5cm}  
		\centering
		\subfloat[Three distinct CNNs for KUBC, PBC, and SUBC.] 
		{\includegraphics[width=15.0cm, angle=0, clip=]{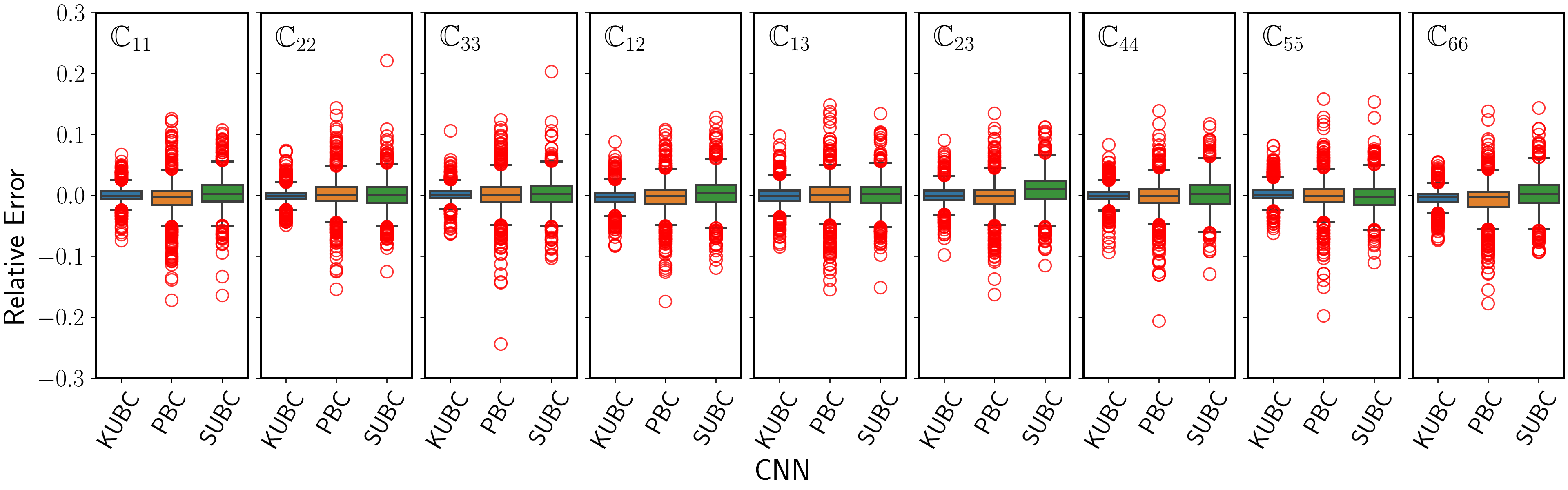}} 
		\\
		\subfloat[One single CNN for all BCs.] 
		{\includegraphics[width=15.0cm, angle=0, clip=]{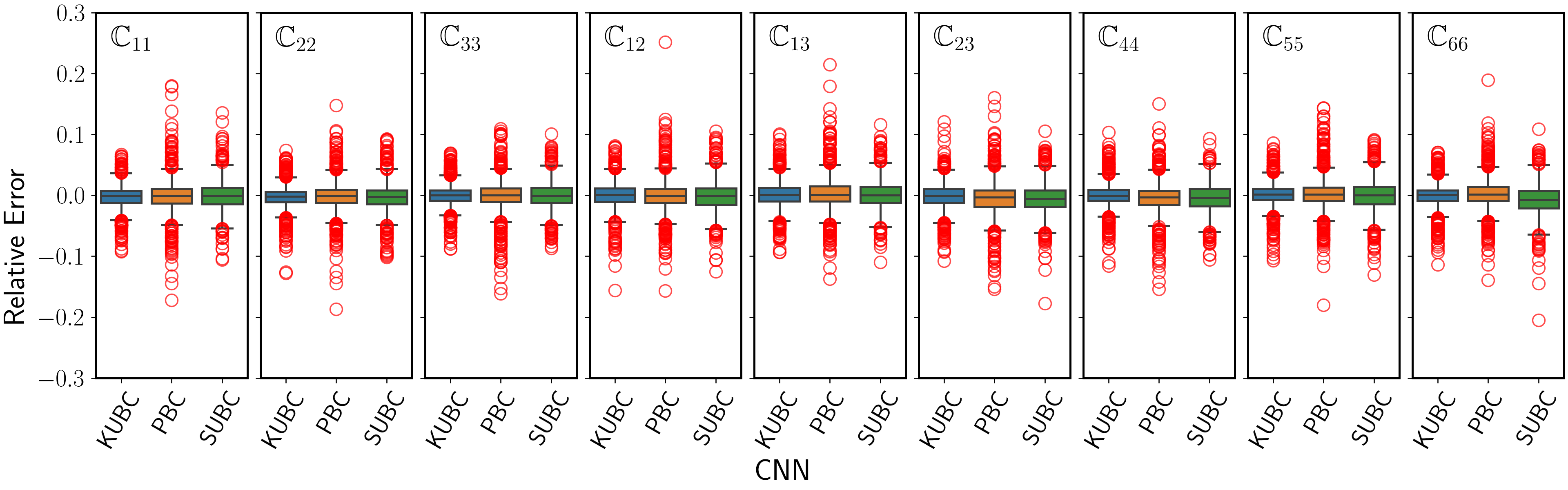}}
	\end{minipage}
	\caption{{\bf Boxplots.} Errors $e_{\text{rel}}$ for the test set with $M_{\mathcal{A}}=10^3$. \label{fig:CijComp-vs-CNN-4-AllBCs}}
\end{figure}

Relative errors according to
\begin{equation}
e_{\text{rel}}:= 
\left({C}_{ij}^{(\text{target})}-{C}_{ij}^{(\text{prediction})}\right)/ \, {C}_{ij}^{(\text{target})} 
\label{eq:err_rel}
\end{equation} 
are shown in the boxplots of Fig.~\ref{fig:CijComp-vs-CNN-4-AllBCs}; in (a) for the case of three distinct CNNs each for one BC, in (b) for one CNN covering all BCs. The accuracy is best for KUBC, lowest for SUBC, for PBC in between. The observed property of ''the stiffer, the more accurate'' manifests in the median, the upper and lower quartile defining the interquartile range IQR, and is equally visible by the whiskers marking 1.5 of the IQR. Outliers above 10\% are rare in either case. The accuracy of one CNN for all BCs is virtually indistinguishable from the separate treatment of each BC case. 

\begin{table}[htbp]
	\vspace*{-4mm}
	\begin{tabular}{| >{\centering\arraybackslash} m{2mm}  >{\centering\arraybackslash} m{7.2cm} >{\centering\arraybackslash} m{7.4cm}  |}  
		\hline 
		& &  \\[-2mm]
		& Three distinct CNNs       &  One CNN        \\[0mm]
		& &  \\[-4mm]
		\rotatebox[origin=c]{90}{KUBC}   & \includegraphics[width=7.2cm, angle=0, clip=]{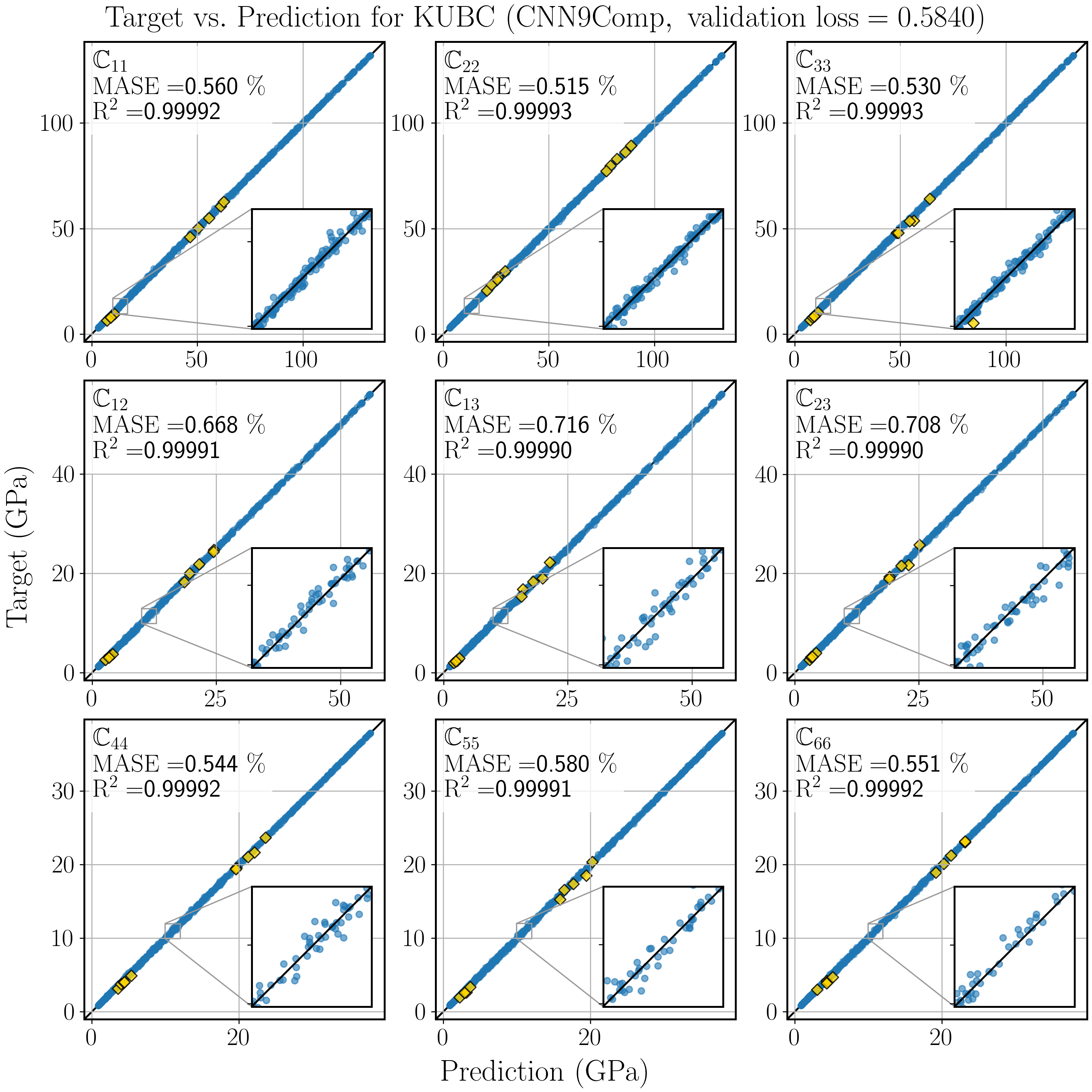}
		& \includegraphics[width=7.2cm, angle=0, clip=]{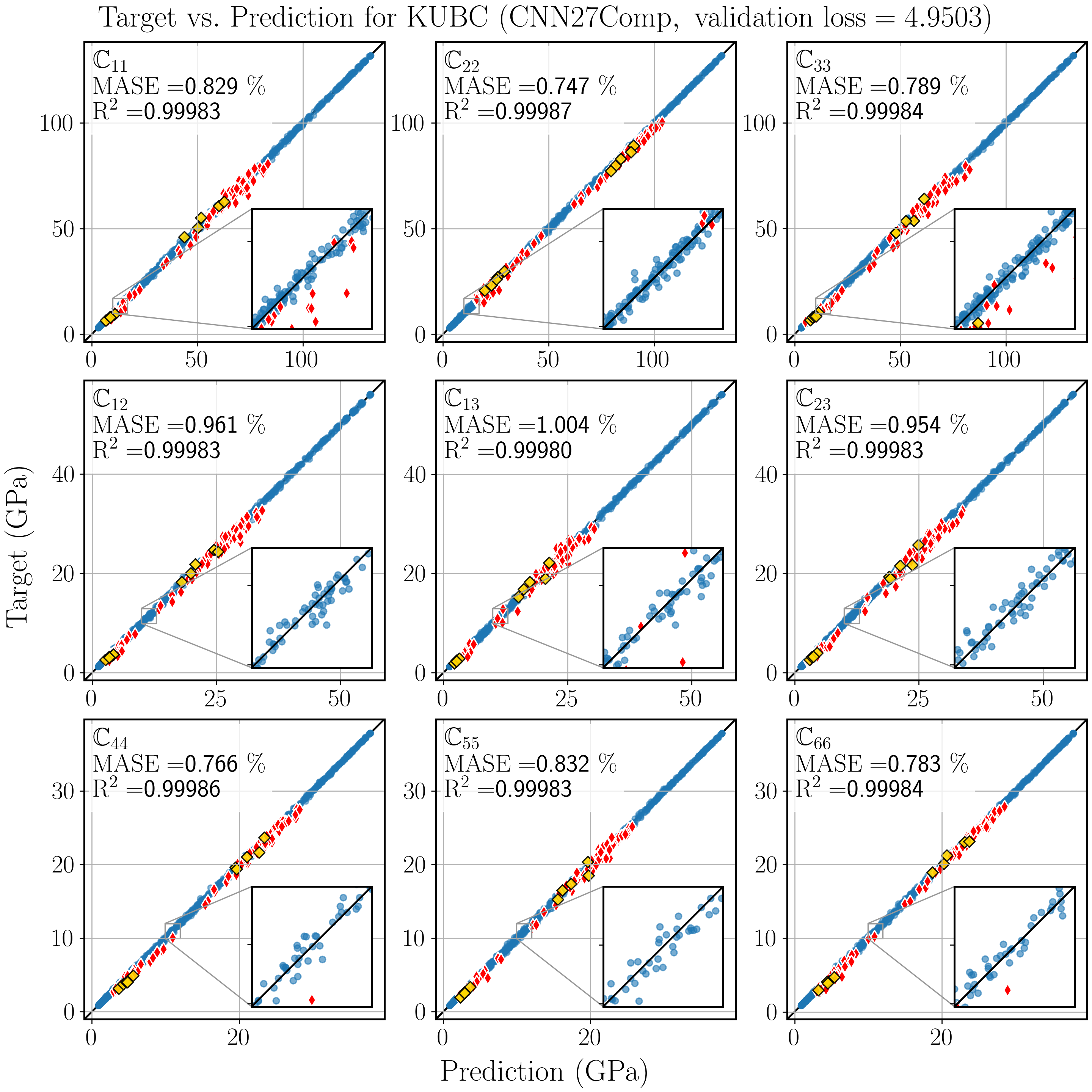} \\[1mm]
		& & \\[-4mm]
		\rotatebox[origin=c]{90}{PBC}  & \includegraphics[width=7.2cm, angle=0, clip=]{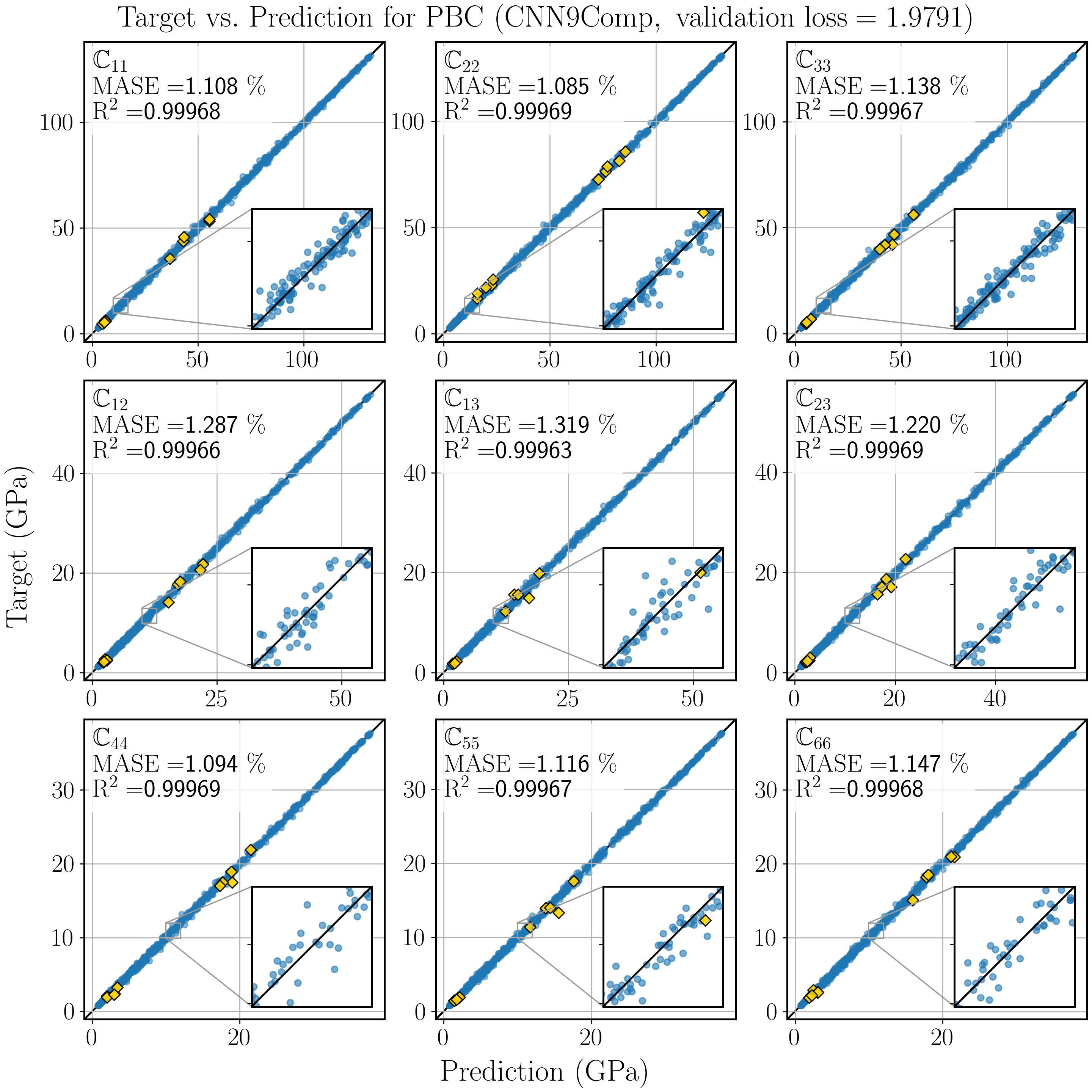}
		& \includegraphics[width=7.2cm, angle=0, clip=]{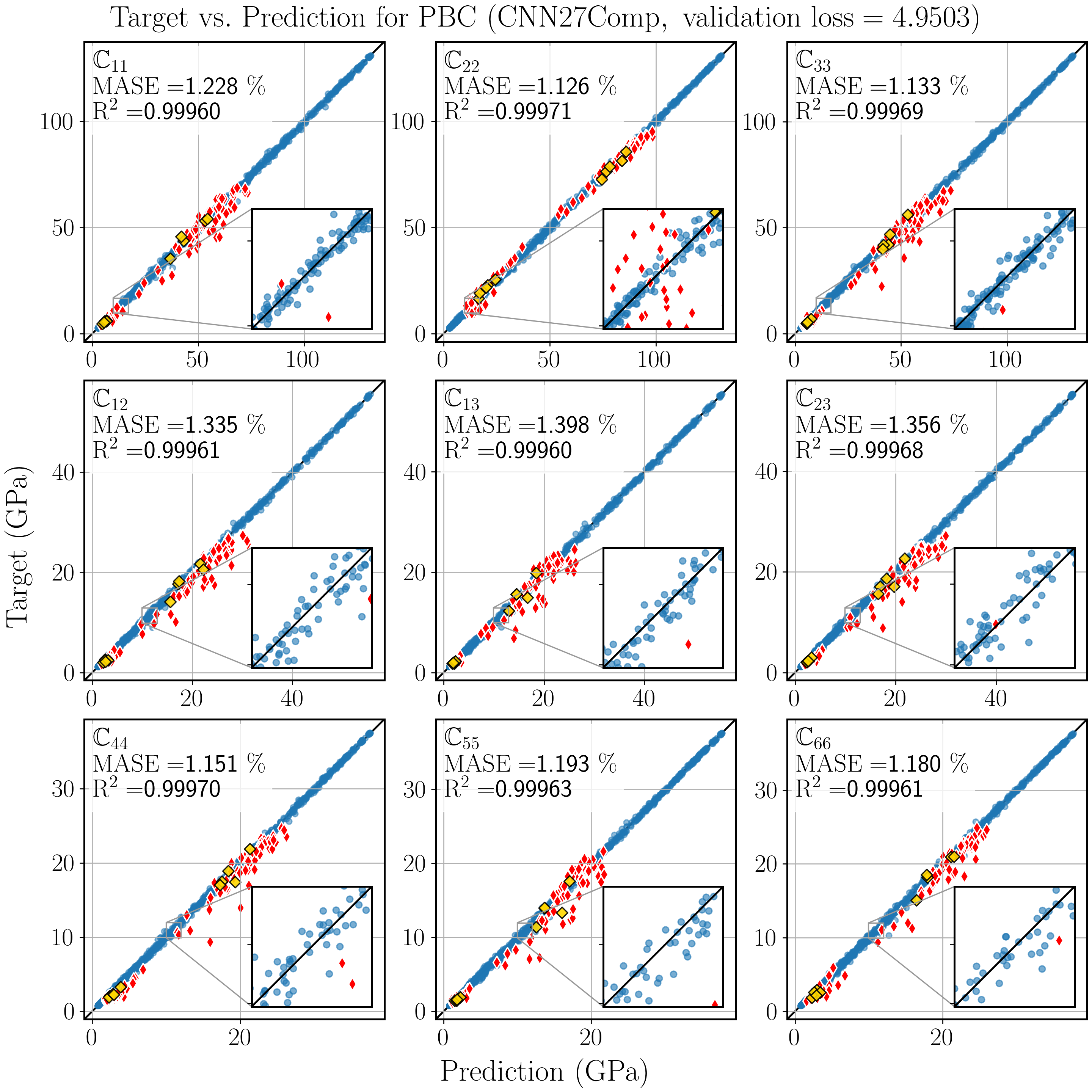} \\[1mm]
		& & \\[-4mm]
		\rotatebox[origin=c]{90}{SUBC} & \includegraphics[width=7.2cm, angle=0, clip=]{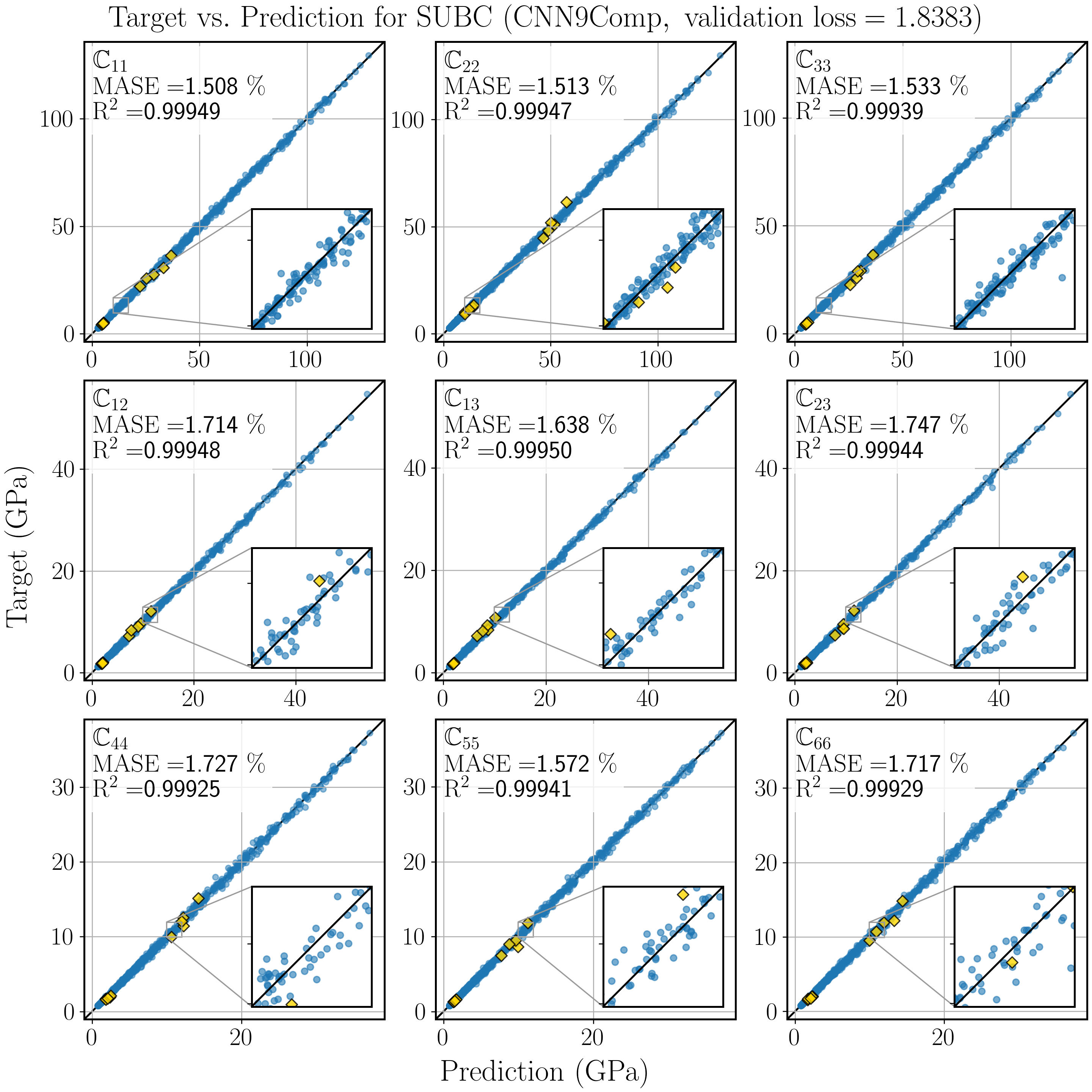}
		& \includegraphics[width=7.2cm, angle=0, clip=]{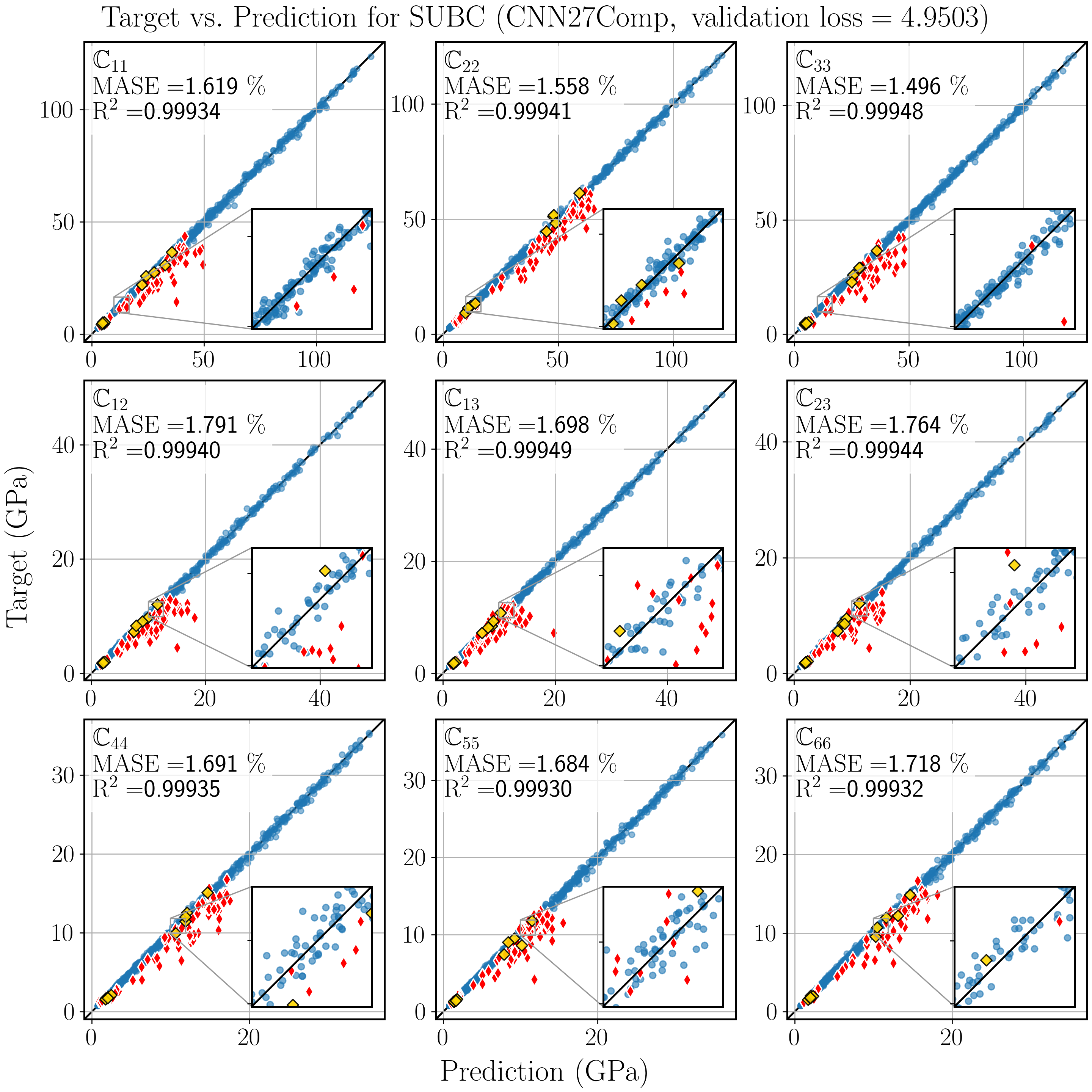} \\[1mm]
		\hline 		
	\end{tabular}
	\caption{{\bf Target versus CNN prediction for $C_{ij}$.} The results refer to three distinct CNNs (left column) and one CNN for all BCs (right column). The diamond/$\beta$-SiC results are marked by yellow diamond symbols for 200 voxels per edge, for 100 voxels in red.} \label{tab:Target-vs-CNN-3vs1-KUBCPBCSUBC-withDiaSiC}
\end{table}

\subsubsection{Target stiffness versus CNN prediction}

The results in Tab.~\ref{tab:Target-vs-CNN-3vs1-KUBCPBCSUBC-withDiaSiC} underpin the high accuracy of the CNN predictions for all BCs and both cases of one and three CNNs; it requires a zoom-in to recognize differences between target predicted values of ${C}_{ij}$. 

\section{Assessing the CNNs by Diamond/$\beta$-SiC Composite Samples}
\label{sec_DiamondSiC}
The CNN predictions for the virtual microstructures achieved high accuracy. Next, the real microstructure of a diamond/$\beta$-SiC thin film shall assess the potential to generalize.

\begin{figure}[htbp]
	\begin{minipage}{16.5cm}  
		\centering   
		\subfloat[Dark: $\beta$-SiC, bright: diamond.] 
		{\includegraphics[width=6.0cm, angle=0, clip=]{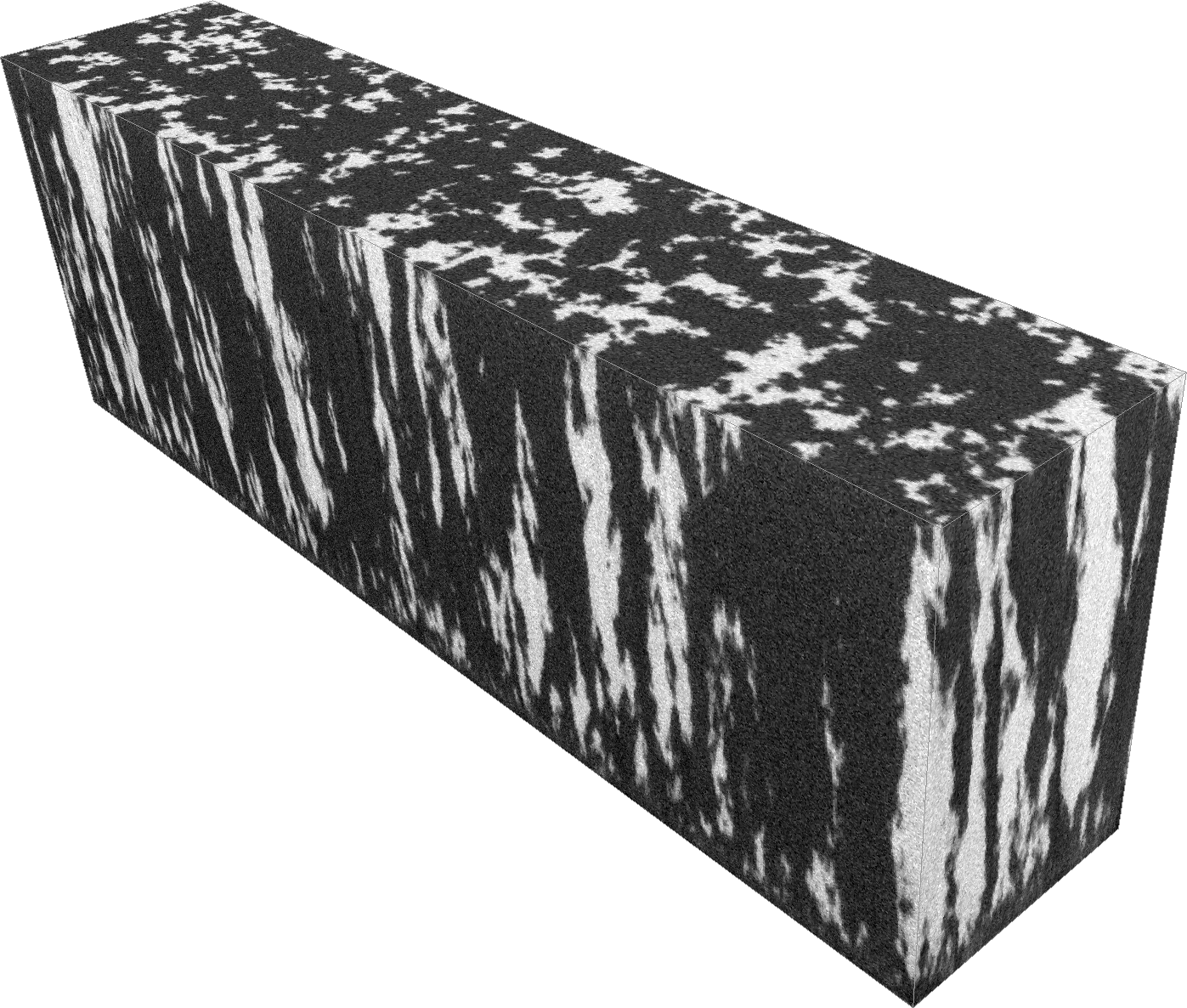}} 
		\subfloat[Diamond phase.] 
		{\includegraphics[width=6.0cm, angle=0, clip=]{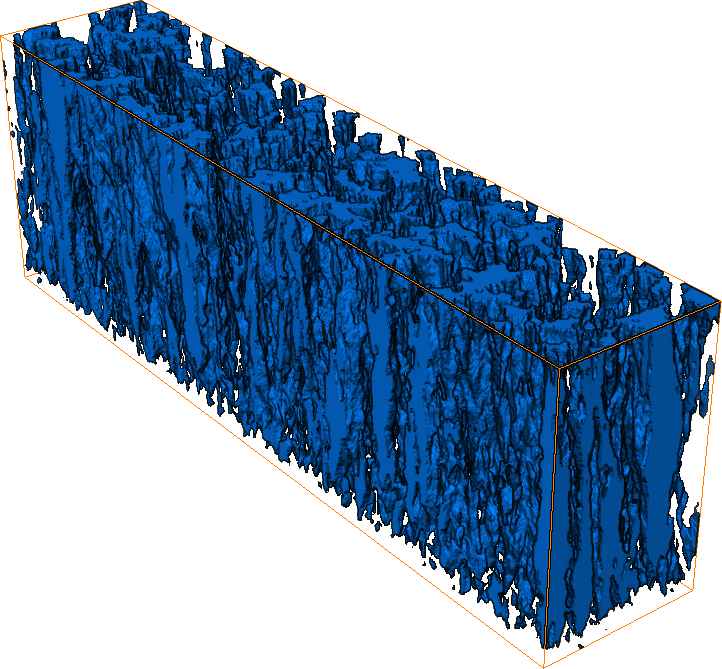}}  
		
		\subfloat[Cubic test samples of length 2~$\mu$m and resolution of 200$^3$ voxels.] 
		{\includegraphics[width=11cm, angle=0, clip=]{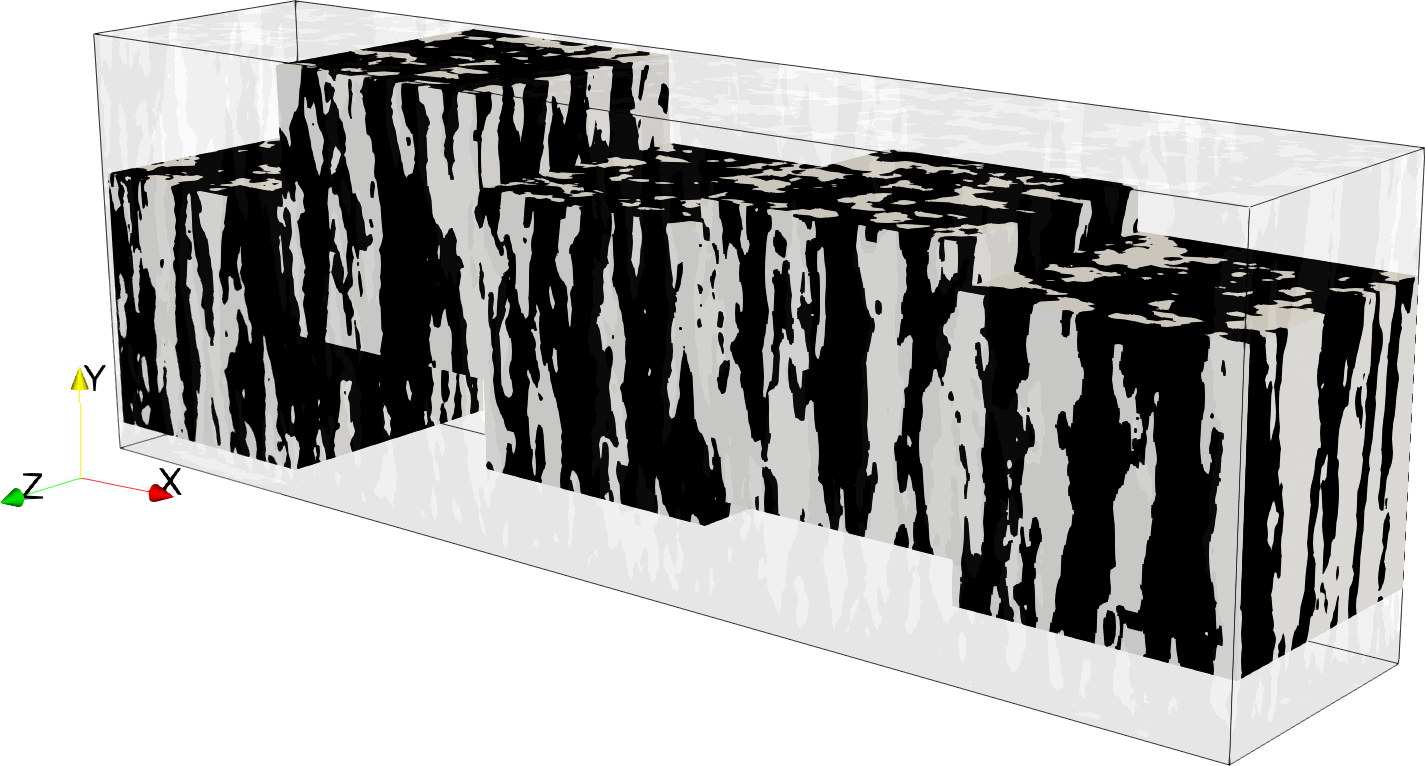}}
	\end{minipage}
	\caption{{\bf Diamond/$\beta$-SiC.} The specimen exhibits the length, height, and width of $10 \times 3.3 \times 2 \mu$m and is resolved by $1000 \times 330 \times 200$ voxels. Hence, a voxel exhibits the length of 10 nm. Notice in (a) and (b), that the bottom layer as the interface to the substrate is purely $\beta$-SiC. The coordinate system in (c) refers to (a) and (b) likewise. For the indices of the elastic moduli it holds $1 \leftarrow x$, $2 \leftarrow y$, $3 \leftarrow z$. \label{fig:Diamond-SiC}} 
\end{figure} 

Diamond thin films are of interest for their outstanding mechanical properties such as high hardness, low friction coefficient and high wear resistance. However, their application as protective coatings is inhibited by the poor adhesion on many substrates. The reason are high stresses at the interface of diamond to substrate, which are induced by different thermal expansion coefficients. An additional problem is a catalytic effect, which results in soot and graphite formation in the context of iron-, cobalt- and nickel-based materials. A solution provide nanocrystalline diamond/$\beta$-SiC composite films as transition layers; they are effective as adhesion layers and, serving as a barrier, they prevent the catalytic effect of the substrate elements. Diamond/$\beta$-SiC composites are made by chemical vapor deposition (CVD).  For synthesis, characterization and applications we refer to \cite{Zhuang.2010,Wang.2014,Yang.2015}. 

The test samples are obtained by cutting out cubes of edge length 100 and 200 voxels, respectively, which renders 60 and 5 non-intersecting VEs. By flipping the phase properties the number of samples is doubled. The 200-voxel samples are reduced to 100-voxel resolution in order to fit the CNN specifications. Figure \ref{fig:Diamond-SiC} (c) indicate that the 200$^3$ voxel cubes are cut out in some distance to the bottom of the CVP-grown specimen, since the bottom layer is made of pure $\beta$-SiC for the reasons mentioned.  

For the test, the diamond-SiC samples adopt the elastic properties of the phases as used in the training.  

\begin{figure}[htbp]
	\begin{minipage}{16.5cm}  
		\centering
		{\includegraphics[width=15.0cm, angle=0, clip=]{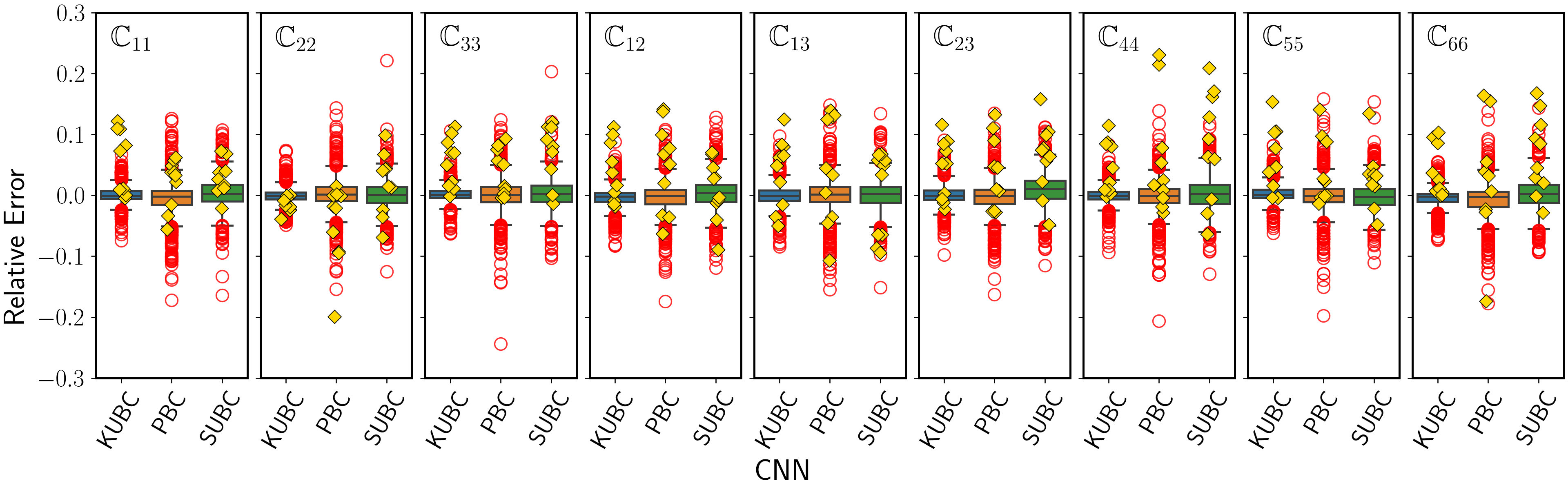}} 
	\end{minipage}
	\caption{{\bf Boxplots.} CNN errors $e_{\text{rel}}$ for the 200-voxel diamond/$\beta$-SiC samples in comparison to the standard test set. The results refer to three distinct CNNs.\label{fig:Boxchart-9-with-DiamondSiC}}
\end{figure} 

The results displayed in the left column of Tab.~\ref{tab:Target-vs-CNN-3vs1-KUBCPBCSUBC-withDiaSiC} indicate, 
that the CNN predictions for the diamond/$\beta$-SiC samples of the 200 voxel size (marked by yellow diamonds) are very accurate and thereby come close to the accuracy for the synthetic microstructures of the standard test set. More descriptive for the assessment are the boxcharts in Fig.~\ref{fig:Boxchart-9-with-DiamondSiC}.  
The predictions for the VEs of edge length 100 voxels (marked by red diamonds), which are displayed in the right column of Tab.~\ref{tab:Target-vs-CNN-3vs1-KUBCPBCSUBC-withDiaSiC}, are less accurate than for the 200 voxel case. Here, the predictions for the stiff KUBC case are much better than for the cases of PBC and SUBC. For the latter BCs the general trend to underestimate the true elastic stiffness for all components can be observed. An explanation for these findings is that the 100 voxel case exhibits very large phase domains with respect to the VE size, a case, which is hardly covered by the training set with variances having an upper limit of 8. 
 
If a particular materials system is envisaged, the approach with training data into the particular direction of that target material system is a reasonable choice as, e.g., in \cite{Li.2019} aiming at elastic predictions for shale. Here in contrast, the generation of VEs described in Sec.~\ref{sec:microstructure-generator} was not restricted to the type of CVP-grown, two-phase microstructures. Notwithstanding, the sample in Fig.~\ref{fig:microstructures} (g) resembles the real microstructure samples in Fig.~\ref{fig:Diamond-SiC} (c) in its parallel alignment of phases by the choice of corresponding variance parameters $s_y > s_x, s_z$. This type of microstructure induces elastic properties very close to transversal isotropy as shown for diamond/$\beta$-SiC in \cite{Eidel.2019}.  
  
\section{Discussion and Conclusions}
\label{sec:Conclusions}

In the present work, CNNs are constructed and trained to link 3D two-phase microstructures with arbitrary phase fractions to their elasticity tensors from homogenization with different BCs. The CNN predicts the apparent elastic stiffness for VE subject to PBC along with its lower bound by SUBC and its upper bound by KUBC in one thrust. Thereby it indicates, whether the VE is large enough to serve as an RVE. 

The pre-analysis of the randomly generated microstructures has revealed a considerable stiffness deviation for KUBC from SUBC for the majority of the generated VEs. Beyond the CNN-related focus of this paper the results underpin, that PBC for random, non-periodic heterogeneous matter imply an uncertainty of the true elastic stiffness and can lead to gross errors, if the VE is not sufficiently large. This is of general interest, but in particular for the design of digital twins of heterogeneous materials. 

The CNN predictions are in excellent agreement with the target of numerical homogenization for the standard test track. The CNN covering all three boundary conditions shows only very minor accuracy losses compared to three distinct CNNs. Remarkably, the CNN predictions turned out to be similarly accurate for real, CVP-grown diamond/$\beta$-SiC microstructures, a material of industrial relevance for its use as protective and biological coating. Since the CNNs were not trained into the particular direction of these microstructures, and learning was based merely on synthetic samples, the CNNs of this work can be seen as a step towards universal CNNs covering several classes of microstructures with a real-world impact. Similarly, more generality is achieved by arbitrary phase fractions of the composite.

The proposed CNNs, which embody the learned lessons about elastic homogenization in their parameters of weights and biases, can be used in two different types of applications. 
First, for real microstructure samples  obtained by tomography and image processing; since in that case of one or a few snapshots the statistics are missing, the prediction of the apparent stiffness and the necessary RVE size must be based on the PBC-results along with the KUBC- and SUBC-bounds. 
Second, the CNNs can be used for statistical analysis of an ensemble of (artificially generated) microstructures in order to determine effective properties following the approach of Kanit et al. \cite{Kanit.2003}.  
 
In either case of application, the predictions are fast and cheap, since they are independent of finite element simulations similar to the off-line stage of Reduced Basis Methods (RBM). Furthermore, the overall CNN-size of \cite{Yang.2018b} could be reduced by approximately 80\% by a size reduction of the FC layers.

Compared to the wealth of experience with the design and optimization of CNNs in image identification and classification, homogenization-CNNs are still in their infancy.
   
\bigskip

{\bf Acknowledgements.} The author acknowledges support by the Deutsche Forschungsgemeinschaft (DFG) within the Heisenberg program (grant no. EI 453/5-1). Simulations were performed with computing resources granted by RWTH Aachen University under project ID BUND0005 and by the University of Siegen. The author thanks Prof. Xin Jiang for providing the diamond-SiC sample and Dr. Lorenz Holzer for images thereof by FIB-tomography from another joint project. Assistance of Ajinkya Gote and Mohamed Imran Peer Mohamed in simulations and postprocessing is gratefully acknowledged.


\begin{appendix}

\addcontentsline{toc}{section}{Appendix}
\renewcommand{\thesubsection}{\Alph{section}.\arabic{subsection}}
\renewcommand{\theequation}{\Alph{section}.\arabic{equation}}
\renewcommand{\thefigure}{\Alph{section}.\arabic{figure}}
\renewcommand{\thetable}{\Alph{section}.\arabic{table}}
\newcommand {\ssectapp}{
                        \setcounter{equation}{0}
                        \setcounter{figure}{0}
                        \setcounter{table}{0}
		                \subsection
                        }

\setcounter{equation}{0}


\end{appendix}

\bibliographystyle{abbrv} 
\bibliography{octree,machlearn,diamondSiC}

\begin{thebibliography}{10}

\bibitem{Abdulle.2006}
A.~Abdulle.
\newblock {Analysis of the {H}eterogeneous {M}ultiscale {FEM} for problems in
  elasticity}.
\newblock {\em {Mathematical Models and Methods in Applied Sciences}},
  16(04):615--635, 2006.

\bibitem{Abdulle.2009b}
A.~Abdulle.
\newblock {The {Finite Element Heterogeneous Multiscale Method}: A
  computational strategy for multiscale {PDE}s}.
\newblock {\em {GAKUTO Int. Ser. Math. Sci. Appl.}}, 31:133--181, 2009.

\bibitem{Abdulle.2012}
A.~Abdulle, W.~E, B.~Engquist, and E.~Vanden-Eijnden.
\newblock {The {Heterogeneous Multiscale Method}}.
\newblock {\em {Acta Numerica}}, 21:1--87, 2012.

\bibitem{Abdulle.2005}
A.~Abdulle and C.~Schwab.
\newblock {{Heterogeneous Multiscale FEM} for Diffusion Problems on Rough
  Surfaces}.
\newblock {\em {Multiscale Modeling {\&} Simulation}}, 3(1):195--220, 2005.

\bibitem{Andra.2013}
H.~Andr{\"a}, N.~Combaret, J.~Dvorkin, E.~Glatt, J.~Han, M.~Kabel, Y.~Keehm,
  F.~Krzikalla, M.~Lee, C.~Madonna, M.~Marsh, T.~Mukerji, E.~H. Saenger,
  R.~Sain, N.~Saxena, S.~Ricker, A.~Wiegmann, and X.~Zhan.
\newblock {Digital rock physics benchmarks---{Part II}: Computing effective
  properties}.
\newblock {\em {Computers {\&} Geosciences}}, 50:33--43, 2013.

\bibitem{Berner.2021}
J.~Berner, P.~Grohs, G.~Kutyniok, and P.~Petersen.
\newblock {The Modern Mathematics of Deep Learning: arXiv preprint
  arXiv:2105.04026}, 2021.

\bibitem{Breuer.2021}
K.~Breuer and M.~Stommel.
\newblock {Prediction of Short Fiber Composite Properties by an Artificial
  Neural Network Trained on an RVE Database}.
\newblock {\em {Fibers}}, 9(2):8, 2021.

\bibitem{Cucker.2007}
F.~Cucker and D.-X. Zhou.
\newblock {\em {Learning theory: An approximation theory viewpoint}}, volume
  v.24 of {\em {Cambridge Monographs on Applied and Computational
  Mathematics}}.
\newblock {Cambridge University Press}, Cambridge, 2007.

\bibitem{Drugan.1996}
W.~J. Drugan and J.~R. Willis.
\newblock {A micromechanics-based nonlocal constitutive equation and estimates
  of representative volume element size for elastic composites}.
\newblock {\em {Journal of the Mechanics and Physics of Solids}},
  44(4):497--524, 1996.

\bibitem{E.2005}
W.~E, P.~Ming, and P.~Zhang.
\newblock {Analysis of the {Heterogeneous Multiscale Method} for elliptic
  homogenization problems}.
\newblock {\em {Journal of the American Mathematical Society}},
  18(01):121--157, 2005.

\bibitem{Eidel.2018}
B.~Eidel and A.~Fischer.
\newblock {The heterogeneous multiscale finite element method for the
  homogenization of linear elastic solids and a comparison with the {FE$^2$}
  method}.
\newblock {\em {Computer Methods in Applied Mechanics and Engineering}},
  329:332--368, 2018.

\bibitem{Eidel.2021}
B.~Eidel, A.~Fischer, and A.~Gote.
\newblock {From image data towards microstructure information -- Accuracy
  analysis at the digital core of materials}.
\newblock {\em {ZAMM}}, 101(6), 2021.

\bibitem{Eidel.2019}
B.~Eidel, A.~Gote, M.~Ruby, L.~Holzer, L.~Keller, and X.~Jiang.
\newblock {Estimating the effective elasticity properties of a diamond/
  {$\beta$}-{S}i{C} composite thin film by {3D} reconstruction and numerical
  homogenization}.
\newblock {\em {Diamond and Related Materials}}, 97:107406, 2019.

\bibitem{Feyel.2000}
F.~Feyel and J.-L. Chaboche.
\newblock {{FE$^2$} multiscale approach for modelling the elastoviscoplastic
  behaviour of long fibre {S}i{C}/{T}i composite materials}.
\newblock {\em {Computer Methods in Applied Mechanics and Engineering}},
  183(3-4):309--330, 2000.

\bibitem{Fischer.2019b}
A.~Fischer and B.~Eidel.
\newblock {Convergence and error analysis of FE-HMM/{FE$^2$} for energetically
  consistent micro-coupling conditions in linear elastic solids}.
\newblock {\em {European Journal of Mechanics - A/Solids}}, 77:103735, 2019.

\bibitem{Fischer.2020}
A.~Fischer and B.~Eidel.
\newblock {Error analysis for quadtree-type mesh coarsening algorithms adapted
  to pixelized heterogeneous microstructures}.
\newblock {\em {Computational Mechanics}}, 16(04):615, 2020.

\bibitem{Gonzalez.2018}
R.~C. Gonzalez and R.~E. Woods.
\newblock {\em {Digital image processing}}.
\newblock Pearson, New York NY, 2018.

\bibitem{Goodfellow.2016}
I.~Goodfellow, Y.~Bengio, and A.~Courville.
\newblock {\em {Deep Learning}}.
\newblock {MIT Press}, 2016.

\bibitem{Gote.2021}
A.~Gote, A.~Fischer, C.~Zhang, and B.~Eidel.
\newblock {Computational Homogenization of Concrete in the Cyber
  Size-Resolution-Discretization (SRD) Parameter Space: arXiv preprint
  arXiv:2103.08957}, 2021.

\bibitem{Graczyk.2020}
K.~M. Graczyk and M.~Matyka.
\newblock {Predicting porosity, permeability, and tortuosity of porous media
  from images by deep learning}.
\newblock {\em {Scientific reports}}, 10(1):21488, 2020.

\bibitem{Higham.2019}
C.~F. Higham and D.~J. Higham.
\newblock {Deep Learning: An Introduction for Applied Mathematicians}.
\newblock {\em {SIAM Review}}, 61(3):860--891, 2019.

\bibitem{Hill.1963}
R.~Hill.
\newblock {Elastic properties of reinforced solids: Some theoretical
  principles}.
\newblock {\em {Journal of the Mechanics and Physics of Solids}},
  11(5):357--372, 1963.

\bibitem{Huet.1990}
C.~Huet.
\newblock {Application of variational concepts to size effects in elastic
  heterogeneous bodies}.
\newblock {\em {Journal of the Mechanics and Physics of Solids}},
  38(6):813--841, 1990.

\bibitem{Jeulin.2000}
D.~Jeulin.
\newblock {Random texture models for material structures}.
\newblock {\em {Statistics and Computing}}, 10(2):121--132, 2000.

\bibitem{Kamrava.2020}
S.~Kamrava, P.~Tahmasebi, and M.~Sahimi.
\newblock {Linking Morphology of Porous Media to Their Macroscopic Permeability
  by Deep Learning}.
\newblock {\em {Transport in Porous Media}}, 131(2):427--448, 2020.

\bibitem{Kanit.2003}
T.~Kanit, S.~Forest, I.~Galliet, V.~Mounoury, and D.~Jeulin.
\newblock {Determination of the size of the representative volume element for
  random composites: statistical and numerical approach}.
\newblock {\em {International Journal of Solids and Structures}},
  40(13-14):3647--3679, 2003.

\bibitem{Keskar.2017}
N.~S. Keskar, D.~Mudigere, J.~Nocedal, M.~Smelyanskiy, and P.~T.~P. Tang.
\newblock {On Large-Batch Training for Deep Learning: Generalization Gap and
  Sharp Minima: arXiv-preprint arXiv:1609.04836}, 2017.

\bibitem{Khan.2020}
A.~Khan, A.~Sohail, U.~Zahoora, and A.~S. Qureshi.
\newblock {A survey of the recent architectures of deep convolutional neural
  networks}.
\newblock {\em {Artificial Intelligence Review}}, 53(8):5455--5516, 2020.

\bibitem{Kingma.2014}
D.~P. Kingma and J.~Ba.
\newblock {Adam: A Method for Stochastic Optimization: arXiv preprint
  arXiv:1412.6980}, 2014.

\bibitem{Kondo.2017}
R.~Kondo, S.~Yamakawa, Y.~Masuoka, S.~Tajima, and R.~Asahi.
\newblock {Microstructure recognition using convolutional neural networks for
  prediction of ionic conductivity in ceramics}.
\newblock {\em {Acta Materialia}}, 141:29--38, 2017.

\bibitem{Kouznetsova.2001}
V.~Kouznetsova, W.~A.~M. Brekelmans, and F.~P.~T. Baaijens.
\newblock {An approach to micro-macro modeling of heterogeneous materials}.
\newblock {\em {Computational Mechanics}}, 27(1):37--48, 2001.

\bibitem{Krizhevsky.2012}
A.~Krizhevsky, I.~Sutskever, and G.~E. Hinton.
\newblock {ImageNet Classification with Deep Convolutional Neural Networks}.
\newblock In F.~Pereira, C.~J.~C. Burges, L.~Bottou, and K.~Q. Weinberger,
  editors, {\em {Advances in Neural Information Processing Systems 25}}, pages
  1097--1105. {Curran Associates, Inc}, 2012.

\bibitem{LeCun.1989}
Y.~LeCun, B.~Boser, J.~S. Denker, D.~Henderson, R.~E. Howard, W.~Hubbard, and
  L.~D. Jackel.
\newblock {Backpropagation Applied to Handwritten Zip Code Recognition}.
\newblock {\em {Neural Computation}}, 1:541--551, 1989.

\bibitem{Li.2019}
X.~Li, Z.~Liu, S.~Cui, C.~Luo, C.~Li, and Z.~Zhuang.
\newblock {Predicting the effective mechanical property of heterogeneous
  materials by image based modeling and deep learning}.
\newblock {\em {Computer Methods in Applied Mechanics and Engineering}},
  347:735--753, 2019.

\bibitem{Michel.1999}
J.~C. Michel, H.~Moulinec, and P.~Suquet.
\newblock {Effective properties of composite materials with periodic
  microstructure: a computational approach}.
\newblock {\em {Computer Methods in Applied Mechanics and Engineering}},
  172(1-4):109--143, 1999.

\bibitem{Miehe.1999}
C.~Miehe, J.~Schr{\"o}der, and J.~Schotte.
\newblock {Computational homogenization analysis in finite plasticity
  simulation of texture development in polycrystalline materials}.
\newblock {\em {Computer Methods in Applied Mechanics and Engineering}},
  171(3-4):387--418, 1999.

\bibitem{Peric.2011}
D.~Peri{\'c}, E.~A. de~{Souza Neto}, R.~A. Feij{\'o}o, M.~Partovi, and A.~J.~C.
  Molina.
\newblock {On micro-to-macro transitions for multi-scale analysis of non-linear
  heterogeneous materials: unified variational basis and finite element
  implementation}.
\newblock {\em {International Journal for Numerical Methods in Engineering}},
  87(1-5):149--170, 2011.

\bibitem{Rao.2020}
C.~Rao and Y.~Liu.
\newblock {Three-dimensional convolutional neural network (3D-CNN) for
  heterogeneous material homogenization}.
\newblock {\em {Computational Materials Science}}, 184:109850, 2020.

\bibitem{Rumelhart.1986}
D.~E. Rumelhart, G.~E. Hinton, and R.~J. Williams.
\newblock {Learning Internal Representations by Error Propagation}.
\newblock In D.~E. Rumelhart and J.~L. Mcclelland, editors, {\em {Parallel
  Distributed Processing: Explorations in the Microstructure of Cognition,
  Volume 1: Foundations}}, pages 318--362. {MIT Press}, Cambridge, MA, 1986.

\bibitem{Saeb.2016}
S.~Saeb, P.~Steinmann, and A.~Javili.
\newblock {Aspects of Computational Homogenization at Finite Deformations: A
  Unifying Review From {R}euss' to {V}oigt's Bound}.
\newblock {\em {Applied Mechanics Reviews}}, 68(5):050801, 2016.

\bibitem{Schmidhuber.2015}
J.~Schmidhuber.
\newblock {Deep Learning in Neural Networks: An Overview}.
\newblock {\em {Neural Networks}}, 61(3):85--117, 2015.

\bibitem{Schroder.2014}
J.~Schr{\"o}der.
\newblock {A numerical two-scale homogenization scheme: the {FE$^2$}-method}.
\newblock In F.~Pfeiffer, F.~G. Rammerstorfer, E.~Guazzelli, B.~Schrefler,
  P.~Serafini, J.~Schr{\"o}der, and K.~Hackl, editors, {\em {Plasticity and
  Beyond}}, volume 550 of {\em {CISM International Centre for Mechanical
  Sciences}}, pages 1--64. {Springer Vienna}, Vienna, 2014.

\bibitem{Tian.2020}
J.~Tian, C.~Qi, Y.~Sun, and Z.~M. Yaseen.
\newblock {Surrogate permeability modelling of low-permeable rocks using
  convolutional neural networks}.
\newblock {\em {Computer Methods in Applied Mechanics and Engineering}},
  366:113103, 2020.

\bibitem{Ting.1996}
T.~C.~T. Ting.
\newblock {\em {Anisotropic Elasticity: Theory and Applications}}.
\newblock {Oxford University Press, USA}, 1996.

\bibitem{Vannucci.2018}
P.~Vannucci.
\newblock {\em {Anisotropic Elasticity}}, volume~85.
\newblock {Springer Singapore}, Singapore, 2018.

\bibitem{Wang.2014}
T.~Wang, S.~Handschuh-Wang, Y.~Yang, H.~Zhuang, C.~Schlemper, D.~Wesner,
  H.~Sch{\"o}nherr, W.~Zhang, and X.~Jiang.
\newblock {Controlled surface chemistry of {diamond/$\beta$}-SiC composite
  films for preferential protein adsorption}.
\newblock {\em {Langmuir : the ACS journal of surfaces and colloids}},
  30(4):1089--1099, 2014.

\bibitem{Wu.2019}
H.~Wu, W.-Z. Fang, Q.~Kang, W.-Q. Tao, and R.~Qiao.
\newblock {Predicting Effective Diffusivity of Porous Media from Images by Deep
  Learning}.
\newblock {\em {Scientific reports}}, 9(1):20387, 2019.

\bibitem{Wu.2018}
J.~Wu, X.~Yin, and H.~Xiao.
\newblock {Seeing permeability from images: fast prediction with convolutional
  neural networks}.
\newblock {\em {Science Bulletin}}, 63(18):1215--1222, 2018.

\bibitem{Yang.2019}
C.~Yang, Y.~Kim, S.~Ryu, and G.~X. Gu.
\newblock {Using convolutional neural networks to predict composite properties
  beyond the elastic limit}.
\newblock {\em {MRS Communications}}, 9(2):609--617, 2019.

\bibitem{Yang.2015}
N.~Yang, editor.
\newblock {\em {Novel Aspects of Diamond}}.
\newblock {Topics in Applied Physics}. {Springer International Publishing},
  Cham, 2015.

\bibitem{Yang.2018}
Z.~Yang, X.~Li, L.~{Catherine Brinson}, A.~N. Choudhary, W.~Chen, and
  A.~Agrawal.
\newblock {Microstructural Materials Design Via Deep Adversarial Learning
  Methodology}.
\newblock {\em {Journal of Mechanical Design}}, 140(11), 2018.

\bibitem{Yang.2018b}
Z.~Yang, Y.~C. Yabansu, R.~Al-Bahrani, W.-k. Liao, A.~N. Choudhary, S.~R.
  Kalidindi, and A.~Agrawal.
\newblock {Deep learning approaches for mining structure-property linkages in
  high contrast composites from simulation datasets}.
\newblock {\em {Computational Materials Science}}, 151:278--287, 2018.

\bibitem{Zeiler.2014}
M.~D. Zeiler and R.~Fergus.
\newblock {Visualizing and Understanding Convolutional Networks}.
\newblock In D.~Fleet, T.~Pajdla, B.~Schiele, and T.~Tuytelaars, editors, {\em
  {Computer Vision -- ECCV 2014}}, volume 8689 of {\em {Lecture Notes in
  Computer Science}}, pages 818--833. {Springer International Publishing},
  Cham, 2014.

\bibitem{Zhuang.2010}
H.~Zhuang, B.~Song, V.~V. S.~S. Srikanth, X.~Jiang, and H.~Sch{\"o}nherr.
\newblock {Controlled Wettability of {Diamond/$\beta$}-SiC Composite Thin Films
  for Biosensoric Applications}.
\newblock {\em {The Journal of Physical Chemistry C}}, 114(47):20207--20212,
  2010.

\end{thebibliography}
\begin{appendix}
\end{appendix}
\end{document}